\documentclass[12pt,letterpaper]{article}

\usepackage{mystyle}

\newcounter{small_constant}
\newcommand{\nextsc}{\addtocounter{small_constant}{1}  c_{\arabic{small_constant}} }
\newcommand{\nextscnu}{c_{\arabic{small_constant}}}

\begin{document}

\title{Secure Degrees of Freedom Regions of Multiple Access and Interference Channels: The Polytope Structure\thanks{This work was supported by NSF Grants CNS 09-64632, CCF 09-64645, CCF 10-18185 and CNS 11-47811, and presented in part at the Asilomar Conference on Signals, Systems and Computers, Pacific Grove, CA, November 2013.}}

\author{Jianwei Xie \qquad Sennur Ulukus\\
\normalsize Department of Electrical and Computer Engineering\\
\normalsize University of Maryland, College Park, MD 20742 \\
\normalsize {\it xiejw@umd.edu} \qquad {\it ulukus@umd.edu}}

\maketitle

\vspace{-0.8cm}

\begin{abstract}
The sum secure degrees of freedom (s.d.o.f.) of two fundamental multi-user network structures, the $K$-user Gaussian multiple access (MAC) wiretap channel and the $K$-user interference channel (IC) with secrecy constraints, have been  determined  recently as $\frac{K(K-1)}{K(K-1)+1}$ \cite{xie_ulukus_isit_2013_mac, xie_sdof_networks_in_prepare} and $\frac{K(K-1)}{2K-1}$ \cite{xie_ulukus_isit_2013_kic, xie_unified_kic},
respectively. In this  paper, we determine the \emph{entire \sdof regions} of these two channel models. The converse for the MAC follows from a middle step in the converse of \cite{xie_ulukus_isit_2013_mac, xie_sdof_networks_in_prepare}. The converse for the IC includes constraints both due to secrecy as well as due to interference.
Although the portion of the region close to the optimum sum \sdof point is governed by the upper bounds due to  secrecy constraints, the
 other portions of the region are  governed by the upper bounds  due to   interference  constraints.
Different from the existing literature, in order to fully understand the characterization of the \sdof region of the IC, one has to study the $4$-user case, i.e., the  $2$ or $3$-user cases do not illustrate the generality of the problem. In order to prove the achievability, we use the polytope structure of the converse region. In both MAC and IC cases, we develop explicit schemes that achieve the extreme points of the polytope region given by the converse. Specifically, the extreme points of the MAC region are achieved by an $m$-user MAC wiretap channel with $K-m$ helpers, i.e., by setting $K-m$ users' secure rates to zero and utilizing them as pure (structured) cooperative jammers. The extreme points of the IC region are achieved by a $(K-m)$-user IC with confidential messages, $m$ helpers, and $N$ external eavesdroppers, for $m\ge1$ and a finite $N$. A byproduct of our results in this paper is that the sum \sdof is achieved \emph{only at one extreme point} of the \sdof region, which is the symmetric-rate extreme point, for both MAC and IC channel models.
\end{abstract}

\newpage

\section{Introduction}

In this paper, we consider two fundamental multi-user network structures under secrecy constraints: $K$-user multiple access channel (MAC) and $K$-user interference channel (IC). Information-theoretic security of communication was first considered by Shannon in \cite{Shannon:1949} via a noiseless wiretap channel. Noisy wiretap channel was introduced by Wyner who showed that information-theoretically secure communication was possible if the eavesdropper was degraded with respect to the legitimate receiver \cite{wyner}. Csiszar and Korner generalized Wyner's result to arbitrary, not necessarily degraded, wiretap channels, and showed that information-theoretically secure communication was possible even when the eavesdropper was not degraded \cite{csiszar}. Leung-Yan-Cheong and Hellman extended Wyner's setting to a Gaussian channel, which is degraded \cite{gaussian}. This line of research has been extended to many multi-user scenarios, for both general and Gaussian channel models, see e.g., \cite{secrecy_ic, xu_bounds_bc_cm_it_09, fading1, bagherikaram_bc_2008, ersen_bc_asilomar_08, ersen_eurasip_2009, he_outerbound_gic_cm_ciss_09, koyluoglu_ic_external,tekin_gmac_w, cooperative_jamming, ersen_mac_allerton_08, liang_mac_cm_08, ersen_mac_book_chapter, wiretap_channel_with_one_helper, relay_1, relay_2, relay_3, relay_4, he_untrusted_relay, ersen_crbc_2011, compound_wiretap_channel, ersen_ulukus_degraded_compound}. The secrecy capacity regions of most of these multi-user channels remain open problems even in simple Gaussian settings. In the absence of exact secrecy capacity regions, the behaviour of the secrecy rates at high signal-to-noise ratio (SNR) regimes have been studied by focusing on the secure degrees of freedom (s.d.o.f.), which is the pre-log of the secrecy rates, in \cite{he_k_gic_cm_09, koyluoglu_k_user_gic_secrecy, xie_k_user_ia_compound, secrecy_ia_new, xiang_he_thesis, secrecy_ia5,raef_mac_it_12, secrecy_ia1, interference_alignment_compound_channel, xie_gwch_allerton, xie_sdof_networks_in_prepare, xie_blind_cj_ciss_2013, xie_layered_network_journal, xie_unified_kic, xie_ulukus_isit_2013_kic, xie_ulukus_isit_2013_mac, khisti_arti_noise_alignment, mohamed_yener_allerton_2013, mohamed_yener_globalsip_2013}. In this paper, we focus on the $K$-user Gaussian MAC wiretap channel and the $K$-user Gaussian IC with secrecy constraints. The secrecy capacity regions of both of these models remain open. The sum s.d.o.f.~of both of these models have been determined recently as $\frac{K(K-1)}{K(K-1)+1}$ \cite{xie_ulukus_isit_2013_mac, xie_sdof_networks_in_prepare} and $\frac{K(K-1)}{2K-1}$ \cite{xie_ulukus_isit_2013_kic, xie_unified_kic}, respectively. In this paper, we determine the \emph{entire s.d.o.f.~regions} of these channel models.

\begin{figure}[t]
\centerline{\includegraphics[scale=0.7]{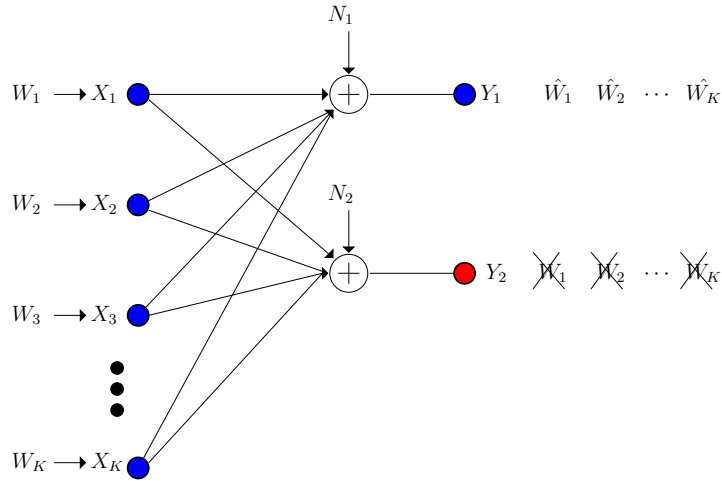}}
\caption{$K$-user multiple access (MAC) wiretap channel.}
\label{fig:sdofregion:mac_k}
\end{figure}

We start with the MAC wiretap channel, where multiple legitimate transmitters wish to have secure communication with a legitimate receiver in the presence of an eavesdropper; see Figure~\ref{fig:sdofregion:mac_k}. The converse for the sum \sdof is developed in \cite{xie_ulukus_isit_2013_mac, xie_sdof_networks_in_prepare} using two lemmas: the \emph{secrecy penalty} lemma and the \emph{role of a helper} lemma, which, respectively, quantify the rate penalty due to the existence of an eavesdropper, and quantify the impact of a helper (interferer) on the rate of another legitimate transmitter. The achievability for the sum \sdof in \cite{xie_ulukus_isit_2013_mac, xie_sdof_networks_in_prepare} is based on real interference alignment \cite{real_inter_align_exploit, real_inter_align} and structured cooperative jamming \cite{cooperative_jamming} with an emphasis on simultaneous alignments at both the legitimate receiver and the eavesdropper. We develop the converse for the \emph{entire region} by starting from a middle step in the converse proof of \cite{xie_ulukus_isit_2013_mac, xie_sdof_networks_in_prepare}. While \cite{xie_ulukus_isit_2013_mac, xie_sdof_networks_in_prepare} developed asymmetric upper bounds for the secure rates, since the sum \sdof was achieved by symmetric rates, \cite{xie_ulukus_isit_2013_mac, xie_sdof_networks_in_prepare} summed up the asymmetric upper bounds to get a single symmetric upper bound to match the achievability.  We revisit the converse proof in \cite{xie_ulukus_isit_2013_mac, xie_sdof_networks_in_prepare} and develop a converse for the entire region by keeping the developed asymmetric upper bounds. Therefore, the converse proofs developed in \cite{xie_ulukus_isit_2013_mac, xie_sdof_networks_in_prepare} to obtain a converse for the sum \sdof suffice
to obtain a tight converse for the entire region.

The converse region for the \sdof problem has a general \emph{polytope} structure, as opposed to the non-secrecy counterpart for the MAC which has a \emph{polymatroid} structure \cite{tse_polymatroid}. Polytope is a bounded polyhedron, which is an intersection of a finite number of half-spaces. Such definition is called a half-space representation, which is exactly the way our converse is expressed. In order to show the achievability of the polytope region, we need to show the achievability of  boundaries of all of the half-spaces, which is inefficient. We use Minkowski theorem \cite[Theorem 2.4.5]{convex_polytopes} which states that the polytope region discussed in this paper can be represented by the convex hull of all of its extreme points, which there are only finitely many. We, therefore, first determine the extreme points of this converse (polytope) region, and then develop an achievable scheme for each extreme point of the converse region; the achievability of the entire region then follows from time-sharing. In particular, each extreme point of the converse region is achieved by an $m$-user MAC wiretap channel with $K-m$ helpers, for $m=1,\ldots, K$, i.e., by setting $K-m$ users' secure rates to zero and utilizing them as pure (structured) cooperative jammers.

\begin{figure}[t]
\centering
\includegraphics[scale=0.75]{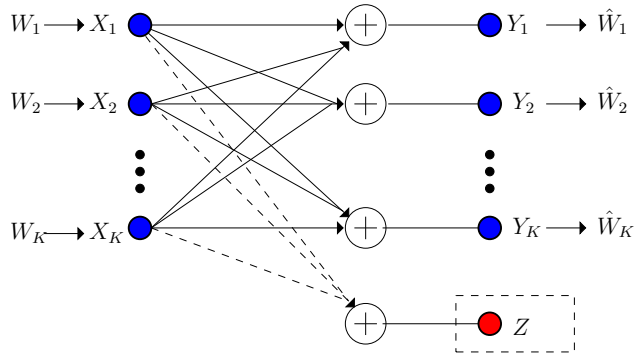}
\caption{$K$-user  interference channel (IC) with secrecy constraints.}
\label{fig:sdofregion:kic-general}
\end{figure}

We then consider the IC with secrecy constraints; see Figure~\ref{fig:sdofregion:kic-general}. In particular, we consider three different secrecy constraints in a unified framework as in \cite{xie_ulukus_isit_2013_kic, xie_unified_kic}: 1) $K$-user IC with one external eavesdropper (IC-EE), where $K$ transmitter-receiver pairs wish to have secure communication against an external eavesdropper. 2) $K$-user IC with confidential messages (IC-CM), where there are no external eavesdroppers, but each transmitter-receiver pair wishes to secure its communication against the remaining $K-1$ receivers. 3) $K$-user IC with confidential messages and one external eavesdropper (IC-CM-EE), which is a combination of the previous two cases, where each transmitter-receiver pair wishes to secure its communication against the $K-1$ receivers and the external eavesdropper. The converse for the sum \sdof (the sum \sdof is the same for all three models) was developed in \cite{xie_ulukus_isit_2013_kic, xie_unified_kic} by using the \emph{secrecy penalty} lemma and the \emph{role of a helper} lemma in a certain way, and then by summing up the obtained asymmetric upper bounds into a single symmetric upper bound. The achievability for the sum \sdof in \cite{xie_ulukus_isit_2013_kic, xie_unified_kic} is based on asymptotical real interference alignment \cite{real_inter_align_exploit} to enable simultaneous alignment at multiple receivers.

In order to develop a converse for the \emph{entire region} for the IC case, similar to the MAC case, we start by re-examining the converse proof in \cite{xie_ulukus_isit_2013_kic, xie_unified_kic} for the sum \sdof However, unlike the MAC case, the original steps used for the sum \sdof are not tight for the characterization of the entire region. There are two reasons for this: First, in the case of the MAC wiretap channel, since there is a single legitimate receiver, each transmitter (helper/interferer) impacts the total rate of all other legitimate transmitters at the legitimate receiver, and therefore, there is a single manner in which the \emph{role of a helper} lemma is applied. In the IC case, there are many different ways in which the \emph{role of a helper} lemma can be invoked as there are multiple receivers. In this case, by pairing up helpers (interferers) and the receivers we obtain $(K-1)^K$ upper bounds; even after removing the redundancies, we get $\big({K \choose K-1}\big) =  {2K-2 \choose K-1}$ upper bounds. In order to obtain the tightest subset of these upper bounds, we choose the most binding pairing of the helpers/interferers and the receivers. In particular, we do not apply the \emph{next one} (i.e., $k=i-1$ and $k=i+1$) selection of helpers/interferers as we have done in  \cite[Eqns.~(24) and (45)]{xie_ulukus_isit_2013_kic}. Instead, we choose all of the transmitters as interfering with a single transmitter-receiver pair; see \eqn{eqn:sdofregion:ic_ee_interfere_single} and \eqn{eqn:sdofregion:ic_cm_step_2_gives_converse_1} in this paper. This yields the tightest upper bounds. Second, we observe that, when we study the \sdof region, we need to consider the non-secrecy upper bounds for the underlying IC \cite{multiplexing_gain_of_networks, interference_alignment} as additional upper bounds. We note that such upper bounds are not binding for the case of MAC wiretap channel \sdof region, or the MAC and IC sum \sdof converses. In fact, such non-secrecy upper bounds for the IC are not binding even for the cases of $K=2$ or $K=3$. We observe that these upper bounds are needed for the IC with secrecy constraints starting with $K\geq 4$. To the best of our knowledge, this is the first time in network information theory that $K=2$ or $K=3$ do not capture the most generality of the problem, and we need to study $K=4$ to observe a certain multi-user phenomenon to take effect.

The converse region for the IC with secrecy constraints has a \emph{polytope} structure as well, and similar to the MAC wiretap channel case, we need to determine the extreme points of this polytope region. However, different from the MAC wiretap channel case, the converse region consists of two classes of upper bounds, due to secrecy and due to interference. This makes it difficult to identify the extreme points of the converse polytope. Finding the extreme points is related to finding full-rank sub-matrices from an overall matrix of size $2K+K(K-1)/2$. Since there are approximately $K^K$ such matrices, an exhaustive search is intractable, and therefore we investigate the consistency of the upper bounds, which reduces the possible number of sub-matrices to examine. After determining the extreme points of the converse polytope, we develop an achievable scheme for each extreme point. In particular, each extreme point of the converse region is achieved by a $(K-m)$-user IC-CM with $m$ helpers and $N$ independent external eavesdroppers, for $m\ge1$ and finite $N$.

Finally, after characterizing the entire \sdof regions of the MAC and IC  with secrecy constraints, as a byproduct of our results in this paper, we note that the sum \sdof is achieved \emph{only at one extreme point} of the \sdof region, which is the symmetric-rate extreme point, for both MAC and IC channel models.

\section{System Model, Definitions and the Result}

\subsection{$K$-user Gaussian MAC Wiretap Channel}
\label{sec:sdofregion:model_mac}

The $K$-user Gaussian MAC wiretap channel (see \fig{fig:sdofregion:mac_k}) is:
\begin{align}
Y_1 & = \sum_{i=1}^{K} h_i X_i + N_1 \\
Y_2 & = \sum_{i=1}^{K} g_i X_i + N_2
\end{align}
where $Y_1$ is the channel output of the legitimate receiver, $Y_2$ is the
channel output of the eavesdropper,
$X_i$ is the channel input of transmitter $i$,
$h_i$ and $g_i$ are the channel gains of transmitter $i$ to the
legitimate receiver and the eavesdropper, respectively, and $N_1$ and
$N_2$ are independent Gaussian random variables with zero-mean and
unit-variance. All the channel gains are  independently drawn from
continuous distributions, and are time-invariant throughout the communication session.
We further assume that all $h_{i}$  and $g_i$ are non-zero. All
channel inputs satisfy average power constraints,
$\E\left[X^2_{i}\right] \le P$, for $i=1,\ldots, K$.

Each transmitter $i$ has a message $W_i$ intended for the legitimate receiver. For each $i$, message $W_i$ is uniformly and independently chosen from set $\mathcal{W}_i$. The rate of message $i$ is $R_i\defn\frac{1}{n}\log|\mathcal{W}_i|$. Transmitter $i$ uses a stochastic function $f_i: \mathcal{W}_i \to \bfX_i$ where the $n$-length vector $\bfX_i\defn X_i^n$ denotes the $i$th user's channel input in $n$ channel uses. All messages are needed to be kept secret from the eavesdropper.
A secrecy rate tuple $(R_1,  \ldots, R_K)$ is said to be achievable if for any $\epsilon>0$ there exist $n$-length codes such that the legitimate receiver can decode the messages reliably, i.e., the probability of decoding error is less than $\epsilon$
\begin{equation}
\pr\left[(W_1,\ldots, W_K)\neq(\hat{W}_1,\ldots,\hat{W}_K)\right] \le \epsilon
\end{equation}
and the messages are kept information-theoretically secure against the eavesdropper
\begin{equation}
\frac{1}{n} H(W_1, \ldots, W_K|\bfY_2) \ge
\frac{1}{n} H(W_1, \ldots, W_K) - \epsilon
\end{equation}
where $\hat{W}_1,\ldots,\hat{W}_K$ are the estimates of the messages based on observation $\bfY_1$, where $\bfY_1\defn Y_1^n$ and $\bfY_2\defn Y_2^n$.

The \sdof region is defined as:
\begin{align}
\label{eqn:sdofregion:defn_sdofregion}
D = \Big\{ & \mathbf{d} : (R_1, \ldots, R_K) \textrm{ is achievable}  \textrm{ and } d_i \defn \lim_{P\to\infty} \frac{R_i}{\frac{1}{2} \log P}, ~ i = 1,\ldots,K \Big\}
\end{align}
The sum \sdof is defined as:
\begin{equation}
\label{eqn:sdofregion:defn_sum_sdof}
D_{s,\Sigma} \defn \lim_{P\to\infty} \sup \frac{\sum_{i=1}^{K} R_i}{\frac{1}{2} \log P}
\end{equation}
where the supremum is over all achievable secrecy rate tuples $(R_{1}, \ldots, R_{K})$. The sum \sdof of the $K$-user Gaussian MAC wiretap channel is characterized in the following theorem.

\begin{theorem}[{{\!\!\cite[Theorem $\mathbf{1}$]{xie_ulukus_isit_2013_mac}}}]
\label{theo:sdof_mac_sum_capacity}
The sum \sdof of the $K$-user Gaussian MAC wiretap channel is $\frac{K(K-1)}{K(K-1)+1}$ for almost all channel gains.
\end{theorem}

In this paper, we characterize the \sdof region of the $K$-user Gaussian MAC wiretap channel in the following theorem.

\begin{theorem}
\label{theo:sdof_mac_capacity_region}
The s.d.o.f.~region $D$ of the $K$-user Gaussian MAC wiretap channel
is the set of all $\mathbf{d}$ satisfying
\begin{align}
   K d_i + (K-1) \sum_{j=1,j\neq i}^K d_j  &\le K-1, & i=1,\ldots,K
\label{eqn:sdof_region_mac_theorem_1} \\
 d_i &\ge 0, & i=1,\dots,K
\label{eqn:sdof_region_mac_theorem_2}
\end{align}
for almost all channel gains.
\end{theorem}

\subsection{$K$-user Gaussian IC with Secrecy Constraints}
\label{sec:sdofregion:model_kic}

The  $K$-user Gaussian IC
 with secrecy constraints (see \fig{fig:sdofregion:kic-general}) is:
\begin{align}
  \label{eqn:sdofregion:ic_chanel_model_1}
  Y_i & = \sum_{j=1}^K h_{ji} X_j + N_i, \qquad i =1,\ldots,K \\
  \label{eqn:sdofregion:ic_chanel_model_2}
  Z   & = \sum_{j=1}^K g_{j} X_j + N_Z
\end{align}
where $Y_i$ is the channel output of receiver $i$, $Z$ is the channel
output of the external eavesdropper (if there is any), $X_i$ is the
channel input of transmitter $i$, $h_{ji}$ is the channel gain of the
$j$th transmitter to the $i$th receiver, $g_j$ is the channel gain of
the $j$th transmitter to the eavesdropper (if there is any), and
$\{N_1,\ldots,N_K,N_Z\}$ are mutually independent zero-mean
unit-variance Gaussian random variables. All the channel gains are
independently drawn from continuous distributions, and are time-invariant throughout the communication session.  We further assume that all $h_{ji}$ are
non-zero, and all $g_j$ are non-zero if there is an external
eavesdropper. All channel inputs satisfy average power constraints,
$\E\left[X^2_{i}\right] \le P$, for $i=1,\ldots, K$.

Each transmitter $i$ intends to send a message $W_i$, uniformly chosen from a set $\mathcal{W}_i$, to receiver $i$. The rate of  message $i$ is $R_i\defn\frac{1}{n}\log|\mathcal{W}_i|$, where $n$ is the number of channel uses. Transmitter $i$ uses a stochastic function $f_i: \mathcal{W}_i\to \bfX_i$ to encode the message, where $\bfX_i\defn X_i^n$ is the $n$-length channel input of user $i$.  The legitimate receiver $j$ decodes the message as $\hat{W}_j$ based
on its observation $\mathbf{Y}_j $. A secrecy rate tuple $(R_1,\ldots,R_K)$ is said to be achievable if for any $\epsilon>0$, there exist joint $n$-length codes such that each receiver $j$ can decode the corresponding message reliably, i.e., the probability of decoding error is less than $\epsilon$ for all messages,
\begin{equation}
 \max_{j}\pr\left[W_j\neq\hat{W}_j\right] \le \epsilon
\end{equation}
and the corresponding secrecy requirement is satisfied. We consider three different secrecy requirements:
\begin{figure}[t]
\centerline{\begin{tabular}{ccc}
\subfigure{\includegraphics[scale=0.65]{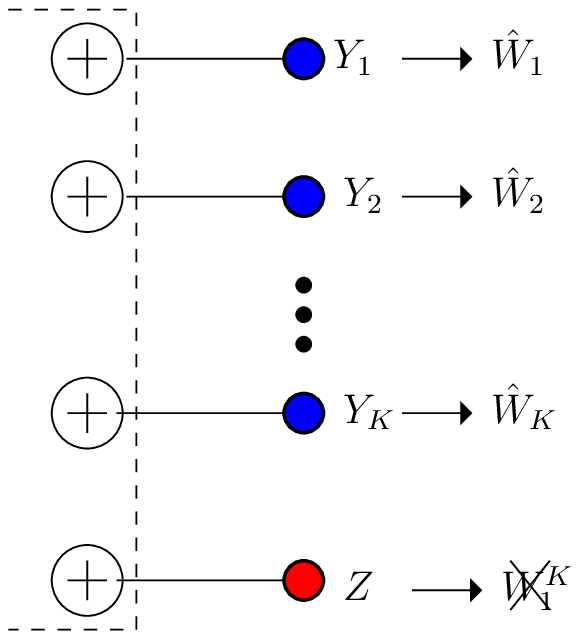}} \hspace*{0.4in}&
\subfigure{\includegraphics[scale=0.65]{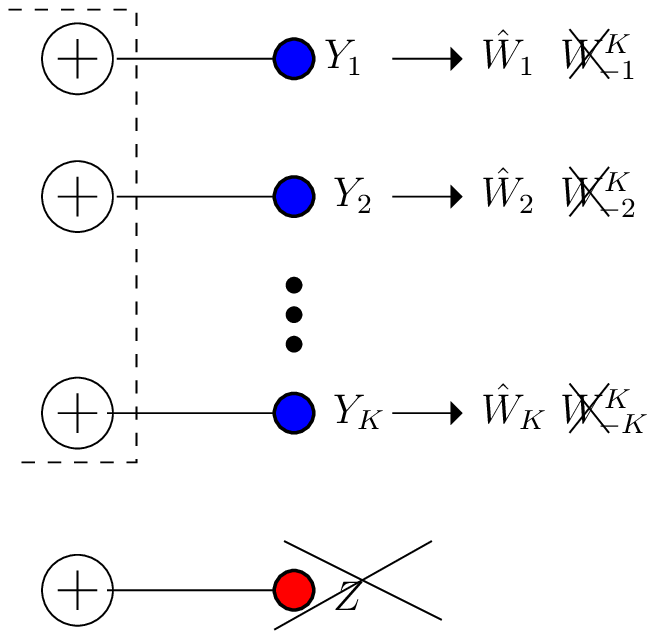}} \hspace*{0.4in}&
\subfigure{\includegraphics[scale=0.65]{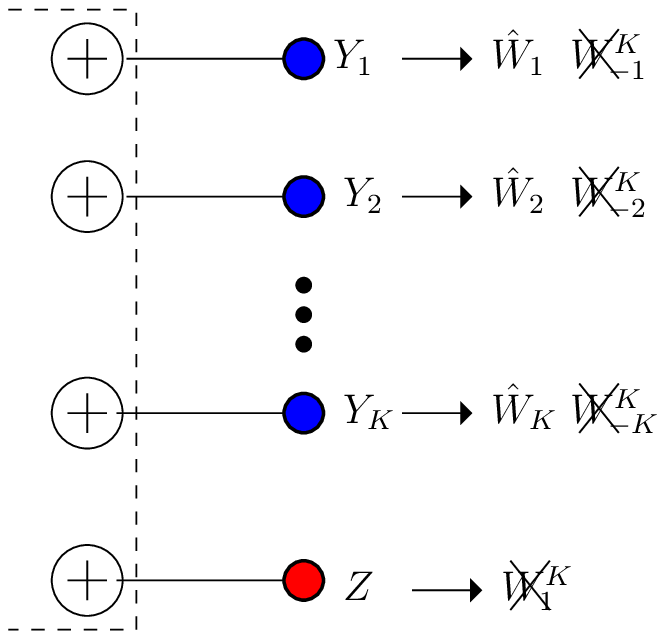}}\\
(a) & (b) & (c)
\end{tabular}}
\caption{The receiver sides of the three channel models: (a) $K$-user
  IC-EE, (b) $K$-user IC-CM, and
  (c) $K$-user IC-CM-EE, where $W_{-i}^K \defn
  \{W_1,\ldots,W_{i-1},W_{i+1},\ldots, W_K\}$.}
\label{fig:sdofregion:kic-subfigs}
\end{figure}
\begin{enumerate}
\item[{1)}] In IC-EE, \fig{fig:sdofregion:kic-subfigs}(a), all of the messages are kept information-theoretically secure against the external eavesdropper,
\begin{align}
  \label{eqn:sdofregion:kic-secrecy-constraint-cmee-1}
  \frac{1}{n}H(W_1,\ldots, W_K|\bfZ) & \ge\frac{1}{n} H(W_1,\ldots, W_K) - \epsilon
\end{align}
\item[{2)}] In IC-CM, \fig{fig:sdofregion:kic-subfigs}(b), all unintended messages are kept
  information-theoretically secure against each receiver,
\begin{align}
  \label{eqn:sdofregion:kic-secrecy-constraint-cmee-2}
  \frac{1}{n} H(W_{-i}^K|\bfY_i) & \ge\frac{1}{n} H(W_{-i}^K) - \epsilon,
  \qquad i=1,\ldots,K
\end{align}
where $W_{-i}^K \defn \{W_1,\ldots,W_{i-1},W_{i+1},\ldots, W_K\}$.
\item[{3)}] In IC-CM-EE, \fig{fig:sdofregion:kic-subfigs}(c), all of the messages are kept information-theoretically secure against both the $K-1$ unintended receivers and the eavesdropper, i.e., we impose both secrecy constraints in \eqn{eqn:sdofregion:kic-secrecy-constraint-cmee-1} and
  \eqn{eqn:sdofregion:kic-secrecy-constraint-cmee-2}.
\end{enumerate}

The \sdof region and the sum \sdof are defined as in \eqn{eqn:sdofregion:defn_sdofregion} and \eqn{eqn:sdofregion:defn_sum_sdof}.
The sum \sdof of the $K$-user IC-EE, IC-CM, and IC-CM-EE is characterized in the following theorem.

\begin{theorem}[{{\!\!\cite[Theorem $\mathbf{1}$]{xie_ulukus_isit_2013_kic}}}]
\label{theo:sdof_region:sdof_kic_sum_capacity}
The sum \sdof of the $K$-user Gaussian IC-EE, IC-CM, and IC-CM-EE is $\frac{K(K-1)}{2K-1}$ for almost all channel gains.
\end{theorem}

In this paper, we characterize the \sdof region of the $K$-user IC-EE, IC-CM, and IC-CM-EE in the following theorem.

\begin{theorem}
\label{theo:sdofregion:ic_ee_sdof_region}
The s.d.o.f. region $D$ of $K$-user IC-EE,  IC-CM, and IC-CM-EE
is the set of all $\mathbf{d}$ satisfying
\begin{align}
  K d_i +  \sum_{j=1,j\neq i}^K d_j & \le K-1,  & i=1,\ldots,K
\label{eqn:sdofregion:ic_ee_converse_1}  \\
\sum_{i\in V} d_i & \le 1, & \forall \ V \subseteq \{1,\ldots,K\}, \ |V|=2
\label{eqn:sdofregion:ic_ee_converse_2} \\
d_i & \ge 0, &  i=1,\ldots,K
\label{eqn:sdofregion:ic_ee_converse_3}
\end{align}
for almost all channel gains.
\end{theorem}

\section{Preliminaries}
\subsection{Polytope Structure and Extreme Points}
\label{sec:sdof_region:preliminaries_polytope_and_ep}
Let $X \subseteq R^n$. The \emph{convex hull} of $X$, $\convhull(X)$, is the set of all convex combinations of the points in $X$:
\begin{align}
\convhull(X) \defn \left\{\sum_i \lambda_i \mathbf{x}_i \,\,  | \,\, \mathbf{x}_i \in X, ~\sum_i \lambda_i =1, ~\lambda_i \in R, \textrm{ and } \lambda_i \ge0, ~\forall i \right\}
\end{align}
A set $P \subseteq R^n$ is a \emph{polyhedron} if there is a system of finitely many inequalities $\mathbf{H} \mathbf{x} \le \mathbf{h}$ such that
\begin{align}
P =\big\{\mathbf{x} \in R^n \,\,|\,\, \mathbf{H} \mathbf{x} \le \mathbf{h}\big\}
\end{align}
A set $P \subseteq R^n$ is a \emph{polytope} if there is a finite set $X \subseteq R^n$ such that $P = \convhull(X)$. Then, we have the following theorem.

\begin{theorem}[{{\!\!\cite[Theorem 3.1.3]{convex_polytopes}}}]
Let $P \subseteq R^n$. Then, $P$ is a bounded polyhedron if only if $P$ is a polytope.
\end{theorem}

Therefore, if $P \subseteq R^n$  is a polytope, then it is a convex hull of some finite set $X$. By the properties of the convex hull of a finite set $X$,  $P$ is a bounded, closed, convex set. Since $P$ is a subset of the Euclidean space, $P$ is a compact convex set. An {extreme  point} is formally defined as follows.

\begin{definition}[Extreme point]
\label{defn:sodf_region_extreme_point}
Let $P\subseteq R^n$. An $\mathbf{x}\in P$ is an extreme point if there are no $\mathbf{y},\mathbf{z} \in P \setminus \{\mathbf{x} \}$ such that $\mathbf{x} = \lambda \mathbf{y} + (1-\lambda) \mathbf{z}$
for any $\lambda \in (0,1)$. Then, $\expoints(P)$ is the set of all extreme points of $P$.
\end{definition}

\begin{theorem}[{{Minkowski, 1910. \cite[Theorem 2.4.5]{convex_polytopes}}}]
\label{theo:sdofregion:minkowski}
Let $P \subseteq R^n$ be a compact convex set. Then,
\begin{equation}
P = \convhull(\expoints(P)).
\end{equation}
\end{theorem}
Minkowski theorem plays an important role in this paper, since it tells that, instead of studying the polytope $P$ itself, for certain problems, e.g., achievability proofs, we can simply concentrate on all extreme points $\expoints(P)$.
Finally, the following theorem helps us find all extreme points of a polytope $P$ efficiently:  We  select any $n$ linearly independent {active/tight} boundaries and check whether they give a point in the polytope $P$.

\begin{theorem}[{{\!\!\cite[Theorem 7.2(b)]{linear_optimization_and_extension}}}]
\label{theo:sdofregion:polyhedron_ep_rank}
$\mathbf{x}\in R^n$ is an extreme point of polyhedron $P(\mathbf{H}, \mathbf{h})$ if and only if $\mathbf{H} \mathbf{x} \le \mathbf{h}$ and $\mathbf{H}' \mathbf{x} = \mathbf{h}'$ for some $n\times (n+1)$ sub-matrix $(\mathbf{H}',\mathbf{h}')$ of $(\mathbf{H},\mathbf{h})$ with $\rank(\mathbf{H}')=n$.
\end{theorem}

\subsection{Real Interference Alignment}

In this subsection, we review pulse amplitude modulation (PAM) and real interference alignment \cite{real_inter_align_exploit, real_inter_align}, similar to the review in \cite[Section~III]{interference_alignment_compound_channel}. The purpose of this subsection is to illustrate that by using real interference alignment, the transmission rate of a PAM scheme can be made to approach the Shannon achievable rate at high SNR. This provides a universal and convenient way to design capacity-achieving signalling schemes at high SNR by using PAM for different channel models as will be done in later sections.

\subsubsection{Pulse Amplitude Modulation}
For a point-to-point scalar Gaussian channel,
\begin{equation}
Y = X + Z
\end{equation}
with additive Gaussian noise $Z$ of zero-mean and variance $\sigma^2$, and
an input power constraint $\mathe{X^2} \le P$, assume that
the input symbols are drawn from a PAM constellation,
\begin{equation}
C(a,Q) = a \left\{ -Q, -Q+1, \ldots, Q-1,Q\right\}
\label{eqn:sdofregion:constel}
\end{equation}
where $Q$ is a positive integer and $a$ is a real number to normalize
the transmit power. Note that, $a$ is also the minimum distance
$d_{min}(C)$ of this constellation, which has the probability
of error
\begin{equation}
\pe(e) = \pe\left[ X \neq \hat X\right] \le \exp\left(  - \frac{d_{min}^2}{8 \sigma^2}\right)  = \exp\left(  - \frac{a^2}{8 \sigma^2}\right)
\end{equation}
where $\hat X$  is an estimate for $X$ obtained by choosing the closest point in the constellation $C(a,Q) $ based on observation $Y$.

The transmission rate of this PAM scheme is
\begin{equation}
R =  \log( 2 Q + 1)
\label{eqn:sdofregion:pam_rate}
\end{equation}
since there are $2Q+1$ signalling points in the constellation. For any small enough $\delta>0$, if we choose $Q = P^{\frac{1-\delta}{2}}$ and $a=\gamma P^{\frac{\delta}{2}}$, where
$\gamma$ is a constant independent of $P$, then
\begin{equation}
\pe(e)
\le \exp\left( -\frac{\gamma^2 P^{{ \delta}}}{8\sigma^2} \right)
\qquad \hbox{and} \qquad
R
 \ge \frac{1-\delta}{2} \log P
\label{eqn:sdofregion:simul_rate_relib}
\end{equation}
and we can have $\pe(e)\to 0$ and $R\to\frac{1}{2}\log P$ as
$P\to\infty$. That is, we can have reliable communication at rates
approaching $\frac{1}{2}\log P$.

Note that the PAM scheme has small probability of error (i.e., reliability) only when $P$ goes to infinity. For arbitrary $P$, the probability of error $\pe(e)$ is a finite number. Similar to the steps in \cite{etkin_discontinuous_2009, real_inter_align_exploit}, we connect the PAM transmission rate to the Shannon rate in the following derivation. We note that Shannon rate of $I(X;Y)$ is achieveable with arbitrary reliability using a random codebook:
\begin{align}
R'
& = I(X;Y) \\
& \ge I(X;\hat X) \\
&= H(X) - H(X|\hat X) \\
& = \log( 2 Q + 1) - H(X|\hat{X}) \\
& \ge \log( 2 Q + 1)  - 1 - \pe(e) \log( 2 Q + 1) \\
& \ge \Big[ 1- \pe(e)\Big] \frac{1-\delta}{2} \log P - 1
\label{eqn:sdofregion:ria__shannon_rate_final-eqn}
\end{align}
where we use the Markov chain $X\rightarrow Y\rightarrow\hat{X}$ and bound $H(X|\hat{X})$ using Fano's inequality. Therefore, we can achieve the rate in (\ref{eqn:sdofregion:ria__shannon_rate_final-eqn}) with arbitrary reliability, where for any fixed $P$, $\pe(e)$ in (\ref{eqn:sdofregion:ria__shannon_rate_final-eqn}) is the probability of error of the PAM scheme given in (\ref{eqn:sdofregion:simul_rate_relib}), which is a well-defined function of $P$. For a finite $P$, while $\pe(e)$ may not be arbitrarily small, the rate achieved in (\ref{eqn:sdofregion:ria__shannon_rate_final-eqn}), which is smaller than the rate of PAM in (\ref{eqn:sdofregion:pam_rate}), is achieved arbitrarily reliably. We finally note that as $P$ goes to infinity $\pe(e)$ goes to zero exponentially, and from (\ref{eqn:sdofregion:ria__shannon_rate_final-eqn}), both PAM transmission rate and the Shannon achievable rate have the same asymptotical performance, i.e., PAM transmission rate has $1$ Shannon d.o.f.

\subsubsection{Real Interference Alignment}

This PAM scheme for the point-to-point scalar channel can be
generalized to multiple data streams.  Let the transmit signal be
\begin{equation}
x = \mb{a}^T \mb{b} = \sum^L_{i=1} a_i b_i
\end{equation}
where $a_1,\ldots, a_L$ are rationally independent real
numbers\footnote{ $a_1, \ldots, a_L$ are rationally independent if
whenever $q_1,\ldots,q_L$ are rational numbers  then
$\sum^L_{i=1} q_i a_i =0$  implies $q_i=0$ for all $i$.  } and  each
$b_i$ is drawn independently from the constellation $C(a,Q)$ in (\ref{eqn:sdofregion:constel}).
The real value $x$ is a combination of  $L$ data streams, and the constellation observed at
the receiver consists of $(2 Q+1)^L$ signal points.

By using the Khintchine-Groshev theorem of Diophantine approximation
in number theory, \cite{real_inter_align_exploit,real_inter_align}
bounded the minimum distance $d_{min}$ of points in the receiver's
constellation: For any $\delta>0$, there exists a
constant $k_\delta$, such that
\begin{equation}
\label{eqn:sdoregion:ria_lb_of_d}
d_{min} \ge \frac{ k_\delta  a}{Q^{L-1+\delta}}
\end{equation}
for almost all rationally independent $\{a_i\}_{i=1}^L$, except for a set
of Lebesgue measure zero. Since the minimum distance of the receiver
constellation is lower bounded, with proper choice of $a$ and $Q$, the
probability of error can be made arbitrarily small, with rate $R$ approaching
$\frac{1}{2} \log P$.  This result is stated in the following lemma, as in \cite[Proposition~3]{interference_alignment_compound_channel}.

\begin{lemma}[\!\! \cite{real_inter_align_exploit,real_inter_align}]
\label{lemma:sdofregion:ria_real_alignment}
For any small enough $\delta>0$, there exists a positive constant
$\gamma$, which is independent of $P$, such that if we choose
\begin{equation}
Q = P^{\frac{1-\delta}{2(L+\delta)}}
\qquad \mbox{and} \qquad
a=\gamma \frac{P^{\frac{1}{2}}}{Q}
\end{equation}
then the average power constraint is satisfied, i.e., $\mathe{X^2}\le P $,
and for almost all $\{a_i\}_{i=1}^L$, except for a set of Lebesgue measure zero,
the probability of error is bounded by
\begin{equation}
  \mathrm{Pr}(e)
\le
\exp\left( - \eta_\gamma P^{{ \delta}} \right)
\end{equation}
where $\eta_\gamma$ is a positive constant which is
independent of $P$.
\end{lemma}

Furthermore, as a simple extension, if $b_i$ are sampled independently
from different constellations $C_i(a,Q_i)$, the lower bound in
\eqn{eqn:sdoregion:ria_lb_of_d} can be modified as
\begin{equation}
d_{min} \ge \frac{ k_\delta  a}{(\max_i Q_i)^{L-1+\delta}}
\end{equation}

\section{S.d.o.f. Region of $K$-User MAC Wiretap Channel}

In this section, we study  the $K$-user MAC wiretap channel defined in
Section \ref{sec:sdofregion:model_mac} and prove the  \sdof region
stated in Theorem \ref{theo:sdof_mac_capacity_region}. We first
illustrate the regions for $K=2$ and $K=3$ cases as examples. We then provide the converse
 in Section \ref{sec:sdofregion:mac_converse}, investigate the converse region in terms of its extreme points in
Section \ref{sec:sdofregion:mac_polytope_and_ep}, and show the
achievability of each extreme point in Section
\ref{sec:m_user_mac_region_section_achievability}.

\begin{figure}[t]
\centerline{\includegraphics[scale=0.7]{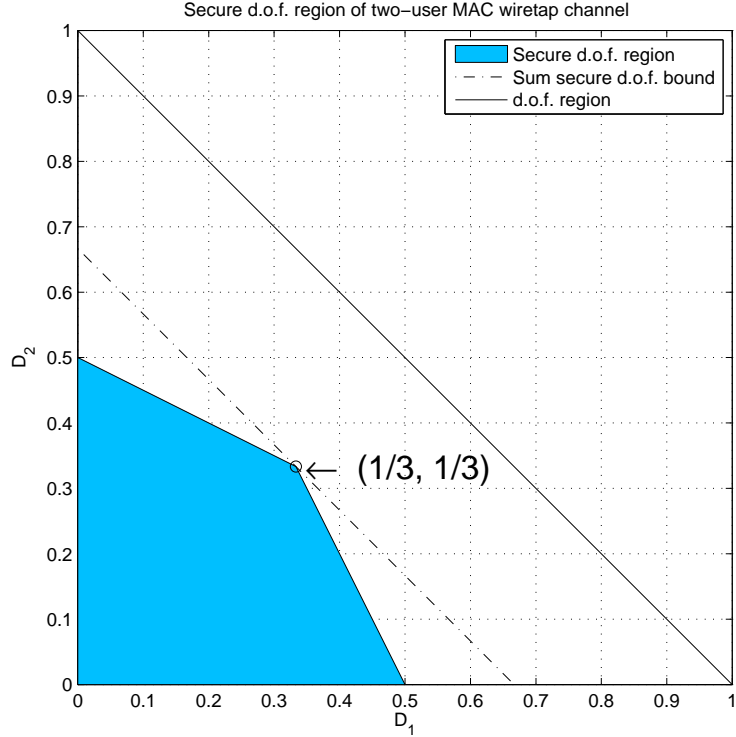}}
\caption{The \sdof region of the $K=2$-user MAC wiretap channel.}
\label{fig:sdof_region_2_user_mac}
\end{figure}

For $K=2$, the  \sdof region  in Theorem
\ref{theo:sdof_mac_capacity_region} becomes
\begin{align}
\label{eqn:sdofregion:mac_region_K_is_two}
D= \Big\{\mathbf{d} : ~  2 d_1 +  d_2 & \le 1,\nl
                          d_1 + 2 d_2 & \le 1, \nl
                              d_1,d_2 & \ge 0\Big\}
\end{align}
and is shown in \fig{fig:sdof_region_2_user_mac}. The extreme points of this region are:  $(0,0), (\frac{1}{2},0), (0,\frac{1}{2})$, and $(\frac{1}{3}, \frac{1}{3})$. In order to provide the
achievability of the region,  it suffices to provide the achievability of these extreme points. In fact the achievabilities of $(\frac{1}{2},0),
(0,\frac{1}{2})$ were proved in \cite{xie_gwch_allerton} in the helper setting and the achievability of $(\frac{1}{3}, \frac{1}{3})$ was proved in \cite{xie_ulukus_isit_2013_mac, xie_sdof_networks_in_prepare}. Note that $(\frac{1}{3}, \frac{1}{3})$ is the only sum \sdof optimum point.

\begin{figure}[t]
\centerline{\includegraphics[scale=0.7]{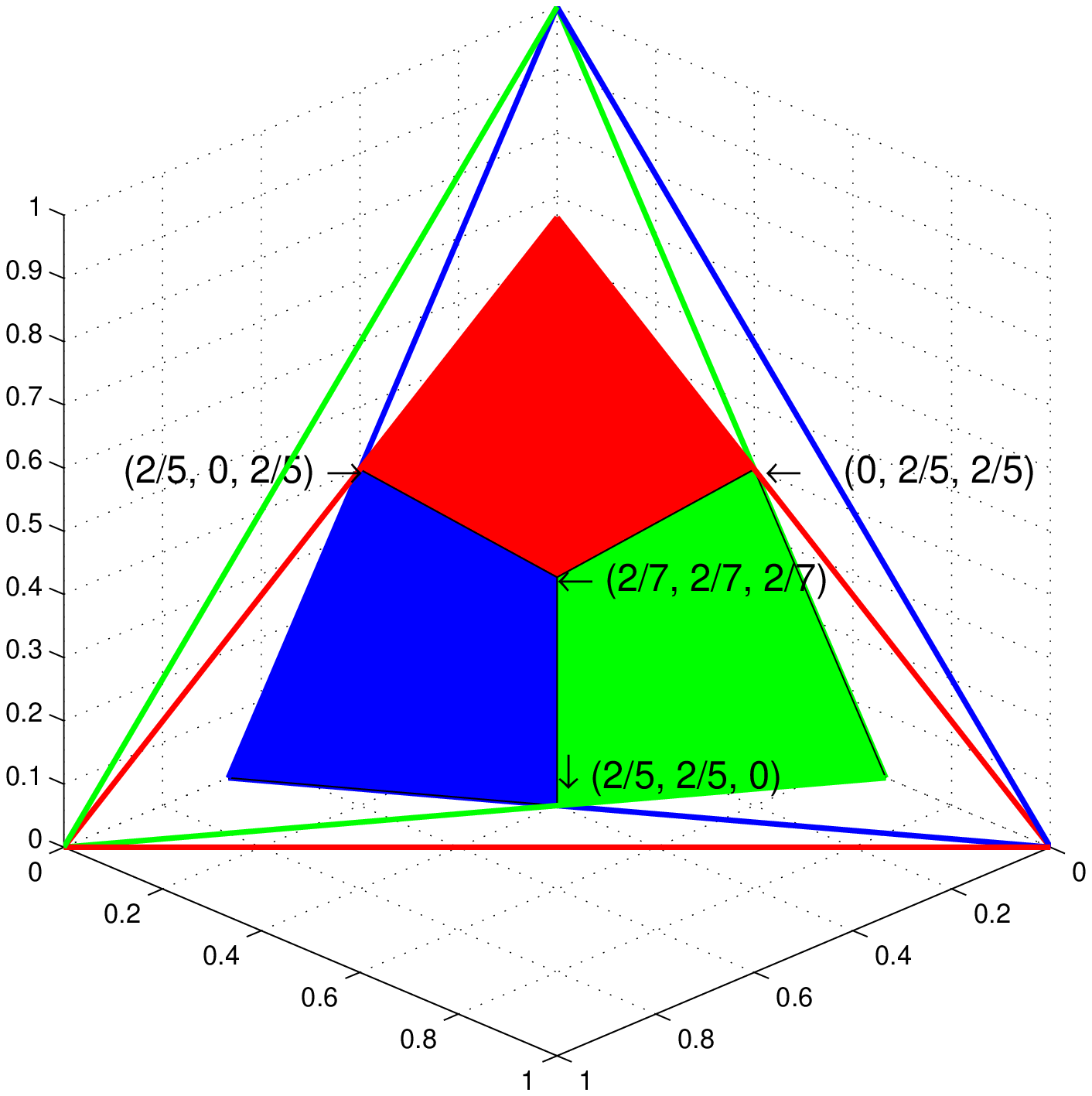}}
\caption{The \sdof region of the $K=3$-user MAC wiretap channel.}
\label{fig:sdof_region_3_user_mac}
\end{figure}

For $K=3$, the \sdof region  in Theorem
\ref{theo:sdof_mac_capacity_region}  becomes
\begin{align}
  D = \Big\{ \mathbf{d} : ~
    3 d_1 + 2d_2 + 2d_3 & \le 2, \nl
 2 d_1 + 3d_2 + 2d_3  & \le 2, \nl
 2 d_1 + 2d_2 + 3d_3  & \le 2, \nl
        d_1,d_2, d_3  & \ge 0
\Big\}
\end{align}
and is shown in \fig{fig:sdof_region_3_user_mac}.
The extreme points of this region are:
\begin{align}
\begin{aligned}
& \left(0,0,0\right)\\
& \left(\frac{2}{3}, 0  , 0  \right), \left(0  , \frac{2}{3}, 0  \right), \left( 0  , 0, \frac{2}{3}  \right)\\
& \left(\frac{2}{5}, \frac{2}{5}, 0  \right), \left(\frac{2}{5}, 0, \frac{2}{5} \right), \left(0, \frac{2}{5},  \frac{2}{5} \right) \\
& \left(\frac{2}{7}, \frac{2}{7}, \frac{2}{7} \right)
\end{aligned}
\label{eqn:sdofregion:mac_example_K_three}
\end{align}
which correspond to the maximum individual \sdof (see Gaussian wiretap
channel with two helpers \cite{xie_gwch_allerton}), the maximum sum of
pair of \sdof (see two-user Gaussian MAC wiretap channel with one
helper, proved in
Section~\ref{sec:m_user_mac_region_section_achievability}), and the
maximum sum \sdof (see three-user Gaussian MAC wiretap channel
\cite{xie_ulukus_isit_2013_mac, xie_sdof_networks_in_prepare}). Note that $(\frac{2}{7},
\frac{2}{7}, \frac{2}{7})$ is the only sum \sdof optimum point.

\subsection{Converse}
\label{sec:sdofregion:mac_converse}

The converse simply follows from a key inequality in the proof in \cite{xie_ulukus_isit_2013_mac}.
We re-examine  \cite[Eqn.~(41)]{xie_ulukus_isit_2013_mac}:
\begin{align}
n R_i + (K-1) \sum_{j=1}^K n R_j \le (K-1) h(\bfY_1) + n c_i, \quad\quad i =1,\ldots,K
\label{eqn:sodf_region_mac_sum-this}
\end{align}
where all $\{c_i \}$ in this paper are constants independent of $P$.

Clearly, \eqn{eqn:sodf_region_mac_sum-this} is not symmetric. However, the lower bound derived in \cite{xie_ulukus_isit_2013_mac} was achieved by a symmetric scheme. Therefore, in \cite{xie_ulukus_isit_2013_mac}, in order to obtain a matching upper bound, we summed up \eqn{eqn:sodf_region_mac_sum-this} for all $i$ to obtain:
\begin{align}
\left[K (K-1) +1 \right]\sum_{j=1}^K n R_j & \le K (K-1) h(\bfY_1) + n c'\\
& \le K(K-1) \frac{n}{2} \log P + n c''
\end{align}
which provided the desired upper bound for the sum \sdof
\begin{align}
D_{s,\Sigma} \le  \frac{ K(K-1)}{K(K-1) +1}
\end{align}
which is the converse for Theorem~\ref{theo:sdof_mac_sum_capacity}.

In fact, \eqn{eqn:sodf_region_mac_sum-this} provides more information than what is needed for the sum \sdof only. In this paper, we start from \eqn{eqn:sodf_region_mac_sum-this}
\begin{align}
n R_i + (K-1) \sum_{j=1}^K n R_j \le (K-1) \left( \frac{n}{2} \log P  \right) + n c_i, \quad\quad i =1,\ldots,K
\end{align}
divide by $\frac{n}{2}\log P$ and take the limit ${P\to\infty}$ on both sides to obtain,
\begin{align}
d_i + (K-1) \sum_{j=1}^K d_j \le K-1, \quad\quad i =1,\ldots,K
\end{align}
that is,
\begin{equation}
K d_i + (K-1) \sum_{j=1,j\neq i}^K d_j \le K-1, \quad\quad i =1,\ldots,K
\end{equation}
which concludes the converse proof of Theorem~\ref{theo:sdof_mac_capacity_region}.

\subsection{Polytope Structure and Extreme Points}
\label{sec:sdofregion:mac_polytope_and_ep}
To prove that the region $D$ in Theorem
\ref{theo:sdof_mac_capacity_region} is tight (i.e., achievable), we first express it in terms of its \emph{extreme points}, explicitly characterize all of its extreme points, and develop a scheme to achieve each of its extreme points.

The region in Theorem
\ref{theo:sdof_mac_capacity_region} is a polytope, which is a convex hull of some finite set $X$, as discussed in Section \ref{sec:sdof_region:preliminaries_polytope_and_ep}. By the properties of the convex hull of a finite set $X$,  $D$ is a bounded, closed, convex set. Since $D\subset R^K$, $D$ is a compact convex set. From Minkowski theorem, the polytope $D$ in Theorem
\ref{theo:sdof_mac_capacity_region}  is a convex hull of its extreme points. Then, in order to prove that $D$ is tight, it suffices to prove that each extreme point of $D$ is achievable. Then, from convexification through time-sharing, all points in $D$ are achievable.

In order to speak of the polytope, we re-write the constraints in
\eqn{eqn:sdof_region_mac_theorem_1} and \eqn{eqn:sdof_region_mac_theorem_2} as
\begin{align}
  K d_i + (K-1) \sum_{j=1,j\neq i}^K d_j & \le K-1, & i=1,\ldots,K
 \label{eqn:sdofregion:mac_polytope_constraint_1}\\
- d_i & \le 0 & i=1,\ldots,K\label{eqn:sdofregion:mac_polytope_constraint_2}
\end{align}
Then, we  write all the left hand sides of \eqn{eqn:sdofregion:mac_polytope_constraint_1} and \eqn{eqn:sdofregion:mac_polytope_constraint_2} as an $N \times K $ matrix $\mathbf{H}$ with corresponding right hand sides forming an $N$-length column vector $\mathbf{h}$, i.e., all points $\mathbf{d}$ in $D$ satisfy
\begin{equation}
\mathbf{H} \mathbf{d} \le \mathbf{h}
\end{equation}
where $N \defn 2 K $. By Theorem \ref{theo:sdofregion:polyhedron_ep_rank}, exploring all extreme points of $D$ is equivalent to finding all sub-matrices $(\mathbf{H}_J, \mathbf{h}_J)$ of $(\mathbf{H}, \mathbf{h})$, such that
\begin{equation}
\rank(\mathbf{H}_J) = K
\end{equation}
and
\begin{equation}
\label{eqn:sdofregion:mac_region_ep_constraint}
\mathbf{H}_J \mathbf{d} = \mathbf{h}_J,\quad\textrm{ and }\quad \mathbf{H} \mathbf{d} \le \mathbf{h}
\end{equation}
where $\mathbf{H}_J$ is a  sub-matrix of $\mathbf{H}$ with rows indexed by the index set $J$, and  $\mathbf{h}_J$ is  the sub-vector of $\mathbf{h}$ with rows indexed by $J$.

Let $\mathbf{d}\in D$ be a non-zero extreme point of $D$. Define a subset $S\subseteq \{1,\ldots,N\}$ as
\begin{align}
\label{eqn:sodf_region:rank_discussion_s}
S \defn \Big\{ s_i\defn s(i):
  \mathbf{H}_{s_i} \mathbf{d} = \mathbf{h}_{s_i}\textrm{ is }
  K d_i + (K-1) \sum_{j=1,j\neq i}^K d_j
  = K-1, ~~ i =1,\ldots,K
 \Big\}
\end{align}
where $s(i)$ is a function of the coordinate $i$  with the value as
the row index of $\mathbf{H}$ corresponding to the active
boundaries in \eqn{eqn:sdofregion:mac_polytope_constraint_1}.
Similarly, define the set  $Z\subseteq \{1,\ldots,N\}$ as
\begin{align}
\label{eqn:sodf_region:rank_discussion_z}
Z \defn \Big\{z_i\defn z(i) :
\mathbf{H}_{z_i} \mathbf{d} = \mathbf{h}_{z_i}\textrm{ is }
 d_i = 0, ~~ i =1,\ldots,K
\Big\}
\end{align}
where $z(i)$ is a function of the coordinate $i$ with the value as the row index of $\mathbf{H}$ corresponding to the active boundaries in \eqn{eqn:sdofregion:mac_polytope_constraint_2}.
Clearly, $S$ and $Z$ are disjoint, i.e.,
\begin{equation}
S \cap Z = \phi
\end{equation}
For any row index set $J$, which corresponds to a set of active
boundaries for $\mathbf{d}$, we have
\begin{equation}
J = S \cup Z
\end{equation}

For example, for the three-user case, $K=3$, according to
\eqn{eqn:sdofregion:mac_polytope_constraint_1} and
\eqn{eqn:sdofregion:mac_polytope_constraint_2}, we have  $\mathbf{H}$ and
$\mathbf{h}$ as
\begin{equation}
\mathbf{H} = \left[
\begin{array}{rrr}
3 & 2 & 2 \\
2 & 3 & 2 \\
2 & 2 & 3 \\
-1 & 0 & 0 \\
0 & -1 & 0 \\
0 & 0 & -1
\end{array}
\right],
\quad\quad\quad
\mathbf{h} = \left[
\begin{array}{r}
2 \\
2 \\
2 \\
0 \\
0 \\
0
\end{array}
\right]
\end{equation}
If the equalities with $i=1,2$ hold in
\eqn{eqn:sdofregion:mac_polytope_constraint_1} and the equality with
$i=3$ holds
in \eqn{eqn:sdofregion:mac_polytope_constraint_2}, then the
corresponding sets $S$,
$Z$, $J$ are
\begin{align}
S =  \{ s_1,s_2 \} = \{ 1,2 \},\quad
Z = \{z_3\} = \{6\},\quad
J = S \cup Z = \{1,2,6\}
\end{align}
with the row-index functions
\begin{align}
  s_i & = s(i) = i \\
  z_i & = z(i) = i + 3
\end{align}
In this example, it is easy to check that
\begin{equation}
\rank(\mathbf{H}_J) = \rank\left( \left[
\begin{array}{rrr}
3 & 2 & 2 \\
2 & 3 & 2 \\
0 & 0 & -1
\end{array}
\right] \right) = 3 = K
\end{equation}
and the solution given by $\mathbf{H}_J \mathbf{d} = \mathbf{h}_J$ is
\begin{equation}
\mathbf{d} = \left(\frac{2}{5},\frac{2}{5},0 \right)
\end{equation}
which satisfies
\eqn{eqn:sdofregion:mac_region_ep_constraint}. Therefore, this is an extreme point.

For the general case, we have the following theorem.

\begin{theorem}
\label{prop:sodf_region_extreme_point}
A point $\mathbf{d}\in D$  of Theorem
\ref{theo:sdof_mac_capacity_region} is an extreme point if and only if it is equal to, up to element reordering, 
\begin{equation}
\Big(\underbrace{\Delta, \ldots, \Delta}_{m \textrm{ items}},
        \underbrace{0,\ldots,0}_{(K-m)  \textrm{ items}} \Big), \quad\quad 0\le m \le K
\label{eqn:sdofregion:mac_ep_structure}
\end{equation}
where
\begin{equation}
\label{eqn:sodf_region_delta_final_value}
\Delta = \frac{K-1}{m(K-1) + 1}
\end{equation}
\end{theorem}

\begin{Proof}
First, for any $m$, $0\le m \le K$, let the point $\mathbf{d}$ be as in \eqn{eqn:sdofregion:mac_ep_structure}. It is easy to check that  the sub-matrix $(\mathbf{H}_J,\mathbf{h}_J)$, where
\begin{equation}
J = \Big\{s_i : 1\le i \le m \Big\} \cup \Big\{z_j : m+1 \le j \le K\Big\}
\end{equation}
satisfies all the conditions in Theorem \ref{theo:sdofregion:polyhedron_ep_rank}, which means that  $\mathbf{d}$ is an extreme point.

In order to show the other direction, we need to show that any extreme point $\mathbf{d}$ has the structure in \eqn{eqn:sdofregion:mac_ep_structure} for some $m$, $0\le m\le K$. To this end, we find the sub-matrix in Theorem \ref{theo:sdofregion:polyhedron_ep_rank}.

If $|Z|=K$, due to
\eqn{eqn:sdofregion:mac_polytope_constraint_2}, the sub-matrix
$\mathbf{H}_Z$ is simply a diagnoal  matrix with $-1$s on the
diagonal, and consequently, $\rank(\mathbf{H}_Z)=K$. Then, the solution of $\mathbf{H}_Z \mathbf{d} = \mathbf{h}_Z$ is $\mathbf{0}$, which satisfies \eqn{eqn:sdofregion:mac_region_ep_constraint}. This extreme point corresponds to the case $m=0$ in Theorem \ref{prop:sodf_region_extreme_point}.

In the rest of the proof, we focus on non-zero extreme points, i.e.,
  $|Z|<K$.
Due to  \eqn{eqn:sdofregion:mac_polytope_constraint_1}, it is easy to verify that $\mathbf{H}_S$ has $|S|$ rows with
$
\rank(\mathbf{H}_S) = |S|
$
where $S$ is defined in \eqn{eqn:sodf_region:rank_discussion_s}.
In order to make $\rank(\mathbf{H}_{J}) = \rank( \mathbf{H}_{S\cup Z}) =K$, we need at least $K-|S| $ more rows from $\mathbf{H}$, i.e., $|Z|\ge K-|S| $.
If $S$ is empty, then $|Z|\ge K$, which contradicts the assumption $ |Z|<K$. Therefore, $S$ is non-empty, i.e., $|S|\ge1$.

First, we claim that
\begin{align}
\label{eqn:sdofregion:mac_d_i_is_same_to_d_k_in_S}
  d_i = d_k, \quad \forall s_i,s_k \in S
\end{align}
If $|S|=1$, there is nothing to prove, and we are done with the proof of
\eqn{eqn:sdofregion:mac_d_i_is_same_to_d_k_in_S}. If $|S|>1$, consider any $s_i,s_k\in
S$, $i\neq k$. By the definition of $S$, we have
\begin{align}
  (K-1) d_k +  {K} d_i  + (K-1) \sum_{l\neq i,k} d_l & = K-1  \\
  (K-1) d_i +  {K} d_k  + (K-1) \sum_{l\neq i,k} d_l & = K-1
\end{align}
which implies that $d_i=d_k$ for any $s_i,s_k\in S$, proving \eqn{eqn:sdofregion:mac_d_i_is_same_to_d_k_in_S} for
$|S|\ge1$.

Next, we claim
\begin{equation}
  \label{eqn:sdofregion:mac_d_i_is_positive}
  d_i>0, \quad\forall s_i \in S
\end{equation}
If $|S|=K$, due to
\eqn{eqn:sdofregion:mac_d_i_is_same_to_d_k_in_S},
\eqn{eqn:sdofregion:mac_d_i_is_positive} is trivially true since we are focusing on a non-zero
extreme point.  If $|S|<K$, then we observe that
\begin{align}
  \label{eqn:sdofregion:mac_d_i_larger_than_d_j}
  d_i \ge d_j, \quad\quad \forall s_i \in S, s_j \not\in S
\end{align}
which indicates that for any $s_i\in S$ the corresponding element in
vector $\mathbf{d}$ is the largest one, i.e., $d_i =\max_{k} d_k$,
which implies \eqn{eqn:sdofregion:mac_d_i_is_positive}. Hence, it now
suffices to show
\eqn{eqn:sdofregion:mac_d_i_larger_than_d_j}. We prove it by contradiction. Assume
that there exists a coordinate $j$ such that $s_j\not\in S$ and $d_j$ is strictly larger
than $d_i$ for any $s_i \in S$. By the definition of $S$ in
\eqn{eqn:sodf_region:rank_discussion_s}, we have
\begin{align}
K -1 & =  K d_i + (K-1) d_j  + (K-1) \sum_{l=1,l\neq i,j}^K d_l \\
     & < K d_i + (K-1) d_j  + (K-1) \sum_{l=1,l\neq i,j}^K d_l
      + (d_j - d_i) \\
     & =  K d_j + (K-1) d_i  + (K-1) \sum_{l=1,l\neq i,j}^K d_l \\
     & =  K d_j +  (K-1) \sum_{l=1,l\neq j}^K d_l
\end{align}
which contradicts the constraint
\eqn{eqn:sdofregion:mac_polytope_constraint_1}. Therefore,  we must have
\eqn{eqn:sdofregion:mac_d_i_larger_than_d_j} and consequently
\eqn{eqn:sdofregion:mac_d_i_is_positive}.

Finally, denote $m\defn |S|$, and, without loss of generality, assume that $ S = \{s_i: 1\le i \le m\}$.
By \eqn{eqn:sdofregion:mac_d_i_is_positive} and the
definition of $Z$ in \eqn{eqn:sodf_region:rank_discussion_z}, we note that $z_j\in Z$  only if $s_j \not\in S$. Together with the constraint $|Z|\ge K-|S|=K-m$, we conclude that we must
have
$
Z = \{ z_j:  m+1\le j \le K\}
$, i.e., $d_j=0$ for $m+1\le j\le K$.
Thus, $\rank(\mathbf{H}_{S\cup Z}) =K$, and, by \eqn{eqn:sdofregion:mac_d_i_is_same_to_d_k_in_S},  the solution given by the corresponding equations can be characterized as \eqn{eqn:sdofregion:mac_ep_structure},
which satisfies \eqn{eqn:sdofregion:mac_region_ep_constraint}, completing the proof.
\end{Proof}

\subsection{Achievability}
\label{sec:m_user_mac_region_section_achievability}
The previous section showed that the converse region is a polytope with extreme points which have $m$ coordinates all equal to $\Delta$ given in \eqn{eqn:sodf_region_delta_final_value}, and the remaining $K-m$ coordinates all equal to zero. It is clear that zero vector is an extreme point in $D$ and is trivially achievable. The rest of the achievability proof focuses on non-zero extreme points. In this section, we prove that each of these extreme points is achievable. Without loss of generality, we prove that the \sdof point of
\begin{equation}
\label{eqn:sodf_region_m_Kminusm_d}
\mathbf{d} = \Big(\underbrace{\Delta,\ldots, \Delta}_{m \textrm{ items}}, \underbrace{0, \ldots, 0}_{(K-m) \textrm{ items}}\Big)
\end{equation}
is achievable for all $1<m<K$ with $\Delta$ in \eqn{eqn:sodf_region_delta_final_value}. By symmetry, this proves the achievability of all extreme points. Note that $m=K$ is shown in \cite{xie_ulukus_isit_2013_mac, xie_sdof_networks_in_prepare}, and $m=1$ is shown in \cite{xie_gwch_allerton}.

\begin{theorem}
\label{theo:sodf_region_m_Kminusm_helpers}
The extreme point $\mathbf{d}\in D$ given in \eqn{eqn:sodf_region_m_Kminusm_d} is achieved by $m$-user  Gaussian MAC wiretap channel with $K-m$ helpers for almost all channel gains.
\end{theorem}
\begin{Proof}
Consider the $m$-user Gaussian MAC wiretap channel with $K-m$ helpers where transmitter $i$, $i=1,\ldots,m$, has confidential message $W_i$ intended for the legitimate receiver and the remaining $K-m$ transmitters serve as independent helpers without messages of their own.

In order to achieve the extreme point $\mathbf{d}$ in
\eqn{eqn:sodf_region_m_Kminusm_d}, transmitter $i$, $i=1,\ldots,m$,
divides its message into $K-1$ mutually independent sub-messages. Each
transmitter sends a linear combination of signals that carry the
sub-messages. In addition to message carrying signals, all
transmitters also send  cooperative jamming signals $U_i$,
$i=1,\ldots,K$, respectively. The messages are sent in such a way that
all of the cooperative jamming signals are aligned in a single
dimension at the legitimate receiver, occupying the smallest possible
space at the legitimate receiver, and hence allowing for the reliable
decodability of the message carrying signals. In addition, each
cooperative jamming signal is aligned with at most $K-1$ message
carrying signals at the eavesdropper to limit the information leakage
rate to the eavesdropper. An example of $K=3$, $m=2$, and $K-m=1$ is
given in \fig{fig:mac_3_pair_rate_ia}.

More specifically, we use a total of $m(K-1) + K$ mutually independent random variables
\begin{align}
& V_{ij},\quad i \in \{1,\ldots,m\}, ~j\in \{1,\cdots,K\} \setminus \{ i \}\\
& U_k ,\quad k\in\{1,\cdots,K\}
\end{align}
where $\{V_{ij}\}_{j\neq i}$ denote the message carrying signals and $U_i$ denotes the cooperative jamming signal sent from transmitter $i$. In particular, $V_{ij}$ carries the $j$th sub-message of transmitter $i$. Each of these random variables is uniformly and independently drawn from the same discrete constellation $C(a,Q)$ given in \eqn{eqn:sdofregion:constel}, where $a$ and $Q$ will be specified later. We choose the input signals of the transmitters as
\begin{align}
X_i & = \sum_{j=1, j\neq i}^K \frac{g_j}{h_j g_i} V_{ij} +\frac{1}{h_i} U_i, \qquad i \in \{1,\ldots,m\} \\
X_j & = \frac{1}{h_j} U_j, \qquad j \in \{m+1, \ldots, K\}
\end{align}

\begin{figure}[t]
\centering
\includegraphics[scale=0.8]{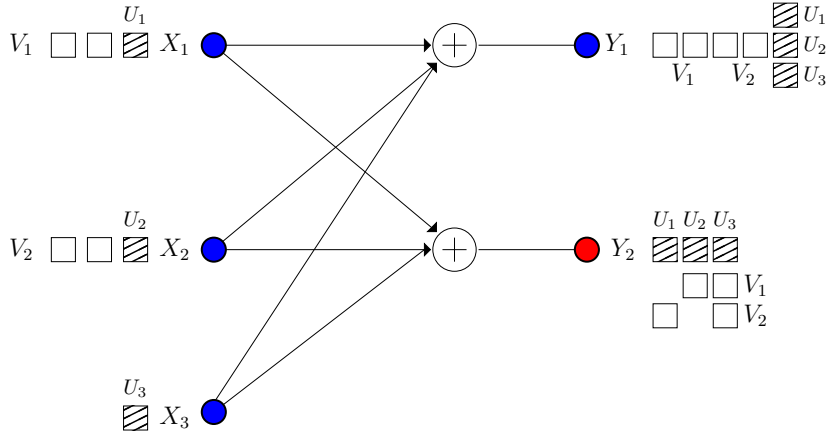}
\caption{Illustration of secure interference alignment for the \sdof triple $(\frac{2}{5}, \frac{2}{5}, 0  )$ for the two-user MAC wiretap channel with one helper; $K=3$ and $m=2$. Here, we define $V_{i}\defn\{V_{ij}: j=1,2,3, j\neq
i\}$ for $i=1,2$.}
\label{fig:mac_3_pair_rate_ia}
\end{figure}

With these input selections, observations of the receivers are
\begin{align}
Y_1 =
&  \left[\sum_{i=1}^m \sum_{j=1, j\neq i}^K \frac{g_j h_i}{h_j g_i} V_{ij}\right]  + \left( \sum_{k=1}^K U_k \right)
+ N_1 \label{eqn:sdof_region_mac_bracket1}
\end{align}
and
\begin{align}
Y_2 =
&\sum_{j=1}^K \frac{g_j }{h_j } \left( U_j + \sum_{i=1, i\neq j}^m V_{ij} \right)
+ N_2 \label{eqn:sdof_region_mac_bracket2}
\end{align}
where the terms inside the parentheses $(\cdot)$ in (\ref{eqn:sdof_region_mac_bracket1}) and (\ref{eqn:sdof_region_mac_bracket2}) are \emph{aligned}.

By \cite[Theorem 1]{secrecy_ia5}, we can achieve the following sum secrecy rate for the $m$ users
\begin{equation}
\sup \sum_{i=1}^{m} R_i  \ge I(\mathbf{V};Y_1) - I(\mathbf{V};Y_2)
\label{eqn:mac-ach}
\end{equation}
where $\mathbf{V} \defn \{V_{ij}: i \in \{1,\ldots,m\}, ~j\in \{1,\cdots,K\} \setminus \{ i \}\}$.

By Lemma \ref{lemma:sdofregion:ria_real_alignment}, for any $\delta>0$, if we choose
$Q = P^{\frac{1-\delta}{2 (m(K-1)+1+\delta)}}$ and $a=\gamma P^{\frac{1}{2}}/Q$, where $\gamma$ is a constant independent of $P$ to meet the average power constraint, then
\begin{align}
\pe\left[\bfV\neq\hat{\bfV}\right]
& \le \exp\left( - \beta P^\delta \right)
\end{align}
for some constant $\beta>0$ (independent of $P$), where $\hat{\bfV}$ is the estimate of $\bfV$ by choosing the closest point in the constellation based on observation $Y_1$.  This means that we can have $\pe[\bfV\neq\hat{\bfV}]\to 0$  as $P\to\infty$.

By Fano's inequality and the Markov chain $\bfV\rightarrow Y_1\rightarrow\hat{\bfV}$, we know that
\begin{align}
 H(\bfV| Y_1) & \le H(\bfV|\hat{\bfV}) \\
&\quad \le 1 +
\exp\left( -\beta P^\delta
\right) \log(2Q+1)^{m(K-1)} \\
& \quad = o(\log P)
\end{align}
where $o(\cdot)$ is the little-$o$ function. This means that
\begin{align}
I(\bfV;Y_1)
&  = H(\bfV) - H(\bfV|Y_1) \\
& =
\log(2Q+1)^{m(K-1)}- H(\bfV|Y_1) \\
& \ge
\log(2Q+1)^{m(K-1)}- o(\log P)
\label{eqn:sdofregion:mac_m_lb_ixy1}
\end{align}
On the other hand, we can bound the second term in \eqn{eqn:mac-ach} as
\begin{align}
  I(\bfV;Y_2)
& \le I\left(\bfV;
Y_2 - N_2
\right) \\
& =
\sum_{j=1}^K  H\left( U_j + \sum_{i=1, i\neq j}^m V_{ij} \right)
   -
H\left(
U_1,\ldots, U_K
\right)\\
& \le K \log\frac{ 2K Q +1 } { 2Q +1 } \label{eqn:sdofregion:mac_k_for_combined_constellation_streams} \\
& \le K \log K \\
& = o(\log P)
\label{eqn:sdofregion:mac_m_lb_ixy2}
\end{align}
where \eqn{eqn:sdofregion:mac_k_for_combined_constellation_streams} is due to the fact that entropy of each $U_j + \sum_{i=1, i\neq j}^m V_{ij}$ is maximized by the uniform distribution which takes values over a set of cardinality $2KQ +1$.

Combining \eqn{eqn:sdofregion:mac_m_lb_ixy1} and \eqn{eqn:sdofregion:mac_m_lb_ixy2}, we obtain
\begin{align}
\sup \sum_{i=1}^m   R_i
&  \ge I(\bfV;Y_1) - I(\bfV;Y_2) \\
&  \ge
\log(2Q+1)^{m(K-1)} - o(\log P)\\
&  ={\frac{{m(K-1)}(1-\delta)}{{m(K-1)}+1+\delta}} \left(\frac{1}{2}\log P\right) + o(\log P)
\end{align}
By choosing $\delta$ arbitrarily small, we can achieve the sum \sdof of $\frac{m(K-1)}{m(K-1)+1}$ for almost all channel gains, which implies that the \sdof tuple of
\begin{equation}
\Bigg(\underbrace{\frac{(K-1)}{m(K-1)+1}, \ldots, \frac{(K-1)}{m(K-1)+1}}_{m \textrm{ item(s)}}, \underbrace{ 0, \ldots, 0}_{(K-m) \textrm{ item(s)}}  \Bigg)
\end{equation}
is achievable by symmetry, which is \eqn{eqn:sodf_region_m_Kminusm_d}.
\end{Proof}

\section{S.d.o.f. Region of $K$-User IC with Secrecy Constraints}
\label{sec:sdofregion:ic_ee_region}

In this section, we study the $K$-user IC with
secrecy constraints defined in Section \ref{sec:sdofregion:model_kic}
and prove the  \sdof region stated in Theorem
\ref{theo:sdofregion:ic_ee_sdof_region}. To this end, we consider
both IC-CM and IC-EE and their combination IC-CM-EE in a unified
framework.
We first illustrate the regions for $K=2,3,4$ cases as examples. The purpose of presenting $K=4$ as an example is to show that, unlike the MAC case, starting with $K=4$ interference constraints become effective and binding.
We then provide converses separately for IC-EE and IC-CM in
Section \ref{sec:sdofregion:kic_converse_ic_ee} and Section
\ref{sec:sdofregion:kic_converse_ic_cm}, respectively, which imply a
converse for IC-CM-EE.  Finally, we show the achievability for IC-CM-EE,
which implies the achievability for IC-EE and IC-CM. Specifically, we investigate the converse region in terms of its extreme points in Section \ref{sec:sdofregion:ic:polytope_structure} and show the general achievability in Section
\ref{sec:sdofregion:achievability_p_m_design}.

For $K=2$, the  \sdof region  in Theorem
\ref{theo:sdofregion:ic_ee_sdof_region} becomes
\begin{align}
\label{eqn:sdofregion:ic_example_k_two}
D= \Big\{\mathbf{d} : ~  2 d_1 +  d_2 & \le 1,\nl
                          d_1 + 2 d_2 & \le 1, \nl
                              d_1,d_2 & \ge 0\Big\}
\end{align}
which is the same as \eqn{eqn:sdofregion:mac_region_K_is_two}, and is shown in \fig{fig:sdof_region_2_user_mac}. Note that \eqn{eqn:sdofregion:ic_ee_converse_2} is not necessary for
the two-user case, since summing the bounds $2 d_1 +  d_2  \le 1$ and
$ d_1 +  2d_2  \le 1$ up gives a new bound
\begin{align}
  d_1 + d_2 \le \frac{2}{3}
\end{align}
which is the result in Theorem
\ref{theo:sdof_region:sdof_kic_sum_capacity} and makes the constraint
in \eqn{eqn:sdofregion:ic_ee_converse_2} strictly loose.

  In order to provide
the achievability,  it suffices to check that the extreme points
$(0,0)$, $(\frac{1}{2},0), (0,\frac{1}{2})$, and $(\frac{1}{3}, \frac{1}{3})$
are achievable. In fact the achievabilities of $(\frac{1}{2},0),
(0,\frac{1}{2})$ are similar to \cite{xie_gwch_allerton} and will be shown in  Section
\ref{sec:sdofregion:ic:polytope_structure}. The
achievability of $(\frac{1}{3}, \frac{1}{3})$ was proved in
\cite{xie_ulukus_isit_2013_kic,xie_unified_kic}. Note that $(\frac{1}{3},
\frac{1}{3})$ is the only sum \sdof optimum point.

For $K=3$, the \sdof region  in Theorem
\ref{theo:sdofregion:ic_ee_sdof_region}  becomes
\begin{align}
  D = \Big\{ \mathbf{d} : ~
    3 d_1 + d_2 + d_3 & \le 2, \nl
  d_1 + 3d_2 + d_3  & \le 2, \nl
  d_1 + d_2 + 3d_3  & \le 2, \nl
        d_1,d_2, d_3  & \ge 0
\Big\}
\label{eqn:sdofregion:ic_three_user_example}
\end{align}
and
\eqn{eqn:sdofregion:ic_ee_converse_2} is not necessary for
the three-user case, either. This is because, due to the positiveness of each element in $\mathbf{d}$, from the first two inequalities in \eqn{eqn:sdofregion:ic_three_user_example},  we have
\begin{align}
    3 d_1 + d_2 \le 3 d_1 + d_2 + d_3 & \le 2
    \label{eqn:sdofregion:ic_example_k_3_ineqn_1} \\
     d_1 + 3d_2 \le d_1 + 3d_2 + d_3  & \le 2
     \label{eqn:sdofregion:ic_example_k_3_ineqn_2}
\end{align}
Summing the left hand sides up of \eqn{eqn:sdofregion:ic_example_k_3_ineqn_1} and \eqn{eqn:sdofregion:ic_example_k_3_ineqn_2}  gives us
\begin{align}
  d_1 + d_2 \le 1
\end{align}
which is \eqn{eqn:sdofregion:ic_ee_converse_2} with $V=\{1,2\}$, and
we have \eqn{eqn:sdofregion:ic_ee_converse_2} for free from \eqn{eqn:sdofregion:ic_three_user_example}.

The extreme points of this region are:
\begin{align}
\begin{aligned}
& \left(0,0,0\right)\\
& \left(\frac{2}{3}, 0  , 0  \right), \left(0  , \frac{2}{3}, 0  \right), \left( 0  , 0, \frac{2}{3}  \right)\\
& \left(\frac{1}{2}, \frac{1}{2}, 0  \right), \left(\frac{1}{2}, 0, \frac{1}{2} \right), \left(0, \frac{1}{2},  \frac{1}{2} \right) \\
& \left(\frac{2}{5}, \frac{2}{5}, \frac{2}{5}\right)
\end{aligned}
\label{eqn:sdofregion:kic_example_K_three}
\end{align}
which correspond to the maximum individual \sdof (see Gaussian wiretap
channel with two helpers \cite{xie_gwch_allerton} and Section \ref{sec:sdofregion:ic:polytope_structure}), the maximum sum of
pair of \sdof (proved in
Section~\ref{sec:sdofregion:ic:polytope_structure}), and the
maximum sum \sdof (see three-user Gaussian IC-CM-EE in
\cite{xie_ulukus_isit_2013_kic, xie_unified_kic}). Note that, $(\frac{1}{2}, \frac{1}{2})$ is the maximum sum \dof for a two-user IC \emph{without} secrecy constraints, and $(\frac{2}{5},
\frac{2}{5}, \frac{2}{5})$ is the only sum \sdof optimum point.
Finally, note the difference of the extreme points of the $3$-user IC in \eqn{eqn:sdofregion:kic_example_K_three}  from the corresponding $3$-user MAC in \eqn{eqn:sdofregion:mac_example_K_three}, even though the \sdof regions and the extreme points of the $2$-user IC and $2$-user MAC in  \eqn{eqn:sdofregion:ic_example_k_two} and \eqn{eqn:sdofregion:mac_region_K_is_two}  were the same.

For $K=4$, the \sdof region  in Theorem
\ref{theo:sdofregion:ic_ee_sdof_region}  becomes
\begin{align}
  D = \Big\{ \mathbf{d} : ~
    4 d_1 + d_2 + d_3 + d_4 & \le 3, \nl
    d_1 + 4 d_2 + d_3 + d_4 & \le 3, \nl
    d_1 + d_2 + 4 d_3 + d_4 & \le 3, \nl
    d_1 + d_2 + d_3 + 4 d_4 & \le 3, \nl
                    d_1+d_2 & \le 1, \nl
                    d_1+d_3 & \le 1, \nl
                    d_1+d_4 & \le 1, \nl
                    d_2+d_3 & \le 1, \nl
                    d_2+d_4 & \le 1, \nl
                    d_3+d_4 & \le 1, \nl
        d_1,d_2, d_3, d_4  & \ge 0
\Big\}
\end{align}

The extreme points of this region are:
\begin{align}
\begin{aligned}
& \left(0,0,0 \right)\\
& \left(\frac{3}{4}, 0  , 0, 0  \right),
  \left(0  , \frac{3}{4}, 0  , 0 \right),
  \left( 0  , 0, \frac{3}{4} , 0 \right),
  \left( 0  , 0, 0, \frac{3}{4} \right)\\
& \left(\frac{2}{3},\frac{1}{3},0,0\right) \quad\quad \textrm{up to element reordering}\\
& \left(\frac{1}{2}, \frac{1}{2}, \frac{1}{2}, 0  \right),
  \left(\frac{1}{2}, \frac{1}{2}, 0, \frac{1}{2} \right),
  \left(\frac{1}{2}, 0, \frac{1}{2}, \frac{1}{2} \right) ,
  \left(0,\frac{1}{2}, \frac{1}{2}, \frac{1}{2} \right)\\
& \left(\frac{3}{7}, \frac{3}{7}, \frac{3}{7}, \frac{3}{7} \right)
\end{aligned}
\label{eqn:sdofregion:kic_example_K_four}
\end{align}
Here, in contrast to the two-user and three-user cases,
\eqn{eqn:sdofregion:ic_ee_converse_2} is absolutely necessary. For example, the
point $(\frac{3}{5}, \frac{3}{5}, 0,0)$ satisfies
\eqn{eqn:sdofregion:ic_ee_converse_1}, but not
\eqn{eqn:sdofregion:ic_ee_converse_2}. In fact, it cannot be achieved, and \eqn{eqn:sdofregion:ic_ee_converse_2} is strictly needed to enforce that fact.

Regarding the region in Theorem \ref{theo:sdofregion:ic_ee_sdof_region}, as illustrated in the examples above,  we provide a few general comments here:
\begin{enumerate}

\item[1)]
Although \eqn{eqn:sdofregion:ic_ee_converse_2} only states the constraints for all pairs of rates, due to the same argument in \cite{interference_alignment}, it can  equivalently be stated as $\sum_{i\in V}d_i\le \frac{|V|}{2}$ for all $|V|\ge2$. We note that,  when $|V|=K$, the corresponding upper bound is strictly loose due to Theorem 1 in \cite{xie_ulukus_isit_2013_kic,xie_unified_kic}, and that is why such bounds were not needed in \cite{xie_ulukus_isit_2013_kic,xie_unified_kic}, where sum \sdof was characterized.

\item[2)] As shown in the examples,  when $K=2$ or $3$, \eqn{eqn:sdofregion:ic_ee_converse_2} is not necessary. When $K\ge4$, we need both \eqn{eqn:sdofregion:ic_ee_converse_1} and \eqn{eqn:sdofregion:ic_ee_converse_2} to completely  characterize the region $D$. Neither of them can be removed from the theorem. For example, the all $\frac{1}{2}$ vector, $(\frac{1}{2},\frac{1}{2},\ldots,\frac{1}{2})$, satisfies \eqn{eqn:sdofregion:ic_ee_converse_2}, but not \eqn{eqn:sdofregion:ic_ee_converse_1}.
On the other hand, the point $(\frac{K-1}{K+1},\frac{K-1}{K+1}$, $0$, $0$, $\ldots$, $0)$, which has only two non-zero elements, satisfies \eqn{eqn:sdofregion:ic_ee_converse_1}, but not \eqn{eqn:sdofregion:ic_ee_converse_2} for any $K\ge4$. Therefore, \eqn{eqn:sdofregion:ic_ee_converse_2} emerges only when $K\ge4$.  To the best of our knowledge, this is the first time that $K=2$ or $K=3$ do not represent the most generality of a multi-user problem, and we need to go up to $K=4$ for this phenomenon to appear.

\item[3)]
Different portions of the region $D$ are governed by different upper bounds. To see this, we can study the structure of  the extreme points of $D$, since $D$ is the convex hull of them.
The sum \sdof tuple, which is symmetric and has no zero elements, is governed by the upper bounds in \eqn{eqn:sdofregion:ic_ee_converse_1} due to  secrecy constraints.
However, as will be shown in Theorem \ref{theo:sdofregion:ic_ee_sdof_region_all_ep} in Section \ref{sec:sdofregion:ic:polytope_structure}, all other extreme points have zeros as some elements, and therefore are  governed by the upper bounds in \eqn{eqn:sdofregion:ic_ee_converse_2} due to   interference  constraints in \cite{multiplexing_gain_of_networks, interference_alignment}.
An explanation can be provided as follows:  When some transmitters do not have messages to transmit, we may employ them as ``helpers''. Even though secrecy constraint is considered in our problem, with the help of the ``helpers'', the effect due to the existence of the eavesdropper in the network can be \emph{eliminated}. Hence,  this portion of the \sdof region is  dominated by the interference constraints.

\end{enumerate}

\subsection{Converse for $K$-User IC-EE}
\label{sec:sdofregion:kic_converse_ic_ee}

The constraint in \eqn{eqn:sdofregion:ic_ee_converse_2} follows from the non-secrecy constraints on the $K$-user IC in \cite{multiplexing_gain_of_networks, interference_alignment}. We note that this same constraint is valid for the converse proof of IC-CM in the next section as well.

In order to prove \eqn{eqn:sdofregion:ic_ee_converse_1} in Theorem
\ref{theo:sdofregion:ic_ee_sdof_region}, we re-examine
 \cite[Eqn.~(23)]{xie_ulukus_isit_2013_kic}. Originally, we applied
  \cite[Lemma~2]{xie_gwch_allerton} in \cite{xie_ulukus_isit_2013_kic} by treating the signal from
transmitter $j$  as the unintended noise to its neighboring
transmitter-receiver pair $j-1$, i.e., for any $i=1,\ldots,K$,
\begin{align}
n\sum_{j=1}^K R_j
& \le
\sum_{j=1,j\neq i}^K h(\tilde\bfX_j)
+ n\nextsc
\\
& \le \left[ h(\bfY_K) - nR_K \right]
+ \left[ h(\bfY_1) - nR_1 \right] + \cdots  + \left[h(\bfY_{i-2}) - n R_{i-2} \right] \nl
&\quad
+ \left[ h(\bfY_{i}) - n R_i \right]+ \cdots+\left[ h(\bfY_{K-1}) - n R_{K-1} \right]
+ n \nextsc
\end{align}
By noting that $h(\bfY_j) \le \frac{n}{2}\log P + n c_j'$
for each $j$, we have
\begin{align}
2 n\sum_{j=1}^K R_j \le (K-1) \frac{n}{2}\log P  + n R_{i} + n \nextsc
\end{align}
Therefore, we have a total of $K$ bounds  for $i=1,\ldots,K$. Summing these $K$
bounds, we obtained:
\begin{align}
(2 K - 1) n\sum_{j=1}^K R_j \le K (K-1) \frac{n}{2}\log P  + n \nextsc
\end{align}
which gave
\begin{align}
  D_{s,\Sigma} \le \frac{ K (K-1)}{2 K - 1}
\end{align}
completing the  converse proof for the sum \sdof of IC-EE in \cite{xie_ulukus_isit_2013_kic} (also Theorem \ref{theo:sdof_region:sdof_kic_sum_capacity} in this paper).

Here, we continue from  \cite[Eqn.~(23)]{xie_ulukus_isit_2013_kic} and re-interpret it as:
\begin{align}
n\sum_{j=1}^K R_j
& \le
\sum_{j=1,j\neq i}^K h(\tilde\bfX_j)
+ n\nextsc
\\
& \le \underbrace{\left[ h(\bfY_i) - nR_i \right] + \cdots  + \left[ h(\bfY_{i}) - n R_{i} \right]}_{K-1 \textrm{ items}}
+ n \nextsc \label{eqn:sdofregion:ic_ee_interfere_single} \\
& = (K-1) h(\bfY_i) - (K-1) n R_i + n\nextscnu \\
& \le (K-1) \left(\frac{n}{2}\log P\right)  - (K-1) n R_i + n\nextsc
\label{eqn:sdofregion:ic_ee_step_gives_converse_1}
\end{align}
where $i\in\{1,\ldots,K\}$ is arbitrary. Here, the second inequality means that we apply  \cite[Lemma~2]{xie_gwch_allerton} by treating the signal from all transmitters $j\neq i$  as the unintended noise to the transmitter-receiver pair $i$.

Rearranging the terms in
\eqn{eqn:sdofregion:ic_ee_step_gives_converse_1}, dividing both sides by
$\frac{n}{2}\log P$, and taking the limit ${P\to\infty}$ on both sides, we obtain
\begin{align}
K d_i + \sum_{j=1,j\neq i}^K d_j \le K-1,\quad\quad i=1,\ldots,K
\end{align}
which is \eqn{eqn:sdofregion:ic_ee_converse_1} in Theorem
\ref{theo:sdofregion:ic_ee_sdof_region}, completing the converse proof for IC-EE.

\subsection{Converse for $K$-User IC-CM}
\label{sec:sdofregion:kic_converse_ic_cm}

When we studied the sum s.d.o.f. of IC-CM, we applied \cite[Lemma~2]{xie_gwch_allerton} to  \cite[Eqn.~(44)]{xie_ulukus_isit_2013_kic}  by treating the signal from
transmitter $j$  as the unintended noise to its neighbor
transmitter-receiver pair $j+1$, i.e., for any $i=1,\ldots,K$
\begin{align}
n\sum_{j=1,j\neq i}^K R_j & \le \sum_{j=1}^K h(\tilde\bfX_j) -h(\bfY_i) + n \nextsc \\
&\le \left[\sum_{j=1}^{K-1} \big[ h(\bfY_{j+1}) - n R_{j+1} \big] \right] + \big[ h(\bfY_{1}) - n R_{1} \big] -h(\bfY_i) + n
\nextsc \\
&= \sum_{j=1}^{K} \big[ h(\bfY_{j}) - n R_{j} \big] -h(\bfY_i) + n
\nextscnu
\end{align}
By noting that $h(\bfY_j) \le \frac{n}{2}\log P + n c_j'$ for each $j$, we have
\begin{align}
n R_i + 2 n\sum_{j=1,j\neq i}^K R_j
& \le  \sum_{j=1,j\neq i}^K  h(\bfY_{j})  + n
\nextscnu  \\
& \le  (K-1) \frac{n}{2}\log P  + n
\nextsc
\end{align}
Therefore, we have a total of $K$ bounds for $i=1,\ldots,K$. Summing these $K$ bounds, we obtained
\begin{align}
(2 K - 1) n\sum_{j=1}^K R_j \le K (K-1) \frac{n}{2}\log P  + n \nextsc
\end{align}
which gave
\begin{align}
  D_{s,\Sigma} \le \frac{ K (K-1)}{2 K - 1}
\end{align}
completing the converse proof for the sum \sdof of IC-CM in \cite{xie_ulukus_isit_2013_kic} (also Theorem \ref{theo:sdof_region:sdof_kic_sum_capacity} in this paper).

Here, we continue from \cite[Eqn.~(44)]{xie_ulukus_isit_2013_kic} and re-interpret it as follows: For any $i\in\{1,\ldots,K\}$, we select
\begin{align}
k\defn \left\{
\begin{array}{ll}
i - 1, & \textrm{if } i \ge 2 \\
K, &  \textrm{if } i = 1
\end{array}
\right.
\end{align}
and then  have
\begin{align}
n\sum_{j=1,j\neq i}^K R_j
& \le
\left[ \sum_{j=1}^K h(\tilde\bfX_j) \right] - h(\bfY_i)
+ n\nextsc
\\
& \le h(\tilde\bfX_{k}) +
\left[ \sum_{j=1, j\neq {k}}^K h(\tilde\bfX_j) \right] - h(\bfY_i)
+ n\nextsc
\\
& \le h(\bfY_i) - nR_i  +
\left[ \sum_{j=1, j\neq {k}}^K h(\tilde\bfX_j) \right] - h(\bfY_i)
+ n\nextsc \label{eqn:sdofregion:ic_cm_step_1_gives_converse_1}
\\
& =
\left[ \sum_{j=1, j\neq {k}}^K h(\tilde\bfX_j) \right] - nR_i
+ n\nextscnu
\\
& \le \underbrace{\left[ h(\bfY_{k}) - nR_{k} \right] + \cdots  + \left[ h(\bfY_{k}) - n R_{k} \right]}_{K-1 \textrm{ items}} - nR_i
+ n \nextsc  \label{eqn:sdofregion:ic_cm_step_2_gives_converse_1} \\
& = (K-1) h(\bfY_{k}) - (K-1) n R_{k} - nR_i + n\nextscnu \\
& \le (K-1) \left(\frac{n}{2}\log P\right)   - (K-1) n R_{k} - nR_i  + n\nextscnu
\end{align}
which is
\begin{align}
(K-1) n R_{k} + n\sum_{j=1}^K R_j \le  (K-1) \left(\frac{n}{2}\log P\right) + n\nextscnu
\label{eqn:sdofregion:ic_cm_step_gives_converse_1}
\end{align}
Here,  inequality \eqn{eqn:sdofregion:ic_cm_step_1_gives_converse_1} means that we apply  \cite[Lemma~2]{xie_gwch_allerton} by treating the signal from transmitter $k$  as the unintended noise to the transmitter-receiver pair $i$. Similarly,  inequality \eqn{eqn:sdofregion:ic_cm_step_2_gives_converse_1} means that we apply  \cite[Lemma~2]{xie_gwch_allerton} by treating the signal from transmitter $j\neq k$  as the unintended noise to the transmitter-receiver pair $k$.

Rearranging the terms in \eqn{eqn:sdofregion:ic_cm_step_gives_converse_1}, dividing both sides by
$\frac{n}{2}\log P$, and taking the limit ${P\to\infty}$ on both sides, we obtain
\begin{align}
K d_k + \sum_{j=1,j\neq k}^K d_j \le K-1,\quad\quad k=1,\ldots,K
\end{align}
which is \eqn{eqn:sdofregion:ic_ee_converse_1}  in Theorem
\ref{theo:sdofregion:ic_ee_sdof_region}, completing the converse proof for IC-CM.

\subsection{Polytope Structure and Extreme Points}
\label{sec:sdofregion:ic:polytope_structure}

Similar to the discussion and approach in the MAC problem in Section \ref{sec:sdofregion:mac_polytope_and_ep}, it is easy to see that the region $D$ characterized by Theorem \ref{theo:sdofregion:ic_ee_sdof_region} is a polytope, which is equal to the convex combinations of all extreme points of $D$ due to Theorem \ref{theo:sdofregion:minkowski}. Therefore, in order to show the tightness of region $D$, it suffices to prove that all extreme points of $D$ are achievable.

We first  assume that $K\ge 3$, and determine the structure of all extreme points of $D$ in the following theorem.
\begin{theorem}
\label{theo:sdofregion:ic_ee_sdof_region_all_ep}
For the $K$-dimensional region $D$, $K\ge 3$, in Theorem
\ref{theo:sdofregion:ic_ee_sdof_region}, any extreme point must be a point with one of the following structures:
\begin{align}
& (0,0,\ldots,0),
\label{eqn:sdofregion:ep_class_zero_vector}  \\
%
%
%
& \Big(\frac{K-1-p}{K-p}, \underbrace{\frac{1}{K-p}, \ldots, \frac{1}{K-p}}_{p \textrm{ items}},
        \underbrace{0,\ldots,0}_{m  \textrm{ items}} \Big), & K-2\ge p \ge0,\ \ m=K-1-p\ge1
\label{eqn:sdofregion:ep_class_k_m_problem}
        \\
& \Big(\underbrace{\frac{1}{2}, \ldots, \frac{1}{2}}_{p' \textrm{ items}},
        \underbrace{0,\ldots,0}_{m'  \textrm{ items}} \Big), & K-2\ge p'\ge3,\ \ m'\ge1,\ \ p'+m'=K\ge5
\label{eqn:sdofregion:ep_class_all_halves}
\\
& \Big(\frac{K-1}{2K-1},\frac{K-1}{2K-1},\ldots,\frac{K-1}{2K-1}\Big)
\label{eqn:sdofregion:ep_class_symmetric}
\end{align}
up to element reordering.
\end{theorem}
The proof of Theorem \ref{theo:sdofregion:ic_ee_sdof_region_all_ep} is provided in Appendix \ref{sec:sdofregion:proof_of_ic_ep_structure}.

Now, in order to show the tightness of region $D$, it suffices to show the achievability for each structure in Theorem \ref{theo:sdofregion:ic_ee_sdof_region_all_ep}.  Clearly, the zero vector in \eqn{eqn:sdofregion:ep_class_zero_vector} is trivially achievable.  The symmetric tuple in \eqn{eqn:sdofregion:ep_class_symmetric} is achievable due to \cite{xie_ulukus_isit_2013_kic, xie_unified_kic}. Therefore, it remains to show the achievability of the structures in  \eqn{eqn:sdofregion:ep_class_k_m_problem}  and \eqn{eqn:sdofregion:ep_class_all_halves}.

In order to address the achievabilities of \eqn{eqn:sdofregion:ep_class_k_m_problem}  and \eqn{eqn:sdofregion:ep_class_all_halves}, we formulate a new channel model as a $(p+1)$-user IC-CM-EE channel with $m$ independent helpers and $N$ independent external eavesdroppers. The formal definition of this channel model is given in Section \ref{sec:sdofregion:achievability_p_m_design}. Then, we have the following theorem.

\begin{theorem}
\label{theo:sdofregion:ic_cm_ee_p_m_n_problem}
For the $(p+1)$-user IC-CM-EE channel with $m$ independent helpers and $N$ independent external eavesdroppers, as far as $p\ge0$, $m\ge1$, and $N$ is finite, the following \sdof tuple is achievable:
\begin{align}
\Big( \frac{m}{m+1},
  \underbrace{ \frac{1}{m+1}, \frac{1}{m+1}, \ldots, \frac{1}{m+1}}_{p \textrm{ items}}
\Big)
\label{eqn:sdofregion:ic_cm_ee_p_m_n_tuple}
\end{align}
for almost all channel gains.
\end{theorem}
The  proof of Theorem \ref{theo:sdofregion:ic_cm_ee_p_m_n_problem} is provided in Section \ref{sec:sdofregion:achievability_p_m_design}.

Here, we provide a few comments about Theorem \ref{theo:sdofregion:ic_cm_ee_p_m_n_problem}.
Theorem \ref{theo:sdofregion:ic_cm_ee_p_m_n_problem} provides quite general results, and subsumes some other known cases:
\begin{enumerate}
\item[1)] The result in \cite{xie_gwch_allerton} is a special case of Theorem \ref{theo:sdofregion:ic_cm_ee_p_m_n_problem} with $p=0,m\ge1,N=1$.
\item[2)]
\eqn{eqn:sdofregion:ep_class_k_m_problem} is a special case of Theorem \ref{theo:sdofregion:ic_cm_ee_p_m_n_problem} with $p\ge0,m=K-1-p\ge1,N=m+1$.
\item[3)]
\eqn{eqn:sdofregion:ep_class_all_halves} is a byproduct  of Theorem \ref{theo:sdofregion:ic_cm_ee_p_m_n_problem}:  By choosing  $p=p'-1,m=1,N=m'+1$, we know that with just one helper, the following \sdof tuple is achievable:
\begin{equation}
\Big(\underbrace{\frac{1}{2},\frac{1}{2},\ldots,\frac{1}{2}}_{p'\textrm{ items}},0\Big)
\end{equation}
Now, if we add $m'-1$ more independent helpers into the network, \eqn{eqn:sdofregion:ep_class_all_halves} can be achieved trivially.
\end{enumerate}

Therefore, with the help of Theorem \ref{theo:sdofregion:ic_cm_ee_p_m_n_problem}, each structure in Theorem \ref{theo:sdofregion:ic_ee_sdof_region_all_ep} can be achieved, which provides the achievability proof for Theorem \ref{theo:sdofregion:ic_ee_sdof_region} for $K\ge3$.

Finally, we address the $K=2$ case. In this case, the region $D$ characterized by \eqn{eqn:sdofregion:ic_ee_converse_1}-\eqn{eqn:sdofregion:ic_ee_converse_3} in Theorem
\ref{theo:sdofregion:ic_ee_sdof_region} is given by \eqn{eqn:sdofregion:ic_example_k_two}.  In order to provide the achievability,  it suffices to prove that the extreme points $(\frac{1}{2},0), (0,\frac{1}{2})$, and $(\frac{1}{3}, \frac{1}{3})$ are achievable.
The achievability of $(\frac{1}{3}, \frac{1}{3})$ was proved  in \cite{xie_ulukus_isit_2013_kic, xie_unified_kic}.
The achievabilities  of $(\frac{1}{2},0), (0,\frac{1}{2})$ are the special cases of Theorem \ref{theo:sdofregion:ic_cm_ee_p_m_n_problem} with $p=0,m=1,N=2$.

\subsection{Achievability}
\label{sec:sdofregion:achievability_p_m_design}

The $(p+1)$-user IC-CM-EE channel with $m$ independent helpers and $N$ independent external eavesdroppers is
\begin{align}
  Y_i & = \sum_{j=1}^{p+1+m} h_{ji} X_j + N_i, \qquad i =1,\ldots,p+1 \\
  Z_k & = \sum_{j=1}^{p+1+m} g_{jk} X_j + N_{z_k}, \qquad k=1,\ldots,N
\end{align}
where $Y_i$ is the channel output of receiver $i$, $Z_k$ is the channel
output of  external eavesdropper $k$, $X_j$ is the
channel input of transmitter $j$, $h_{ji}$ is the channel gain of the
$j$th transmitter to the $i$th receiver, $g_{jk}$ is the channel gain of
the $j$th transmitter to the $k$th eavesdropper, and
$\{N_1,\ldots,N_{p+1},N_{z_1},\ldots,N_{z_N}\}$ are mutually independent zero-mean
unit-variance Gaussian random variables. All the channel gains are
independently drawn from continuous distributions, and are time-invariant
throughout the communication session.  We further assume that all $h_{ji}$ and  $g_{jk}$ are
non-zero. All channel inputs satisfy average power constraints,
$\E\left[X^2_{j}\right] \le P$, for $j=1,\ldots, p+1+m$.

Transmitter $j$, $j=p+2,\ldots,p+1+m$, is an independent helper in the network. On the other hand, each transmitter $i$, $i=1,\ldots,p+1$, has a message $W_i$ intended for the receiver $Y_i$.
A rate tuple $(R_1,\ldots,R_{p+1})$ is said to be achievable if for any $\epsilon>0$, there exist joint $n$-length codes such that each receiver $i$ can decode the corresponding message reliably, i.e., the probability of decoding error is less than $\epsilon$ for all messages,
\begin{equation}
 \max_{i}\pr\left[W_i\neq\hat{W}_i\right] \le \epsilon
\end{equation}
where $\hat{W}_i$ is the estimation based
on its observation $\mathbf{Y}_i $. The secrecy constraints are defined as follows:
\begin{align}
  \frac{1}{n} H(W_{-i}^{p+1}|\bfY_i) & \ge \frac{1}{n} H(W_{-i}^{p+1}) -  \epsilon,
  \qquad i=1,\ldots,p+1 \\
  \frac{1}{n} H(W_1,\ldots, W_{p+1}|\bfZ_k) & \ge \frac{1}{n} H(W_1,\ldots, W_{p+1}) -  \epsilon, \qquad k=1,\ldots,N
\end{align}
where $W_{-i}^{p+1}\defn \{ W_1,\ldots,W_{p+1}\} \backslash \{W_i\}$. A \sdof tuple, $(d_1,\ldots,d_{p+1})$, is achievable if there exists an achievable rate tuple $(R_1,\ldots,R_{p+1})$ such that
\begin{align}
d_i = \lim_{P\to\infty} \frac{R_i}{\frac{1}{2}\log P}
\end{align}
for $i=1,\ldots,p+1$.

Now, we prove Theorem \ref{theo:sdofregion:ic_cm_ee_p_m_n_problem}, i.e., for $p\ge0$, $m\ge1$, and $N$ is finite, the following \sdof tuple is achievable:
\begin{align}
\Big( \frac{m}{m+1},
  \underbrace{ \frac{1}{m+1}, \frac{1}{m+1}, \ldots, \frac{1}{m+1}}_{p \textrm{ items}}
\Big)
\label{eqn:sdofregion:p_m_problem_result_repeat}
\end{align}
for almost all channel gains.

The purpose of Theorem \ref{theo:sdofregion:ic_cm_ee_p_m_n_problem} is to prove the achievability of the structure \eqn{eqn:sdofregion:ep_class_k_m_problem} in Theorem \ref{theo:sdofregion:ic_ee_sdof_region_all_ep}. As shown in \eqn{eqn:sdofregion:ep_class_k_m_problem}, we  partition the transmitters into
three groups: 1) the first group consists of only one transmitter with the largest s.d.o.f.,
$\frac{K-1-p}{K-p}$, which is no smaller than $\frac{1}{2}$, 2) the second group consists of $p\ge0$ transmitters with the same s.d.o.f., $\frac{1}{K-p}$ , which is no larger than $\frac{1}{2}$, and 3) the third group consists of $m\ge1$ transmitters serving as independent helpers. Therefore, in \eqn{eqn:sdofregion:p_m_problem_result_repeat}, we consider the $(p+1)$-user IC with $m$ helpers where $K=p+1+m$.
 Therefore, \eqn{eqn:sdofregion:p_m_problem_result_repeat} and Theorem \ref{theo:sdofregion:ic_cm_ee_p_m_n_problem} show the achievability of \eqn{eqn:sdofregion:ep_class_k_m_problem}. We know from remark 2) above that the achievability of \eqn{eqn:sdofregion:ep_class_all_halves} is a byproduct of Theorem \ref{theo:sdofregion:ic_cm_ee_p_m_n_problem}. Also, \eqn{eqn:sdofregion:ep_class_zero_vector} is trivially achieved, and the achievability of \eqn{eqn:sdofregion:ep_class_symmetric}  is shown in \cite{xie_ulukus_isit_2013_kic, xie_unified_kic}. Therefore, we focus on Theorem \ref{theo:sdofregion:ic_cm_ee_p_m_n_problem}, from this point on.

\begin{figure}[t]
\centerline{\includegraphics[scale=0.8]{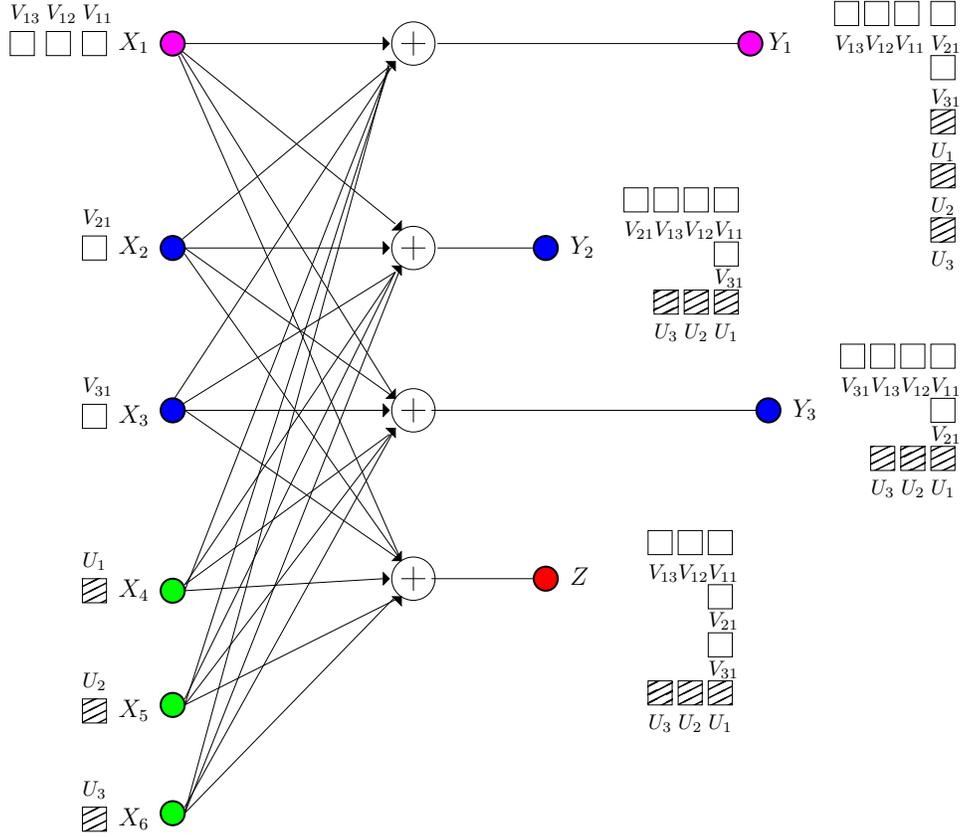}}
\caption{Illustration of secure interference alignment of Theorem \ref{theo:sdofregion:ic_cm_ee_p_m_n_problem} with $m=3,p=2,N=1$.}
\label{fig:sdofregion:kic_6_2_rate_ia}
\end{figure}

The technique we use in the proof of Theorem \ref{theo:sdofregion:ic_cm_ee_p_m_n_problem} is asymptotical interference alignment \cite{real_inter_align_exploit} and structured cooperative jamming \cite{cooperative_jamming}. The  alignment scheme is illustrated in \fig{fig:sdofregion:kic_6_2_rate_ia} with $m=3,p=2,N=1$.
In \fig{fig:sdofregion:kic_6_2_rate_ia}, we partition the transmitters into
three groups, which are $\{X_1\}$ as the first group, $p=2$ other transmitters $\{X_2,X_3\}$ as the second group, and $m=3$ helpers as the third group. From  the perspective of $Y_1$ and the eavesdropper $Z$, due to the existence of independent helpers, the alignment signaling design is similar to that in wiretap channel with helpers in \cite[Fig. 4]{xie_gwch_allerton}. However, from the perspective of $Y_2$, $Y_3$, and the eavesdropper $Z$, the alignment signaling design is similar to that in the interference channel in \cite[Fig. 2]{xie_ulukus_isit_2013_kic} (see the details of the corresponding design in \cite{xie_unified_kic}).
This  suggests that the signalling scheme  that achieves on arbitrary extreme point of the \sdof region is in between the signalling scheme that achieves the sum \sdof of IC-CM-EE in \cite{xie_ulukus_isit_2013_kic,xie_unified_kic} and the signalling scheme used in the helper network in \cite{xie_gwch_allerton}.
Furthermore, if we let $p = 0$, the signaling scheme in \fig{fig:sdofregion:kic_6_2_rate_ia} would be almost identical to \cite[Fig. 4]{xie_gwch_allerton}.
However, we cannot let $m$ be equal to $0$. As far as the number of independent helper(s)  in \fig{fig:sdofregion:kic_6_2_rate_ia},  $m$,  is non-zero, in contrast to the scheme in \cite[Fig.~2]{xie_ulukus_isit_2013_kic}, the legitimate transmitters in the first and second groups do not send cooperative jamming signals by themselves, however, in \cite{xie_ulukus_isit_2013_kic,xie_unified_kic} for IC-CM-EE without helpers, each legitimate transmitter needed to send both message signals and a cooperative signal. Note that in \fig{fig:sdofregion:kic_6_2_rate_ia} here, legitimate transmitters $\{X_1,X_2,X_3\}$ do not send any cooperative jamming signals (no shaded boxes).

\newcommand{\sdofregionlargeconstant}{l}
\newcommand{\sdofregionlargeconstantpower}{\theta}

Here, we give the general achievable scheme. Let
$\sdofregionlargeconstant$ be a large constant. Let us define a set $T_1$ which will represent \emph{dimensions}
as follows:
\begin{align}
\label{eqn:sdofregion:ic_ache_T_1}
  T_1 \defn \left\{
      \left(
        \prod_{(j,k)\in L } h_{jk}^{r_{jk}}
      \right)
      \left(
         \prod_{k=1}^N
         \prod_{j=1}^{p+1+m} g_{jk}^{s_{jk}}
      \right)
      :
      ~r_{jk}, s_{jk} \in \{1,\ldots,\sdofregionlargeconstant\}
    \right\}
\end{align}
where $L$ contains almost all pairs corresponding to the cross-link channel gains
\begin{align}
L
  & = \Big\{ (j,k): j\in \{2,\ldots, p+2\}, k=1 \Big\} \nl
  & \quad \cup \Big\{ (j,k): j\in \{1,\ldots,p+1+m\}, k\in \{2,\ldots,p+1\}, j\neq k\Big\}
  \label{eqn:sdofregion:L_cross_link_channel_gains_set}
\end{align}
Clearly, starting from the second helper $X_{p+3}$, if there exists any, the cross-link channel gains to the first legitimate receiver $Y_1$ are not in the set $L$. Therefore, we define the sets $\{T_j\}_{j=2}^m$
\begin{align}
\label{eqn:sdofregion:ic_ache_T_j}
T_j =\frac{1}{h_{p+1+j,1}} T_1, \quad\quad j=2,\ldots,m
\end{align}
Let $M_i$ be the cardinality of $T_i$, $i=1,\ldots,m$. Note that all $M_i$ are the same, thus we denote them as $M$,
\begin{equation}
  M \defn \sdofregionlargeconstant^{|L| + N (p+1+m)} = \sdofregionlargeconstant^{\sdofregionlargeconstantpower}
\end{equation}
where $\sdofregionlargeconstantpower \defn (p+1+m)p + p+  N (p+1+m) + 1$.

Let $\bft_{ij}$ and $\bft_{(j)}$ be the vector containing all the elements in the set $T_j$ for any possible $i$. 
Therefore, $\bft_{ij}$ and $\bft_{(j)}$ are $M$-dimensional vectors containing $M$ rationally independent real numbers in $T_j$. The sets $\bft_{ij}$ and $\bft_{(j)}$ will represent the \emph{dimensions} along which message signals are transmitted.
In particular, as illustrated in \fig{fig:sdofregion:kic_6_2_rate_ia}, for each legitimate transmitter $i$, $i=1,\ldots,p+1$, the message signal $V_{i1}$ is transmitted in dimensions $\bft_{i1}$.  In order to asymptotically align $U_1$ from the first helper $X_{p+2}$ with all $V_{i1}$s, the cooperative jamming signal $U_{1}$ is transmitted in dimensions $\bft_{(1)}$.
Similarly, for the first transmitter $X_1$,  the message signal $V_{1j}$, $j=2,\ldots,m$, is transmitted in dimensions $\bft_{1j}$. Since we want to align the cooperative jamming signal $U_j$ from the helper $X_{p+1+j}$ with $V_{1j}$ one by one, the jamming  signal $U_{j}$ is transmitted in dimensions $\bft_{(j)}$.

Let us define an $m M$ dimensional vector $\mathbf{b}_1$ by stacking $\bft_{i1}$s as
\begin{align}
\mathbf{b}_1^T
=
\left[
\mathbf{t}_{11}^T,
\mathbf{t}_{12}^T,
\ldots,
\mathbf{t}_{1m}^T
\right]
\end{align}
Then, transmitter $1$ generates a vector $\mathbf{a}_1$, which
contains a total of $mM$ discrete signals each identically and independently
drawn from $C(a,Q)$ given in \eqn{eqn:sdofregion:constel}. For convenience, we partition this transmitted signal as
\begin{align}
\label{eqn:sdofregion:ic_ache_a_1}
\mathbf{a}_1^T
=
\left[
\mathbf{v}_{11}^T,
\mathbf{v}_{12}^T,
\ldots,
\mathbf{v}_{1m}^T
\right]
\end{align}
where $\bfv_{1j}$ represents the information symbols in $V_{1j}$. Each of these vectors has length $M$, and therefore, the total length of $\mathbf{a}_1$ is $mM$. The channel input of transmitter $1$ is
\begin{equation}
\label{eqn:sdofregion:ic_ache_x_1}
x_1 = \bfa_1^T \bfb_1
\end{equation}

Similarly, for the second group transmitters $X_{i}$,
$i=2,\ldots,p+1$, let $\mathbf{b}_i$ be
$  \mathbf{b}_i =  \mathbf{t}_{i1}$.
Then,  transmitter $i$ generates a vector $\mathbf{a}_i=\mathbf{v}_{i1}$, which
contains a total of $M$ discrete signals each identically and independently
drawn from $C(a,Q)$ given in \eqn{eqn:sdofregion:constel}.
The channel input of transmitter $i$ is
\begin{equation}
\label{eqn:sdofregion:ic_ache_x_i}
x_i = \bfa_i^T \bfb_i = \mathbf{v}_{i1}^T \mathbf{t}_{i1}, \quad\quad i=2,\ldots,p+1
\end{equation}

Finally, for the third group transmitters $X_{k}$,
$k=p+2,\ldots,p+1+m$, serving as the helpers, let $\mathbf{b}_{k}$ be
$  \mathbf{b}_{k} =  \mathbf{t}_{(k-p-1)} $.
Then,  helper $k$ generates a vector $\mathbf{u}_{k-p-1}$ representing the cooperative jamming signal in $U_{k-p-1}$, which
contains a total of $M$ discrete signals each identically and independently
drawn from $C(a,Q)$ given in \eqn{eqn:sdofregion:constel}.
The channel input of transmitter $k$ is
\begin{equation}
\label{eqn:sdofregion:ic_ache_x_k}
x_k = \bfu_{k-p-1}^T \bfb_k = \mathbf{u}_{k-p-1}^T \mathbf{t}_{(k-p-1)}, \quad\quad k=p+2,\ldots,p+1+m
\end{equation}

Before we investigate the performance of this signalling scheme, we analyze the structure of
the received signals at the receivers. To see the detailed dimension structure of the received signals at the receivers, let us define $\tilde{T}_i$ as a superset of $T_i$, as follows
\begin{align}
  \tilde{T}_{1} & \defn \left\{
      \left(
        \prod_{(j,k)\in L } h_{jk}^{r_{jk}}
      \right)
      \left(
         \prod_{k=1}^N
         \prod_{j=1}^{p+1+m} g_{jk}^{s_{jk}}
      \right)
      :
      ~r_{jk}, s_{jk} \in \{1,\ldots,\sdofregionlargeconstant+1\}
    \right\} \\
  \tilde{T}_j & =\frac{1}{h_{p+1+j,1}} \tilde{T}_1, \quad\quad j=2,3,\ldots,m
\end{align}
where $L$ is defined in \eqn{eqn:sdofregion:L_cross_link_channel_gains_set} and the cardinalities of all $T_i$ sets are the same and are denoted as $\tilde{M}=(\sdofregionlargeconstant+1)^{\sdofregionlargeconstantpower} $.  Also, it is easy to check that since pair $(p+1+j,1)\not\in L$ for $j\ge2$, we must have
\begin{align}
\label{eqn:sdofregion:ic_ache_t_i_t_j_not_intersaction}
\tilde{T}_i \cap \tilde{T}_j = \phi
\end{align}
for all $i\neq j$.

We first focus on
receiver 1, which has the channel output
\begin{align}
\label{eqn:sdofregion:ic_ache_y_1}
y_1 = \sum_{i=1}^{p+1+m} h_{i1} x_1  + n_1
\end{align}
Substituting \eqn{eqn:sdofregion:ic_ache_x_1}, \eqn{eqn:sdofregion:ic_ache_x_i} and \eqn{eqn:sdofregion:ic_ache_x_k} into \eqn{eqn:sdofregion:ic_ache_y_1}, we get
\begin{align}
y_1 & =
 h_{11} x_1 + \sum_{j=2}^{p+1} h_{j1} x_{j} + \sum_{k=p+2}^{p+1+m} h_{k1} x_k
  +
  n_1 \\
  & = h_{11}
  \left(
    \sum_{i=1}^m \mathbf{v}_{1i}^T \mathbf{t}_{1i}
  \right)
  +
  \left(
    \sum_{j=2}^{p+1} h_{j1}\mathbf{v}_{j1}^T \mathbf{t}_{j1}
  \right)
  +
  \left(
    \sum_{k=p+2}^{p+1+m} h_{k1} \mathbf{u}_{k-p-1}^T \mathbf{t}_{(k-p-1)}
  \right)
  +
  n_1 \\
 & =
  \Big(
     \mathbf{v}_{11}^T h_{11} \mathbf{t}_{11}
  \Big)
  +
  \Big(
     \mathbf{v}_{12}^T h_{11} \mathbf{t}_{12}
  \Big)
  +
  \ldots
  +
  \Big(
     \mathbf{v}_{1m}^T h_{11} \mathbf{t}_{1m}
  \Big)
     \nl
  & \quad\quad +
  \Big(
    \sum_{j=2}^{p+1} h_{j1}\mathbf{v}_{j1}^T \mathbf{t}_{j1}
  +
    \sum_{k=p+2}^{p+1+m} h_{k1} \mathbf{u}_{k-p-1}^T \mathbf{t}_{(k-p-1)}
  \Big)
  +
  n_1
  \label{eqn:sdofregion:ic_ache_y_1_dimension_split}
\end{align}
Since $\mathbf{v}_{ij}$ and $\mathbf{u}_{k-p-1}$ are integer signals in $C(a,Q)$, it suffices to study their dimensions. In addition, note that $ \mathbf{t}_{ij}$ and $\mathbf{t}_{(j)}$ represent the same dimensions in $T_j$ defined in \eqn{eqn:sdofregion:ic_ache_T_1} and \eqn{eqn:sdofregion:ic_ache_T_j}.
It is easy to verify that
\begin{align}
h_{j1} T_1 &\subseteq  \tilde{T}_1, \quad\quad j=2,\ldots,p+1 \\
h_{k1} T_{k-p-1} & \subseteq \tilde{T}_1, \quad\quad k=p+2,\ldots,p+1+m
\end{align}
which implies that except the intended message signals $\mathbf{v}_{1i}$, $i=1,\ldots,m$, all unintended signals including message signals and cooperative jamming signals are all transmitted in the dimensions belonging to $\tilde{T}_1$. On the other hand, for intended signals,
\begin{align}
h_{11}  T_1 & \subset h_{11} \tilde{T}_1 \\
h_{11} T_i & \subseteq h_{11} \tilde{T}_i = \frac{h_{11}}{h_{p+1+i,1}} \tilde{T}_1,\quad\quad i=2,\ldots,m
\end{align}
Note that the pair $(p+1+i,1)\not\in L$ for $i\ge2$ which implies that
\begin{align}
\label{eqn:sdofregion:ic_ache_t_i_j_are_separable}
h_{11} \tilde{T}_i \cap h_{11} \tilde{T}_j =\phi
\end{align}
for all $i,j\in \{1,\ldots,m\}$, $i\neq j$. Furthermore, $(1,1)\not\in L$ either, which implies that
\begin{align}
h_{11} \tilde{T}_i \cap \tilde{T}_1 = \phi,\quad\quad i\in \{1,\ldots,m\}
\end{align}
Together with \eqn{eqn:sdofregion:ic_ache_t_i_j_are_separable}, this indicates that the dimensions are separable as suggested by the parentheses in \eqn{eqn:sdofregion:ic_ache_y_1_dimension_split} and also the $Y_1$ side of \fig{fig:sdofregion:kic_6_2_rate_ia},
which further implies that all the elements in the set
\begin{equation}
  R_1 \defn
  \left(
    \displaystyle \bigcup_{j=1}^m h_{11} \tilde{T}_j
  \right)
    \cup
    \tilde{T}_1
\end{equation}
are rationally independent, and thereby the cardinality of $R_1$ is
\begin{align}
\label{eqn:sdofregion:ic_ache_m_r}
M_R
\defn |R_1|
& = (m+1) \tilde{M} = (m+1) (\sdofregionlargeconstant+1)^{\sdofregionlargeconstantpower}
\end{align}

For the legitimate receivers $Y_i$, $i=2,\ldots,p+1$, without loss of generality, we focus on
receiver 2; by symmetry, a similar structure will exist
at all other receivers.  We observe that
\begin{align}
y_2 & =
 h_{12} x_1 + \sum_{j=2}^{p+1} h_{j2} x_{j} + \sum_{k=p+2}^{p+1+m} h_{k2} x_k + n_2\\
 & =
  h_{12}
  \left(
    \sum_{i=1}^m \mathbf{v}_{1i}^T \mathbf{t}_{1i}
  \right)
  +
  \left(
    \sum_{j=2}^{p+1} h_{j2}\mathbf{v}_{j1}^T \mathbf{t}_{j1}
  \right)
  +
  \left(
    \sum_{k=p+2}^{p+1+m} h_{k2} \mathbf{u}_{k-p-1}^T \mathbf{t}_{(k-p-1)}
  \right)
  +
  n_2 \\
  & =
  \Big(
    h_{22}\mathbf{v}_{21}^T \mathbf{t}_{21}
  \Big)
  +
  \Big(
     \mathbf{v}_{11}^T h_{12}\mathbf{t}_{11}
     +
     \sum_{j=3}^{p+1}\mathbf{v}_{j1}^T  h_{j2}\mathbf{t}_{j1}
     +
    \mathbf{u}_{1}^T   h_{p+2,2}  \mathbf{t}_{(1)}
  \Big)
  \nl
  & \quad\quad +
  \left(
    \mathbf{v}_{12}^T h_{12} \mathbf{t}_{12} +
      \mathbf{u}_{2}^T  h_{p+3,2} \mathbf{t}_{(2)}
  \right)
  +
  \ldots
  +
  \left(
    \mathbf{v}_{1m}^T h_{12} \mathbf{t}_{1m} +
      \mathbf{u}_{m}^T h_{p+1+m,2} \mathbf{t}_{(m)}
  \right)
  +
  n_2
  \label{eqn:sdofregion:ic_ache_y_2_dimension_split}
\end{align}
Similarly, we observe that in the second set of parentheses of \eqn{eqn:sdofregion:ic_ache_y_2_dimension_split}, since $\mathbf{t}_{i1}$ and $\mathbf{t}_{(1)}$ represent the same dimensions in $T_1$ for all $i$, we have
\begin{align}
h_{i2} T_1 & \subseteq \tilde{T}_1,\quad\quad i\in\{1,\ldots,p+2\}, i \neq 2
\end{align}
Starting from the third set of parentheses of \eqn{eqn:sdofregion:ic_ache_y_2_dimension_split}, we have
\begin{align}
h_{12} T_{j} & \subseteq \tilde{T}_j  \\
h_{p+1+j,2}  T_{j} & \subseteq \tilde{T}_j
\end{align}
for all $j=2,\ldots,m$. In addition, since the pair $(2,2)\not\in L$, we can infer that
\begin{align}
h_{22} T_1 \subseteq h_{22} \tilde{T}_1
\end{align}
and
\begin{equation}
 h_{22} \tilde{T}_1 \cap  \tilde{T}_j
\end{equation}
for $j=1,\ldots,m$. Together with \eqn{eqn:sdofregion:ic_ache_t_i_t_j_not_intersaction},
this indicates that the dimensions are separable as suggested by the parentheses in \eqn{eqn:sdofregion:ic_ache_y_2_dimension_split} and also the $Y_2$ side of \fig{fig:sdofregion:kic_6_2_rate_ia},
which further implies that all the elements in the set
\begin{equation}
  R_2 \defn
  \left(
    \displaystyle \bigcup_{j=1}^m \tilde{T}_j
  \right)
    \cup
     h_{22} \tilde{T}_1
\end{equation}
are rationally independent, and thereby the cardinality of $R_2$ is $M_R$ in \eqn{eqn:sdofregion:ic_ache_m_r}.

For the external eavesdropper $Z_k$, we note that
\begin{align}
z_k & =
 g_{1k} x_1 + \sum_{j=2}^{p+1} g_{jk} x_{j} + \sum_{i=p+2}^{p+1+m} g_{ik} x_i + n_{z_k}\\
 & =
  g_{1k}
  \left(
    \sum_{i=1}^m \mathbf{v}_{1i}^T \mathbf{t}_{1i}
  \right)
  +
  \left(
    \sum_{j=2}^{p+1} g_{jk} \mathbf{v}_{j1}^T \mathbf{t}_{j1}
  \right)
  +
  \left(
    \sum_{i=p+2}^{p+1+m} g_{ik} \mathbf{u}_{i-p-1}^T \mathbf{t}_{(i-p-1)}
  \right)
  +
  n_{z_k} \\
  & =
  \Big(
     \mathbf{v}_{11}^T g_{1k} \mathbf{t}_{11}
     +
     \sum_{j=2}^{p+1}\mathbf{v}_{j1}^T  g_{jk} \mathbf{t}_{j1}
     +
    \mathbf{u}_{1}^T   g_{p+2,k}  \mathbf{t}_{(1)}
  \Big)
  \nl
  & \quad\quad +
  \left(
    \mathbf{v}_{12}^T g_{1k} \mathbf{t}_{12} +
      \mathbf{u}_{2}^T  g_{p+3,k} \mathbf{t}_{(2)}
  \right)
  +
  \ldots
  +
  \left(
    \mathbf{v}_{1m}^T g_{1k} \mathbf{t}_{1m} +
      \mathbf{u}_{m}^T g_{p+1+m,k} \mathbf{t}_{(m)}
  \right)
  +
  n_{z_k}
  \label{eqn:sdofregion:ic_ache_y_z_dimension_split}
\end{align}
In the first set of parentheses of \eqn{eqn:sdofregion:ic_ache_y_z_dimension_split}, since $\mathbf{t}_{i1}$ and $\mathbf{t}_{(1)}$ represent the same dimensions in $T_1$ for all $i$, we have
\begin{align}
g_{ik} T_1 & \subseteq \tilde{T}_1,\quad\quad i\in\{1,\ldots,p+2\}
\end{align}
Starting from the second set of parentheses of \eqn{eqn:sdofregion:ic_ache_y_z_dimension_split}, we have
\begin{align}
g_{1k} T_{j} & \subseteq \tilde{T}_j  \\
g_{p+1+j,k}  T_{j} & \subseteq \tilde{T}_j
\end{align}
for all $j=2,\ldots,m$. Due to \eqn{eqn:sdofregion:ic_ache_t_i_t_j_not_intersaction},
this indicates that the dimensions are separable as suggested by the parentheses in \eqn{eqn:sdofregion:ic_ache_y_z_dimension_split} and also the $Z$ side of \fig{fig:sdofregion:kic_6_2_rate_ia},
which further implies that all the elements in the set
\begin{equation}
  R_Z \defn
  \left(
    \displaystyle \bigcup_{j=1}^m \tilde{T}_j
  \right)
\end{equation}
are rationally independent, and thereby the cardinality of $R_Z$ is $M_{R_Z}$
\begin{align}
M_{R_Z} \defn |R_Z| = m \tilde{M} = m (\sdofregionlargeconstant+1)^{\sdofregionlargeconstantpower}
\end{align}

We will compute the secrecy rates achievable with the asymptotic alignment based scheme proposed above by using the following theorem.
\begin{theorem}[{{\!\!\cite[Theorem 2]{xie_unified_kic}}}]
\label{theorem:sdofregion:kic:theo-achievability}
For $K'$-user interference channels with confidential messages, the following rate region is achievable
\begin{equation}
R_i\ge I(V_i;Y_i) - {\max_{j\in\mathcal{K}'_{-i}}} I(V_i; Y'_j|V_{-i}^{K'}), \qquad i=1,\ldots,K'
\label{eqn:sdofregion:kic-lower-bound}
\end{equation}
where $V_{-i}^{K'} \defn \{V_j\}_{j=1,j\neq i}^{K'}$ and $\mathcal{K'}_{-i} = \{1,\ldots, i-1,i+1, \ldots,K'\}$. The auxiliary random variables
$\{V_i\}^{K'}_{i=1}$ are mutually independent, and for each $i$, we
have the following Markov chain $V_i\rightarrow X'_i\rightarrow
(Y'_1,\ldots,Y'_{K'})$.
\end{theorem}

We can reinterpret Theorem \ref{theorem:sdofregion:kic:theo-achievability} as follows: For the $(p+1)$-user IC-CM-EE with $m$ helpers and $N$ external eavesdroppers, since each independent helper's contribution is the same as noise to both items in \eqn{eqn:sdofregion:kic-lower-bound}, which depend only on marginal distributions,  we can treat the $(p+1)$-user IC-CM-EE channel as a $(p+1+N)$-user IC-CM with $N$ new transmitters which keep silent, i.e., $V_i$ and $X'_i$, $i=p+2,\ldots,p+1+N$, are equal to zero, and
\begin{align}
p(y'_1,\ldots,y'_{p+1+N}|x'_1,\ldots,x'_{p+1+N})=p(y_1,\ldots,y_{p+1},z_1,\ldots,z_N|x_1,\ldots,x_{p+1})
\end{align}
where $x'$ and $y'$ are the transmitter and receiver of the $(p+1+N)$-user IC-CM and $x,y,z$ are the entities of the original $(p+1)$-user IC-CM-EE with $m$ helpers and $N$ external eavesdropper.

We thereby first select $V_i$ as
\begin{align}
V_1 & \defn \mathbf{a}_1 \\
V_i & \defn \mathbf{v}_{i1}, \quad\quad i=2,\ldots,p+1
\end{align}
where  $\mathbf{a}_1$ is defined in \eqn{eqn:sdofregion:ic_ache_a_1}.
Then, we evaluate the \eqn{eqn:sdofregion:kic-lower-bound} for $i=1,\ldots,p+1$.

For $i=1$, by Lemma \ref{lemma:sdofregion:ria_real_alignment}, for any $\delta>0$, if we choose $Q =
 P^{\frac{1-\delta}{2(M_R+\delta)}}$ and $a = \frac{\gamma_1
P^{\frac{1}{2}}}{Q}$,
the probability of error of estimating $V_1$ as $ \tilde{V}_1$ based on $Y_1$
can be upper bounded by
\begin{equation}
  \mathrm{Pr}(e_1)
\le
\exp\left( - \eta_{\gamma_1} P^{{ \delta}} \right)
\end{equation}
Furthermore, by Fano's inequality, we can conclude that
\begin{align}
  I(V_1; Y_1) & \le I(V_1; \tilde{V}_1) \\
  & = H(V_1) - H(V_1|\tilde{V}_1)\\
  & \ge
\frac{  m {M} (1-\delta)} { M_R + \delta }
\left(\frac{1}{2} \log P\right) + o(\log P)\\
& = \frac
       { m   (1-\delta)}
       { (m+1) \left( 1 + \frac{1}{\sdofregionlargeconstant}
\right)^{\sdofregionlargeconstantpower} + \frac{\delta}{\sdofregionlargeconstant^{\sdofregionlargeconstantpower}} }
\left(\frac{1}{2} \log P\right) + o(\log P)
\label{eqn:sdofregion:ic_ivy_1_low}
\end{align}
where $o(\cdot)$ is the little-$o$ function. This provides a lower bound for the first term in \eqn{eqn:sdofregion:kic-lower-bound} with $i=1$.

Next, we need to derive an upper bound for the second item in \eqn{eqn:sdofregion:kic-lower-bound}, i.e, the secrecy penalty, for $i=1$. For and $j\in\{2,\ldots,p+1\}$,  by the Markov chain,
\begin{equation}
V_1 \rightarrow \left(\sum_{k=1}^{p+1} h_{kj} X_{kj} , V_{2}^{p+1} \right) \rightarrow Y_j
\end{equation}
we have
\begin{align}
 I(V_1; Y_j | V_{2}^{p+1} )
 & \le I\left( V_1; \sum_{k=1}^{p+1} h_{kj} X_k \Big| V_{2}^{p+1} \right) \\
 & =  H\left( \sum_{k=1}^{p+1} h_{kj} X_k \Big| V_{2}^{p+1} \right)
     - H\left(\sum_{k=1}^{p+1} h_{kj} X_k \Big| V_{1}^{p+1} \right)
 \label{eqn:sdofregion:ic_secrecy_v_1_yj}
\end{align}
The first term in \eqn{eqn:sdofregion:ic_secrecy_v_1_yj} can be rewritten as
\begin{align}
 H\left( \sum_{k=1}^{p+1} h_{kj} X_k \Big| V_{2}^{p+1} \right)
& = H\left[
  \sum_{i=k}^m
    \Big(
     \mathbf{v}_{1k}^T h_{1j}\mathbf{t}_{1k}
     + \mathbf{u}_{k}^T   h_{p+1+k,j}  \mathbf{t}_{(k)}
    \Big)
  \right]
\end{align}
Note that there are in total $m M_R$ rational dimensions each taking value from $C(a,2Q)$.
Regardless of the distribution in each rational dimension, the entropy is maximized by uniform distribution, i.e.,
\begin{align}
H\left( \sum_{k=1}^{p+1} h_{kj} X_k \Big| V_{2}^{p+1} \right)
\le \log \left[ (2Q+1)^{m \tilde{M}} \right]
= \frac{ m \tilde{M} (1-\delta) }{M_R +\delta } \left( \frac{1}{2} \log P \right) + o(\log P)
\label{eqn:sdofregion:ic_secrecy_v_1_yj_1}
\end{align}
The second term in \eqn{eqn:sdofregion:ic_secrecy_v_1_yj} is
\begin{align}
H\left(\sum_{k=1}^{p+1} h_{kj} X_k \Big| V_{1}^{p+1} \right)
& =
H\left[
  \sum_{i=k}^m
    \Big(
      \mathbf{u}_{k}^T   h_{p+1+k,j}  \mathbf{t}_{(k)}
    \Big)
  \right]
=
\log \left[ (2Q+1)^{m M} \right] \\
& = \frac{ m M (1-\delta) }{M_R +\delta } \left( \frac{1}{2} \log P \right) + o(\log P)
\label{eqn:sdofregion:ic_secrecy_v_1_yj_2}
\end{align}
Substituting \eqn{eqn:sdofregion:ic_secrecy_v_1_yj_1} and \eqn{eqn:sdofregion:ic_secrecy_v_1_yj_2} into \eqn{eqn:sdofregion:ic_secrecy_v_1_yj}, we get
\begin{align}
\label{eqn:sdofregion:ic_secrecy_v_1_yj_3}
 I(V_1; Y_j | V_{2}^{p+1} ) \le \frac{ m (\tilde{M} - M) (1-\delta) }{M_R +\delta } \left( \frac{1}{2} \log P \right) + o(\log P)
\end{align}
\newcommand{\asymptoticsmall}{\xi}
We note that
\begin{align}
\asymptoticsmall \defn \frac{  m (\tilde{M} - M) (1-\delta)}{M_R +\delta }
& = \frac{  m (\tilde{M} - M) (1-\delta) }{ (m+1) \tilde{M} +\delta } \\
& = \frac{
      m \left[(\sdofregionlargeconstant+1)^{\sdofregionlargeconstantpower}
      - \sdofregionlargeconstant^{\sdofregionlargeconstantpower}\right] (1-\delta)
    }{
      (m+1) (\sdofregionlargeconstant+1)^{\sdofregionlargeconstantpower} + \delta
    } \\
& = \frac{
      m \left[ \sum_{k=0}^{\sdofregionlargeconstantpower-1 }
        {\sdofregionlargeconstantpower \choose k } l^k\right] (1-\delta)
    }{
      (m+1) (\sdofregionlargeconstant+1)^{\sdofregionlargeconstantpower} + \delta
    }
\end{align}
The maximum power of $\sdofregionlargeconstant$ in the numerator  is $\sdofregionlargeconstantpower-1$ and is less than the power $\sdofregionlargeconstantpower$ of $\sdofregionlargeconstant$ in the denominator. This implies that when $m$ and $\delta$ are fixed, by choosing $\sdofregionlargeconstant$ large enough,
 the factor before the
$\frac{1}{2}\log P$ term in \eqn{eqn:sdofregion:ic_secrecy_v_1_yj_3}, $\asymptoticsmall$, can be made arbitrarily small. Due to the non-perfect (i.e., only asymptotical) alignment, the upper bound for the information leakage rate is not a constant as in \cite{xie_sdof_networks_in_prepare}, but a function which can be made to approach zero d.o.f.

Similarly, we can derive the following
\begin{align}
 I(V_1; Z_k| V_{2}^{p+1} )
 & \le \asymptoticsmall \left( \frac{1}{2} \log P \right) + o(\log P)
 \label{eqn:sdofregion:ic_secrecy_v_1_y_0}
\end{align}
where $Z_k$, $k=1,\ldots,N$, is the external eavesdropper.
Substituting \eqn{eqn:sdofregion:ic_ivy_1_low}, \eqn{eqn:sdofregion:ic_secrecy_v_1_yj_3} and
\eqn{eqn:sdofregion:ic_secrecy_v_1_y_0} into
\eqn{eqn:sdofregion:kic-lower-bound}, we obtain a lower bound for the achievable
secrecy rate $R_1$ as
\begin{equation}
  R_1 \ge \left[ \frac
       { m   (1-\delta)}
       { (m+1) \left( 1 + \frac{1}{\sdofregionlargeconstant}
\right)^{\sdofregionlargeconstantpower} + \frac{\delta}{\sdofregionlargeconstant^{\sdofregionlargeconstantpower}} }
           - \asymptoticsmall \right]
\left(\frac{1}{2} \log P\right) + o(\log P)
\end{equation}
Similarly,  it is easy to derive that
\begin{equation}
  R_i \ge \left[ \frac
       {   (1-\delta)}
       { (m+1) \left( 1 + \frac{1}{\sdofregionlargeconstant}
\right)^{\sdofregionlargeconstantpower} + \frac{\delta}{\sdofregionlargeconstant^{\sdofregionlargeconstantpower}} }
           - \asymptoticsmall' \right]
\left(\frac{1}{2} \log P\right) + o(\log P)
\end{equation}
for $i=2,\ldots,p+1$ and $\asymptoticsmall'$ can be made arbitrarily small.
By choosing  $\sdofregionlargeconstant\to\infty$ and $\delta\to0$, we can achieve a \sdof tuple
arbitrarily close to
\begin{align}
\Big(
  \frac{m}{m+1},
  \underbrace{\frac{1}{m+1}, \ldots, \frac{1}{m+1}}_{p \textrm{ items}},
  \Big)
\end{align}
which is \eqn{eqn:sdofregion:ic_cm_ee_p_m_n_tuple}, completing the proof of Theorem \ref{theo:sdofregion:ic_cm_ee_p_m_n_problem}.

\section{Conclusions}
In this  paper, we determined the \emph{entire \sdof regions} of
$K$-user MAC wiretap channel, $K$-user IC-EE, $K$-user IC-CM, and
$K$-user IC-CM-EE. The converse for MAC directly followed from the results in
\cite{xie_ulukus_isit_2013_mac, xie_sdof_networks_in_prepare}.
The converse for IC was shown to be dominated by secrecy constraints and interference constraints
in different parts. To show the tightness and achieve the regions
characterized by the converses, we provided a general method to
investigate this class of channels, whose \sdof regions have a polytope structure. We provided an equivalence
between the extreme points in the polytope structure and the rank of
sub-matrices containing all active upper bounds associated with each
extreme point. Then, we achieved each extreme point by relating it to
a specific channel model.  More specifically, the extreme
points of the MAC region can be achieved by an $m$-user MAC wiretap
channel with $K-m$ helpers, i.e., by setting $K-m$ users' secure rates
to zero and utilizing them as pure (structured) cooperative jammers.
On the other hand, the
asymmetric  extreme points of the IC region can be achieved by a $(p+1)$-user
IC-CM  with $m$ helpers, and $N$
external eavesdroppers.

\appendix
\section{Proof of Theorem \ref{theo:sdofregion:ic_ee_sdof_region_all_ep}}
\label{sec:sdofregion:proof_of_ic_ep_structure}

Regarding  Theorem \ref{theo:sdofregion:ic_ee_sdof_region_all_ep}, first, we have few comments:
\begin{enumerate}
\item[1)] \eqn{eqn:sdofregion:ep_class_all_halves} will not be possible until $K\ge 5$ due to the constraint $K-2\ge p' \ge 3$.
\item[2)] The point in \eqn{eqn:sdofregion:ep_class_all_halves} with
  $p'=K-1$, i.e., $(\frac{1}{2},\frac{1}{2},\ldots,\frac{1}{2},0)$,  is actually an extreme
  point, but since \eqn{eqn:sdofregion:ep_class_k_m_problem} with
  $p=K-2$ also includes it, we classify it as
  \eqn{eqn:sdofregion:ep_class_k_m_problem} here.
\item[3)] Assume that we allow $p'=2$ in
  \eqn{eqn:sdofregion:ep_class_all_halves} with $K\ge 5$. Then, the
  point becomes
  \begin{equation}
    \mathbf{d}_1 = \left(\frac{1}{2},\frac{1}{2},0,0,\ldots,0\right)
  \end{equation}
  However, this is just the
  middle point of two points in
  \eqn{eqn:sdofregion:ep_class_k_m_problem}. More specifically, by
  choosing $p=1$ in  \eqn{eqn:sdofregion:ep_class_k_m_problem}, we
  have  $\mathbf{d}_1'= (\frac{K-2}{K-1},\frac{1}{K-1},0,0,\ldots,0)$
  and $\mathbf{d}_1''=(\frac{1}{K-1},\frac{K-2}{K-1},0,0,\ldots,0)$
  (by swapping the first two elements in $\mathbf{d}_1'$). Here
  $\mathbf{d}_1'\neq \mathbf{d}_1''$ due to $K\ge 5$, and also it is
  easy to check that $\mathbf{d}_1 = \frac{1}{2}
  (\mathbf{d}_1'+\mathbf{d}_1'')$, which means that $\mathbf{d}_1$ is
  not an extreme point. Therefore,  in
  \eqn{eqn:sdofregion:ep_class_all_halves} $p'$ must satisfy $p'\ge3$.
\end{enumerate}

Now, we start the proof of Theorem \ref{theo:sdofregion:ic_ee_sdof_region_all_ep}. In order to speak of a polytope, we re-write \eqn{eqn:sdofregion:ic_ee_converse_3} as
\begin{align}
- d_i \le 0, \quad\quad i=1,\ldots,K
\label{eqn:sdofregion:ic_ee_converse_4}
\end{align}
Then, we can write all the left hand sides of \eqn{eqn:sdofregion:ic_ee_converse_1}, \eqn{eqn:sdofregion:ic_ee_converse_2}, \eqn{eqn:sdofregion:ic_ee_converse_4} as an $N \times K $ matrix $\mathbf{H}$ with corresponding right hand sides forming an $N$-length column vector $\mathbf{h}$, i.e.,  all points $\mathbf{d}$ in $D$ satisfy
\begin{equation}
\label{eqn:sdofregion:polytope_standard_form_for_ic_problem}
\mathbf{H} \mathbf{d} \le \mathbf{h}
\end{equation}
where $N \defn 2 K +  {K \choose 2} = 2K + K(K-1)/2$. For any extreme point $\mathbf{d}\in D$,  let $ J(\mathbf{d})$ be a set such that
\begin{align}
J(\mathbf{d}) = \Big\{l: \mathbf{H}_l \mathbf{d} = \mathbf{h}_l, \ \  l\in \{1,\ldots,N\} \Big\}
\end{align}
where $\mathbf{H}_l$ is
the $l$th row of $\mathbf{H}$ and $\mathbf{h}_l$ is the $l$th element of $\mathbf{h}$. Therefore, $J(\mathbf{d})$ represents all active boundaries.  The remaining rows  satisfy
\begin{align}
\label{eqn:sdofregion:ep_proof_constraints_on_j_complement}
\mathbf{H}_l \mathbf{d} < \mathbf{h}_l
\end{align}
for $l\not\in J$.

For convenience, denote by $\mathbf{H}_J$ the sub-matrix of $\mathbf{H}$ with rows indexed by $J \defn J(\mathbf{d}) $. Similarly denote by $\mathbf{h}_J$ the sub-vector of $\mathbf{h}$ with rows indexed by $J$.
In order to find all extreme points in $D$, by Theorem \ref{theo:sdofregion:polyhedron_ep_rank} in Section \ref{sec:sdof_region:preliminaries_polytope_and_ep}, we need to find all $K\times (K+1)$ sub-matrices $(\mathbf{H}',\mathbf{h}')$ of $(\mathbf{H},\mathbf{h})$ with $\rank(\mathbf{H}')=K$ such that $\mathbf{H} \mathbf{d} \le \mathbf{h}$ and $\mathbf{H}' \mathbf{d} = \mathbf{h}'$, which is also equivalent to finding all index sets  $J$ representing the active boundaries  such that $\mathbf{H} \mathbf{d} \le \mathbf{h}$, $\mathbf{H}_J \mathbf{d} = \mathbf{h}_J$, and $\rank(\mathbf{H}_J) = K$.

For convenience of presentation, we always partition the set $J$ as a union of mutually exclusive sets $S,  P$ and $Z$, i.e.,
\begin{align}
J = S  \cup P  \cup Z
\end{align}
We denote by $S$ the row indices representing the active boundaries in \eqn{eqn:sdofregion:ic_ee_converse_1}
\begin{align}
\label{eqn:sdofregion:ic_ee_defn_s}
S & \defn \Big\{ s_i\defn s(i): \mathbf{H}_{s_i} \mathbf{d} = h_{s_i} \textrm{ is } (K-1) d_i +  \sum_{j=1}^K d_j  = K-1, ~~ i=1,\ldots,K  \Big\}
\end{align}
where $s_i$ stands for the function $s(i)$  of the coordinate $i$ with the value as the row index of $\mathbf{H}$ corresponding to the active boundaries $(K-1) d_i +  \sum_{j=1}^K d_j  = K-1$. Thus, we have a one-to-one mapping between the row index and the function $ s_i\defn s(i) $, i.e., if the row index $s_i\in J$, we know exactly the $i$th upper bound in \eqn{eqn:sdofregion:ic_ee_converse_1} is active; on the other hand, if we know the coordinate $i$, we can determine the unique corresponding row index in $\mathbf{H}$ by the mapping $s: i\mapsto s_i $.

Similarly, we denote by $P$  the row indices representing the active boundaries in \eqn{eqn:sdofregion:ic_ee_converse_2}
\begin{align}
\label{eqn:sdofregion:ic_ee_defn_p}
P & \defn \Big\{ p_V \defn p(V): \mathbf{H}_{p_V} \mathbf{d} = h_{p_V} \textrm{ is } \sum_{i\in V} d_i =1, ~~V\subseteq \{1,\ldots,K\}, |V|=2   \Big\}
\end{align}
where the value of $p_V$ is the corresponding row index of $\mathbf{H}$.

Finally, we denote by $Z$  the row indices representing the active boundaries in \eqn{eqn:sdofregion:ic_ee_converse_4}
\begin{align}
Z & \defn \Big\{ z_i\defn z(i): \mathbf{H}_{z_i} \mathbf{d} = h_{z_i} \textrm{ is } d_i =0, ~~i=1,\ldots,K \Big\}
\end{align}
where the value of $z_i$ is the corresponding row index of $\mathbf{H}$.

\newcounter{counter:sdofregion:ic_ee_sdof_region_ep}
\newcommand{\increaseandref}[1]{\stepcounter{#1}{\bf \arabic{#1}}}
\newcommand{\increaseandreflabel}[2]{\refstepcounter{#1}\label{#2}\arabic{#1}}
\newcommand{\propertyep}[1]{property \ref{#1}) of Lemma \ref{property:sdofregion:ic_ee_sdof_region_ep}}
\newcommand{\ipropertyep}[1]{property \ref{#1})}

There are approximately in total
\begin{equation}
\label{eqn:sdofregion:approximation_of_matrix_selection}
{N \choose K} \approx \frac{ \left(\frac{K+2}{2}\right)^K e^K } { \sqrt{2\pi K}}
\end{equation}
possible selections of $K$ equations in   \eqn{eqn:sdofregion:polytope_standard_form_for_ic_problem} for large $K$. In order for this search to have a reasonable complexity, we need to investigate the structure of $D$ more carefully. We identify the following simple properties for the extreme points in the following lemmas.
\begin{lemma}
\label{property:sdofregion:ic_ee_sdof_region_ep}
Let $\mathbf{d}$ be a non-zero extreme point in $D$.
Then, it must satisfy the following properties:

\noindent \increaseandreflabel{counter:sdofregion:ic_ee_sdof_region_ep}
{counter:sdofregion:ic_ee_sdof_region_ep_less_than_k_minus_1_over_})
$\max_k d_k \le \frac{K-1}{K}$.

\noindent \increaseandreflabel{counter:sdofregion:ic_ee_sdof_region_ep}
{counter:sdofregion:ic_ee_sdof_region_ep_at_most_one_larger_than_half})
At most one element, if there is any, in  $\mathbf{d}$ is strictly larger than $\frac{1}{2}$.

\noindent \increaseandreflabel{counter:sdofregion:ic_ee_sdof_region_ep}
{counter:sdofregion:ic_ee_sdof_region_ep_half_is_maximum})
If there exists an element, say $d_i$, which is equal to $\frac{1}{2}$, then, $d_j\le d_i = \frac{1}{2}$ for all $j$.

\noindent \increaseandreflabel{counter:sdofregion:ic_ee_sdof_region_ep}
{counter:sdofregion:ic_ee_sdof_region_ep_s_two_ele_then_no_larger_than_half})
If $|S|\ge2$ and $\forall s_i,s_j \in S$, where $i\neq j$, then $0<d_i=d_j\le \frac{1}{2}$.

\noindent \increaseandreflabel{counter:sdofregion:ic_ee_sdof_region_ep}
{counter:sdofregion:ic_ee_sdof_region_ep_s_is_maximum})
If $s_i\in S$, then $d_j\le d_i$ for all $j$. Equivalently, if $|S|\ge 1$ and $s_i\in S$, then $d_i=\max_{j=1,\ldots,K}d_j$. Equivalently, if $|S|\ge1$ and $d_i=\max_{j=1,\ldots,K}d_j$, then $s_i\in S$.


\noindent \increaseandreflabel{counter:sdofregion:ic_ee_sdof_region_ep}
{counter:sdofregion:ic_ee_sdof_region_ep_more_than_half_empty_T})
If $\max_{i} d_i > \frac{1}{2}$, then  $|S|\le 1$.
\end{lemma}

The proof of Lemma \ref{property:sdofregion:ic_ee_sdof_region_ep} is provided in Appendix \ref{sec:sdof_region:proof_of_property_of_ep}. In addition to the properties of the elements of the extreme points, we  also need  some results regarding the rank of the sub-matrices. It is easy to verify that a trivial necessary condition for $\rank(\mathbf{H}_J)=K$ is
$|S|+|P| +|Z|\ge K$. More formally, we have the following lemma.
\begin{lemma}
\label{lemma:sdofregion:ic_ee_sdof_region_ep_rank_constraint}
For an extreme point $\mathbf{d}$, $\rank(\mathbf{H}_J)=K$ only if
\begin{align}
\label{eqn:sdofregion:necessary_cond_rank_h_j_by_equality}
\rank(\mathbf{H}_{S\cup P}) + |Z| \ge K
\end{align}
\end{lemma}

\begin{lemma}
\label{counter:sdofregion:ic_ee_sdof_region_ep_no_t_but_p}
Let $\mathbf{d}$ be a non-zero extreme point of $D$.
If $|P| \ge 1$ and $\max_k d_k > \frac{1}{2}$, then there exists a coordinate $i_*$  such that
\begin{align}
\frac{K-1}{K} \ge d_{i_*} & =\max_{k} d_k > \frac{1}{2}
\label{eqn:sdofregion:ic_ee_sdof_region_ep_no_t_but_p_structure_d_i}
\end{align}
and a non-empty set
\begin{equation}
\label{eqn:sdofregion:ic_ee_sdof_region_ep_no_t_but_p_structure_d_j}
U' \defn \Big\{  j: d_j = 1 - d_{i_*} > 0
\Big\}
\end{equation}
with cardinality $m'\defn |U'| = |P|$ and
\begin{align}
P & = P'\defn \Big\{p_V: V = \{i_*, j\}, j\in U' \Big\}
\label{eqn:sdofregion:ic_ee_sdof_region_ep_no_t_but_p_structure_p}
\end{align}
In addition, $S$ is either empty or
\begin{align}
S & = \{s_{i_*}\}
\label{eqn:sdofregion:ic_ee_sdof_region_ep_no_t_but_p_structure_s}
\end{align}
Futhermore,
\begin{equation}
\rank(\mathbf{H}_{S\cup P}) = |P| + \indicator_{\{|S|\ge1\}}
\end{equation}
where $\indicator_{\{ \cdot \}}$ is the indicator function.
\end{lemma}

\begin{lemma}
\label{counter:sdofregion:ic_ee_sdof_region_ep_largest_v_in_t}
Let $\mathbf{d}$ be a non-zero extreme point of $D$.
If $|P| \ge 1$ and $\max_k d_k \le \frac{1}{2}$, then there exists a non-empty set
\begin{equation}
\label{eqn:sdofregion:ic_ee_sdof_region_ep_p_with_all_halves__structure_u_prime_prime}
U'' = \Big\{  i: d_i =\frac{1}{2}
\Big\}
\end{equation}
with cardinality $m''\defn |U''|$, $2\le m''\le K-1$, and
\begin{align}
\label{eqn:sdofregion:ic_ee_sdof_region_ep_p_with_all_halves__structure_p}
P = P''\defn \Big\{
p_V : V=\{k,j\},\  k\neq j,\mbox{ and } k,j \in U''
\Big\}
\end{align}
with rank
\begin{equation}
\rank(\mathbf{H}_P)= \left\{
\begin{array}{ll}
m'', & |P|>1 \\
1, & |P|=1
\end{array}
\right.
\end{equation}
In addition, $S$ is either empty or
\begin{equation}
\label{eqn:sdofregion:ic_ee_sdof_region_ep_p_with_all_halves__structure_s}
S=\Big\{s_i: i \in U''\Big\}
\end{equation}
Futhermore,
\begin{equation}
\rank(\mathbf{H}_{S \cup P}) = \left\{
\begin{array}{ll}
1, & |P|=1 \textrm{ and }   |S|=0 \\
m''  + \indicator_{\{|S|\ge1\}}, & \mbox{o.w.}
\end{array}
\right.
\end{equation}
where $\indicator_{\{ \cdot \}}$ is the indicator function.
\end{lemma}

The proofs of Lemmas \ref{lemma:sdofregion:ic_ee_sdof_region_ep_rank_constraint}, \ref{counter:sdofregion:ic_ee_sdof_region_ep_no_t_but_p}, and \ref{counter:sdofregion:ic_ee_sdof_region_ep_largest_v_in_t} are provided in Appendix \ref{sec:sdof_region:proof_of_property_of_ep}.

Now, we are ready to prove Theorem \ref{theo:sdofregion:ic_ee_sdof_region_all_ep}.

\newcommand{\subcase}[1]{{\noindent \bf Case: #1}.}


\subcase{$|Z|=K$} Clearly, $\rank(\mathbf{H}_Z)=K$ and only the zero vector satisfies
\begin{align}
\mathbf{H} \mathbf{0} & \le \mathbf{h} \\
\mathbf{H}_Z \mathbf{0} & = \mathbf{h}_Z
\end{align}
Thus, $\mathbf{0}$ is an extreme point of $D$, which is \eqn{eqn:sdofregion:ep_class_zero_vector}. Therefore, in the remaining discussion we focus on non-zero points and $|Z| < K$.


\subcase{$|P| = 0$}
Since $|Z|<K$, by Lemma \ref{lemma:sdofregion:ic_ee_sdof_region_ep_rank_constraint}, $|S|\ge1$.

If $|S|=1$, then again by Lemma \ref{lemma:sdofregion:ic_ee_sdof_region_ep_rank_constraint}, $|Z|=K-1$.
By \propertyep{counter:sdofregion:ic_ee_sdof_region_ep_s_is_maximum}, $S=\{s_i\}$ for some $i$ and $Z=\{z_j:j\neq i\}$. The extreme point $\mathbf{d}$ has the structure \eqn{eqn:sdofregion:ep_class_k_m_problem} with $p=0$.

If $|S|=K$, then by \propertyep{counter:sdofregion:ic_ee_sdof_region_ep_s_two_ele_then_no_larger_than_half}, $Z=\phi$, and the corresponding extreme point is \eqn{eqn:sdofregion:ep_class_symmetric}.

If $2\le|S|\le K-1$, due the positiveness implied by \propertyep{counter:sdofregion:ic_ee_sdof_region_ep_s_two_ele_then_no_larger_than_half} and the cardinality constraint by Lemma \ref{lemma:sdofregion:ic_ee_sdof_region_ep_rank_constraint}, the only consistent $Z$, which gives a solution for $\mathbf{H}_J \mathbf{d} = \mathbf{h}_J$, is
\begin{equation}
Z = \Big\{
z_j: s_j\not\in S
\Big\}
\end{equation}
Denote by $x$ any $d_i$ for $s_i\in S$. Then, we have
\begin{equation}
K x + (|S|-1) x = K-1
\end{equation}
which implies that
\begin{equation}
\label{eqn:sdofregion:ep_proof_p_is_empty_x_value}
x =\frac{K-1}{K-1+|S|}
\end{equation}
Since $P$ is empty, $x$ must satisfy $x<\frac{1}{2}$ due to $|S|\ge2$ and \propertyep{counter:sdofregion:ic_ee_sdof_region_ep_s_two_ele_then_no_larger_than_half}.
Substituting \eqn{eqn:sdofregion:ep_proof_p_is_empty_x_value} into  $x<\frac{1}{2}$ gives $|S|>K-1$, which contradicts the assumption $|S|<K$.
Therefore, the solution given by $\mathbf{H}_{J} \mathbf{d} = \mathbf{h}_{J}$, where $J=S\cup Z$, violates \eqn{eqn:sdofregion:ep_proof_constraints_on_j_complement}.


\subcase{$|P|\ge1$ and $\max_k d_k > \frac{1}{2}$} First of all, due to the positiveness implied by \eqn{eqn:sdofregion:ic_ee_sdof_region_ep_no_t_but_p_structure_d_i} and \eqn{eqn:sdofregion:ic_ee_sdof_region_ep_no_t_but_p_structure_d_j}, the consistent set $Z$ must satisfy
\begin{equation}
Z \subseteq \Big\{ z_k : k\not\in\{ i_*\}\cup U'
\Big\}
\end{equation}
which implies $|Z|\le K - |U'|-1 = K - |P|-1$.

If $S$ is empty, by Lemma \ref{counter:sdofregion:ic_ee_sdof_region_ep_no_t_but_p}, $
\rank(\mathbf{H}_{S\cup P}) = |P|$, which implies
\begin{equation}
\rank(\mathbf{H}_{S\cup P})  + |Z| < K
\end{equation}
which implies that $\rank(\mathbf{H}_J)<K$, which does not give any extreme point, by Lemma \ref{lemma:sdofregion:ic_ee_sdof_region_ep_rank_constraint}.

Therefore, $S$ is non-empty and determined by \eqn{eqn:sdofregion:ic_ee_sdof_region_ep_no_t_but_p_structure_s}. In addition, Lemma \ref{counter:sdofregion:ic_ee_sdof_region_ep_no_t_but_p}  gives
\begin{equation}
\rank(\mathbf{H}_{S\cup P}) = |P|+1
\end{equation}

If $|P|=K-1$, due to \eqn{eqn:sdofregion:ic_ee_sdof_region_ep_no_t_but_p_structure_d_j} and \eqn{eqn:sdofregion:ic_ee_sdof_region_ep_no_t_but_p_structure_s}, we  have the equality in \eqn{eqn:sdofregion:ic_ee_converse_1} hold for $i_*$, i.e.,
\begin{align}
K d_{i_*} + (K-1) (1 - d_{i_*}) = K-1
\end{align}
which leads to $d_{i_*}=0$ contradicting \eqn{eqn:sdofregion:ic_ee_sdof_region_ep_no_t_but_p_structure_d_i}.

Therefore, $|P|<K-1$.  Then, the consistent set $Z$ satisfying Lemma \ref{lemma:sdofregion:ic_ee_sdof_region_ep_rank_constraint} is
\begin{equation}
Z = \Big\{ z_k : k\not\in\{ i_*\}\cup U'
\Big\}
\end{equation}
In addition, due to \eqn{eqn:sdofregion:ic_ee_sdof_region_ep_no_t_but_p_structure_d_j} and \eqn{eqn:sdofregion:ic_ee_sdof_region_ep_no_t_but_p_structure_s}, we have the equality in \eqn{eqn:sdofregion:ic_ee_converse_1} hold for $i_*$, i.e.,
\begin{align}
K d_{i_*} + |P| (1 - d_{i_*}) = K-1
\end{align}
which implies that
\begin{align}
d_{i_*} = \frac{K-1-|P|}{K-|P|}
\end{align}
Since $d_{i_*}=\max_k d_k>\frac{1}{2}$, we have
\begin{equation}
|P|<K-2
\end{equation}
The solution of this choice is exactly
\eqn{eqn:sdofregion:ep_class_k_m_problem} with $1\le p<K-2$, and it satisfies \eqn{eqn:sdofregion:polytope_standard_form_for_ic_problem}.


\subcase{$|P|\ge1$ and $\max_k d_k \le \frac{1}{2}$} If $S$ is empty, then by Lemma \ref{counter:sdofregion:ic_ee_sdof_region_ep_largest_v_in_t},
\begin{equation}
\rank(\mathbf{H}_{S \cup P}) = \left\{
\begin{array}{ll}
 m'', & |P|>1 \\
1, & |P|=1
\end{array}
\right.
\end{equation}
where $m''$ is  the cardinality of  $U''$ defined in \eqn{eqn:sdofregion:ic_ee_sdof_region_ep_p_with_all_halves__structure_u_prime_prime}. Since  $m''\ge2$, for both cases, $\rank(\mathbf{H}_{S \cup P})\le m''$. Due to the positiveness of the elements in $U''$, $|Z|\le K - m''$. Therefore, by Lemma \ref{lemma:sdofregion:ic_ee_sdof_region_ep_rank_constraint}, the cardinality of $Z$ can only take the value $|Z|=K-m''$, i.e.,
\begin{equation}
d_j = 0,\quad\quad \forall j\not\in U''
\end{equation}
Also, Lemma \ref{lemma:sdofregion:ic_ee_sdof_region_ep_rank_constraint} implies that  $|P|>1$ and $m''>2$; otherwise, $\rank(\mathbf{H}_{S \cup P})+|Z| = 1 + |Z| \le 1 + K - m'' \le K-1 < K$.

Therefore, the elements in $\mathbf{d}$ are either $\frac{1}{2}$ or $0$, and
the number of $\frac{1}{2}$s is $m''$.  Note that $S$ is empty.
Therefore, for any $i\in U''$,  we must have the equality in \eqn{eqn:sdofregion:ic_ee_converse_1}  not hold, i.e.,
\begin{equation}
\frac{K}{2} + (m''-1) \frac{1}{2} < K-1
\end{equation}
which indicates that
\begin{equation}
m''<K-1
\end{equation}
Combining  with the condition $m''>2$ gives an extreme point that has the structure \eqn{eqn:sdofregion:ep_class_all_halves}.

It remains to discuss the case where $S$ is non-empty. By Lemma \ref{counter:sdofregion:ic_ee_sdof_region_ep_largest_v_in_t}, $S$ is determined by \eqn{eqn:sdofregion:ic_ee_sdof_region_ep_p_with_all_halves__structure_s} and
\begin{equation}
\rank(\mathbf{H}_{S \cup P}) =
 m''  + 1
\end{equation}

If $m''=K-1$, then the only solution is given by choosing $Z=\{z_j: j\not\in U''\}$ with $|Z|=1$, which is the structure in \eqn{eqn:sdofregion:ep_class_k_m_problem} with $p=K-2$.

If $m''<K-1$, then $\rank(\mathbf{H}_{S \cup P}) <K$. By Lemma \ref{lemma:sdofregion:ic_ee_sdof_region_ep_rank_constraint} and the positiveness implied by $U''$ with cardinality $m''$,  $Z$ must satisfy
\begin{equation}
K-m'' \ge |Z|\ge K - \rank(\mathbf{H}_{S \cup P}) = K - m'' - 1>0
\end{equation}
i.e., $Z$ is not empty and the extreme point $\mathbf{d}$ has either  $K - m'' - 1$ or $K-m''$ zero(s). On the other hand,  $\mathbf{d}$ also has in total $m''$ $\frac{1}{2}$s due to the definition of $U''$ in \eqn{eqn:sdofregion:ic_ee_sdof_region_ep_p_with_all_halves__structure_u_prime_prime}. If $|Z|=K-m''$, then the extreme point $\mathbf{d}$ has the following form
\begin{equation}
d_i = \left\{
\begin{array}{ll}
\frac{1}{2}, & i \in U''\\
0, & i \not \in U''
\end{array}
\right.
\end{equation}
and we must have the equality in \eqn{eqn:sdofregion:ic_ee_converse_1}  hold for some $i\in U''$, i.e.,
\begin{equation}
\frac{K}{2} + (m''-1)\frac{1}{2}= K-1
\end{equation}
which is not valid  since $m''<K-1$. Therefore, the equations corresponding to the selection of $J$ are inconsistent.
On the other hand, if $|Z|=K-m''-1$, then the extreme point $\mathbf{d}$ has the following form
\begin{equation}
d_i = \left\{
\begin{array}{ll}
\frac{1}{2}, & i \in U''\\
0, & z_i \in Z \\
x, & \mbox{o.w.}
\end{array}
\right.
\end{equation}
where $0<x<\frac{1}{2}$. Again, we must have the equality in \eqn{eqn:sdofregion:ic_ee_converse_1}  hold for some $i\in U''$, i.e.,
\begin{equation}
\frac{K}{2} + (m''-1)\frac{1}{2} + x = K-1
\end{equation}
which implies that
\begin{equation}
x = \frac{K-1-m''}{2}
\end{equation}
Substituting this formula into $0<x<\frac{1}{2}$ leads to
\begin{equation}
K-2< m''< K-1
\end{equation}
which is not possible since $m''$ is an integer, which completes the proof of Theorem \ref{theo:sdofregion:ic_ee_sdof_region_all_ep}.

\section{Proofs of Lemma \ref{property:sdofregion:ic_ee_sdof_region_ep} through \ref{counter:sdofregion:ic_ee_sdof_region_ep_largest_v_in_t}}
\label{sec:sdof_region:proof_of_property_of_ep}

\subsection{Proof of Lemma \ref{property:sdofregion:ic_ee_sdof_region_ep}}
We prove all the properties one by one.
\setcounter{counter:sdofregion:ic_ee_sdof_region_ep}{0}

\noindent \increaseandref{counter:sdofregion:ic_ee_sdof_region_ep})
The constraint \eqn{eqn:sdofregion:ic_ee_converse_1}  and the positiveness constraint in \eqn{eqn:sdofregion:ic_ee_converse_3} imply that for any coordinate $i$, we have
\begin{align}
K d_i \le K d_i + \sum_{j\neq i} d_j =K-1
\end{align}
i.e., $d_i\le \frac{K-1}{K}$ for any $i$. Therefore, $\max_k d_k \le \frac{K-1}{K}$.

\noindent \increaseandref{counter:sdofregion:ic_ee_sdof_region_ep})
We prove by contradiction. Assume that we have distinct coordinates, $i,j$, such that $d_i,d_j>\frac{1}{2}$ in $\mathbf{d}$.
Then, the set $V\defn \{i,j\}$ with $|V|=2$ violates the constraint in \eqn{eqn:sdofregion:ic_ee_converse_2}. Therefore, this contradiction implies that at most one element, if  any, in  $\mathbf{d}$ is strictly larger than $\frac{1}{2}$.

\noindent \increaseandref{counter:sdofregion:ic_ee_sdof_region_ep})
Similarly, assume that there exists a $j$ such that $d_j>\frac{1}{2}$. Since $d_i=\frac{1}{2}$ by assumption, $d_i+d_j>1$, which violates  constraint \eqn{eqn:sdofregion:ic_ee_converse_2}. This implies that $d_j\le d_i=\frac{1}{2}$ for all $j$.

\noindent \increaseandref{counter:sdofregion:ic_ee_sdof_region_ep})
Let $ i,j \in S$ and $i\neq j$. Due to the definition of $S$, $s_i,s_j\in S$, i.e., from \eqn{eqn:sdofregion:ic_ee_defn_s}
\begin{align}
 K d_i + d_j + \sum_{k=1,k\neq i,j}^K d_k  & = K-1 \\
 K d_j + d_i + \sum_{k=1,k\neq i,j}^K d_k  & = K-1
\end{align}
which implies $(K-1) d_i = (K-1) d_j$. Since $K-1>0$, $d_i=d_j$.
Furthermore, due to  \ipropertyep{counter:sdofregion:ic_ee_sdof_region_ep_at_most_one_larger_than_half}, both are no larger than $\frac{1}{2}$, and due to \ipropertyep{counter:sdofregion:ic_ee_sdof_region_ep_half_is_maximum}, for any $k$, $d_k\le d_i$. If $d_i=0$, then the point $\mathbf{d}$ is the zero vector, which contradicts the assumption that $\mathbf{d}$ is a non-zero extreme point in $D$. Therefore, $d_i=d_j>0$.

\noindent \increaseandref{counter:sdofregion:ic_ee_sdof_region_ep})
The three equivalent statements in this property are simply from three different perspectives addressing the same fact that the coordinates of $\mathbf{d}$, which are associated with the elements in $S$, are the most significant coordinates. We will prove the first statement and then prove the equivalence of all three statements.

We prove the first statement of \ipropertyep{counter:sdofregion:ic_ee_sdof_region_ep_s_is_maximum} by contraction.  Assume that there exists a $j$ such that $d_j>d_i$. Then, consider the following expression (for $K\ge3$)
\begin{align}
K d_j +  d_i + \sum_{k=1,k\neq i,j}^K d_k
& = d_j +  d_i + (K-1) d_j  + \sum_{k=1,k\neq i,j}^K d_k  \\
& > d_j +  d_i + (K-1) d_i  + \sum_{k=1,k\neq i,j}^K d_k  \\
& = K d_i  + \sum_{k=1,k\neq i}^K d_k  \\
& = K-1
\end{align}
where the last equality is due to the assumption $s_i\in S$. This result violates the constraint \eqn{eqn:sdofregion:ic_ee_converse_1}. Therefore, for all $j$, $d_j\le d_i$.

Next, we prove the second statement of \ipropertyep{counter:sdofregion:ic_ee_sdof_region_ep_s_is_maximum} using the first statement. This is trivially true because the assumption $|S|\ge1$ and $s_i\in S$ imply that, by the first statement, $d_i\ge d_j$ for all $j$, i.e.,  $d_i=\max_j d_j$.

Then, we prove the third statement of \ipropertyep{counter:sdofregion:ic_ee_sdof_region_ep_s_is_maximum} using the second statement. By assumption, let $d_i=\max_{k} d_k$. However, assume that $s_i\not\in S$. This implies that  there exists another coordinate $j$, $j\neq i$ such that $s_j\in S$ (since $|S|\ge1$) and thereby by the second statement $d_j=\max_{k} d_k=d_i$. Then, consider
\begin{align}
K d_i + d_j + \sum_{k=1,k\neq i,j}^K d_k = K d_j + d_i + \sum_{k=1,k\neq i,j}^K d_k = K-1
\end{align}
where the last equality is due to $s_j\in S$. This implies that $s_i$ must belong to $S$ by definition in \eqn{eqn:sdofregion:ic_ee_defn_s}, i.e., $s_i\in S$, which contradicts the assumption that $s_i\not\in S$.

Finally, we prove the first statement of \ipropertyep{counter:sdofregion:ic_ee_sdof_region_ep_s_is_maximum} using the third statement. We prove this by contradiction as well. As stated in the condition of the first statement, $s_i\in S$, this means $|S|\ge1$. Assume that there exists at least one element which is strictly larger than $d_i$. Choose the largest one among them and denote it by $d_j$. Clearly, $j\neq i$ and $d_j=\max_k d_k>d_i$. By the  third statement, $s_j\in S$. Then, $|S|\ge 2$ and by \ipropertyep{counter:sdofregion:ic_ee_sdof_region_ep_s_two_ele_then_no_larger_than_half} $d_i=d_j$, which contradicts the assumption $d_j>d_i$.



\noindent \increaseandref{counter:sdofregion:ic_ee_sdof_region_ep})
We prove $|S|\le 1$  by contraction. Assume that $|S|\ge 2$. Due to \ipropertyep{counter:sdofregion:ic_ee_sdof_region_ep_s_two_ele_then_no_larger_than_half} and the second statement of \ipropertyep{counter:sdofregion:ic_ee_sdof_region_ep_s_is_maximum}, we have two distinct $j,k\in S$ such that $\frac{1}{2}\ge d_j=d_k=\max_i d_i > \frac{1}{2}$, which leads to a contradiction. Thus, $|S|\le 1$.

\subsection{Proof of Lemma \ref{lemma:sdofregion:ic_ee_sdof_region_ep_rank_constraint}}
It is straightforward that
\begin{equation}
\rank(\mathbf{H}_Z) = |Z|
\end{equation}
since there are in total $|Z|$ $1$s in the sub-matrix $\mathbf{H}_Z$ and the row index and column index of any two $1$s are different. Since $(S\cup P) \cap Z = \phi$, we have
\begin{align}
K = \rank(\mathbf{H}_{J}) = \rank(\mathbf{H}_{S \cup P \cup Z}) \le
\rank(\mathbf{H}_{S\cup P}) + \rank(\mathbf{H}_{Z})
\end{align}

\subsection{Proof of Lemma \ref{counter:sdofregion:ic_ee_sdof_region_ep_no_t_but_p}}

If $|P|=1$, then $P=\{p_V\}$ for a unique $V=\{i,j\}$ with $|V|=2$.
If $d_i= d_j$, then $d_i= d_j=\frac{1}{2}$ and $\max_k d_k\le \frac{1}{2}$ due to \propertyep{counter:sdofregion:ic_ee_sdof_region_ep_half_is_maximum}, which contradicts the condition $\max_k d_k>\frac{1}{2}$.
Therefore, $d_i\neq d_j$. Without loss of generality, let $d_i>d_j$, then $d_i>\frac{1}{2}$ and $i$ is the $i_*$ required in Lemma \ref{counter:sdofregion:ic_ee_sdof_region_ep_no_t_but_p} due to \propertyep{counter:sdofregion:ic_ee_sdof_region_ep_at_most_one_larger_than_half}. By \propertyep{counter:sdofregion:ic_ee_sdof_region_ep_less_than_k_minus_1_over_}, $d_j =1 - d_{i_*}>0$, thus  $j\in U'$. If there exists any $k$, $k\neq j$, such that $d_k=1 - d_{i_*}$, then clearly $V'\defn\{i_*,k\} \neq V$, but $p_{V'}\in P$, which contradicts the condition $|P|=1$. Hence, $U'=\{j\}$ and $P$ satisfies \eqn{eqn:sdofregion:ic_ee_sdof_region_ep_no_t_but_p_structure_p}.

If $|P|\ge2$, assume that $V_1=\{i,j\}$, $V_2=\{x,y\}$, $V_1\neq V_2$, and $p_{V_1}, p_{V_2}\in P$.
Without loss of generality, let $d_i=\max_{k\in\{i,j,x,y\}} d_k$.
If $d_i<\frac{1}{2}$, then $d_j+d_i < 1$, which  contradicts $p_{V_1}\in P$.
If $d_i=\frac{1}{2}$, then due to \propertyep{counter:sdofregion:ic_ee_sdof_region_ep_half_is_maximum}, $\max_k d_k\le\frac{1}{2}$, which contradicts the condition $\max_k d_k > \frac{1}{2}$.
Therefore, $d_i=\max_{k\in\{i,j,x,y\}} d_k > \frac{1}{2}$ and $i$ is the $i_*$ required in Lemma \ref{counter:sdofregion:ic_ee_sdof_region_ep_no_t_but_p}.
For any $p_V\in P$, let $V=\{a,b\}$ and assume $d_a\ge d_b$. If $d_a=\frac{1}{2}$, this leads to a contradiction of $d_{i_*}>\frac{1}{2}$ due to \propertyep{counter:sdofregion:ic_ee_sdof_region_ep_half_is_maximum}.
Thus, $d_a>\frac{1}{2}$. Due to  \propertyep{counter:sdofregion:ic_ee_sdof_region_ep_at_most_one_larger_than_half}, the coordinate $a$ must be $i_*$, i.e., $a=i_*$.
Then, $d_b=1-d_{i_*}>0$ and that is true for any $p_V$. Hence, $|P|=|U'|$ and \eqn{eqn:sdofregion:ic_ee_sdof_region_ep_no_t_but_p_structure_p} are trivially true.

If $S$ is empty, we have a sub-matrix which has the form (by removing all columns containing all zeros and rearranging the columns)
\begin{equation}
\mathbf{H}_{S \cup P} = \mathbf{H}_{P} \stackrel{\cdot}{=} \left[
\begin{array}{cccccc}
1 & 1 & 0 & 0 & \hdots & 0 \\
1 & 0 & 1 & 0 & \hdots & 0 \\
\vdots & \vdots & \vdots & \vdots & \ddots & \vdots \\
1 & 0 & 0 & 0 & \hdots & 1
\end{array}
\label{eqn:sdofregion:proof_of_p_with_larger_than_half_matrix_h_P}
\right]
\end{equation}
where the number of rows is $|P|=|U'|$, the  number of columns is $|P|+1$, and the index of the first column corresponds to $i_*$ and the indices of other columns correspond to  $U'$ defined in \eqn{eqn:sdofregion:ic_ee_sdof_region_ep_no_t_but_p_structure_d_j}.
Therefore, $\rank(\mathbf{H}_{S \cup P})=|P|$ and we are done.

If $S$ is not empty, due to \eqn{eqn:sdofregion:ic_ee_sdof_region_ep_no_t_but_p_structure_d_i} and \propertyep{counter:sdofregion:ic_ee_sdof_region_ep_more_than_half_empty_T}, $|S|=1$.
Furthermore, due to \propertyep{counter:sdofregion:ic_ee_sdof_region_ep_s_is_maximum}, $s_{i_*}\in S$, which is \eqn{eqn:sdofregion:ic_ee_sdof_region_ep_no_t_but_p_structure_s}.
Note that $\mathbf{H}_{S}$ is a $K$-length row vector containing no zeros. If $|P|+1<K$, then $\mathbf{H}_{S}$ has more columns than
the sub-matrix on the right hand side of \eqn{eqn:sdofregion:proof_of_p_with_larger_than_half_matrix_h_P}. $\mathbf{H}_{S \cup P}=|P|+1$ is true. If $|P|+1=K$, then
\begin{equation}
\mathbf{H}_{P \cup S} = \left[
\begin{array}{cccccc}
1 & 1 & 0 & 0 & \hdots & 0 \\
1 & 0 & 1 & 0 & \hdots & 0 \\
\vdots & \vdots & \vdots & \vdots & \ddots & \vdots \\
1 & 0 & 0 & 0 & \hdots & 1 \\
K & 1 & 1 & 1 & \hdots & 1 \\
\end{array}
\right] \defn M(K)
\label{eqn:sdofregion:proof_of_p_with_larger_than_half_matrix_h_P_and_S}
\end{equation}
where $M(n)$ is $n\times n$ square matrix as in \eqn{eqn:sdofregion:proof_of_p_with_larger_than_half_matrix_h_P_and_S}, where $n\ge 2$. Therefore, $\mathbf{H}_{P \cup S}=M(K)$.
If we denote $f(n)\defn \det [M(n)]$, then  it is easy to write the recursive formula as
\begin{align}
f(n) & = (-1)^{n} - f(n-1), \quad\quad n\ge 3 \\
f(2) & = 1 - K
\end{align}
which gives that $f(n) = (-1)^{n}(n-K-1)$, i.e., $\det \mathbf{H}_{P \cup S} = \det M(K)=(-1)^{K+1} \neq0$ and $\rank(\mathbf{H}_{P \cup S})=|P|+1=K$, which completes the proof.

\subsection{Proof of Lemma \ref{counter:sdofregion:ic_ee_sdof_region_ep_largest_v_in_t}}

If $\max_k d_K<\frac{1}{2}$, then $|P|=0$, which contradicts the assumption $|P|\ge 1$.
Therefore, $\max_k d_K=\frac{1}{2}$, which implies $|U''|\ge 1$.
Assume that $i_*\in U''$. Due to \propertyep{counter:sdofregion:ic_ee_sdof_region_ep_half_is_maximum}, $d_j\le d_{i_*}=\frac{1}{2}$ for all $j$.
If $\max_{k\neq i_*} d_k<\frac{1}{2}$, then we cannot find a set $V$ such that $|V|=2$ and $\sum_{k\in V} d_k=1$, i.e., $|P|=0$, which contradicts the
assumption $|P|\ge 1$. Thus, $m''\defn |U''|\ge 2$. On the other hand, if $ m''=K$, by definition of $U''$, all elements in $\mathbf{d}$ are $\frac{1}{2}$, which violates the constraint \eqn{eqn:sdofregion:ic_ee_converse_1}. Therefore, $m''\le K-1$.

Next, $P''$ defined in
\eqn{eqn:sdofregion:ic_ee_sdof_region_ep_p_with_all_halves__structure_p}
satisfies $P''\subseteq P$.
On the other hand, for any coordinate pair $(k',j')$ such that $k'\neq
j'$ and $p_{\{k',j'\}}\in P$,  since $d_{k'},d_{j'}\le
\frac{1}{2}$, we must have $d_{k'}=d_{j'}=\frac{1}{2}$, and by
definition of $U''$, $k',j'\in U''$, which implies $p_{\{k',j'\}}\in
P''$. Therefore, $P=P''$.

If $S$ is empty, then $\mathbf{H}_P = 1$ if $|P|=1$ and we are done. If $S$ is empty but $|P|>1$, the index set of the columns of $\mathbf{H}_P$, which contains nonzero elements, is $U''$ due to \eqn{eqn:sdofregion:ic_ee_sdof_region_ep_p_with_all_halves__structure_p}.
Therefore, $\rank(\mathbf{H}_P)\le |U''|=m''$. In order to study the rank, we remove the columns containing all zeros and rearrange the columns. Assume that
\begin{equation}
U''=\Big\{i_1, i_2, \ldots, i_{m''}\Big\}
\end{equation}
where $i_1=i_*$. Then, consider a  $m''\times m''$ sub-matrix of $\mathbf{H}_P$
\begin{equation}
\mathbf{H}_{J''} \stackrel{\cdot}{=}
\left[
\begin{array}{ccccccc}
1 & 1 & 0 & 0 & 0 & \hdots & 0 \\
1 & 0 & 1 & 0 & 0 & \hdots & 0 \\
1 & 0 & 0 & 1 & 0 & \hdots & 0 \\
\vdots & \vdots & \vdots & \vdots & \vdots & \ddots & \vdots \\
1 & 0 & 0 & 0 & 0 & \hdots & 1 \\
0 & 1 & 1 & 0 & 0 & \hdots & 0
\end{array}
\right]
\end{equation}
where
\begin{equation}
\label{eqn:sdofregion:ic_ee_sdof_region_ep_p_with_all_halves__structure_j_prime_prime}
J'' \defn \{p_{V}: V=\{i_*,i_j\}, j=2,\ldots,m''\} \cup \{p_{\{i_2,i_3\}}\}\subseteq P
\end{equation}
It is easy to verify that $\det \mathbf{H}_{J''} = (-1)^{m''} \times 2 \neq 0 $, therefore $\rank(\mathbf{H}_{J''}) = m''$, i.e., $\rank(\mathbf{H}_{P}) = m''$.
This completes the proof of the case where $S$ is empty.

Assume that $|S|\ge1$, by \propertyep{counter:sdofregion:ic_ee_sdof_region_ep_s_is_maximum}, $S$ must have the form of \eqn{eqn:sdofregion:ic_ee_sdof_region_ep_p_with_all_halves__structure_s}. If $|P|=1$, $m''=|U''|=2$. Then, the $3 \times K$ matrix $\mathbf{H}_{P\cup S}$ must have the structure
\begin{equation}
\mathbf{H}_{P\cup S} {=}
\left[
\begin{array}{ccccccc}
1 & 1 & 0 & 0 & 0 & \hdots & 0 \\
K & 1 & 1 & 1 & 1 & \hdots & 1 \\
1 & K & 1 & 1 & 1 & \hdots & 1
\end{array}
\right]
\end{equation}
where the indices of the first two columns belong to  $U''$. Clearly, $\mathbf{H}_{P\cup S}=3=m''+1$ since $m''=2$.

If $|P|>1$, by using the $J''$ in \eqn{eqn:sdofregion:ic_ee_sdof_region_ep_p_with_all_halves__structure_j_prime_prime} and the condition $m''\le K-1$, we have
\begin{equation}
\mathbf{H}_{J''\cup S} = \left[
\begin{array}{ccccccc|cccc}
1 & 1 & 0 & 0 & \hdots & 0 & 0 & 0 & 0 & \hdots & 0 \\
1 & 0 & 1 & 0 & \hdots & 0 & 0 & 0 & 0 & \hdots & 0  \\
\vdots & \vdots & \vdots &\vdots & \ddots & \vdots & \vdots  &   \vdots &  \vdots & \ddots & \vdots  \\
1 & 0 & 0 & 0 & \hdots & 0 & 1 & 0 & 0 & \hdots & 0  \\
0 & 1 & 1 & 0 & \hdots & 0 & 0 & 0 & 0 & \hdots & 0  \\ \hline
K & 1 & 1 & 1 & \hdots &1 & 1 & 1 & 1 & \hdots & 1  \\
1 & K & 1 & 1 & \hdots & 1 &1 & 1 & 1 & \hdots & 1  \\
\vdots & \vdots & \vdots &\vdots & \ddots & \vdots &  \vdots  &   \vdots &  \vdots & \ddots & \vdots  \\
1 & 1 & 1 & 1 & \hdots & K & 1 & 1 & 1 & \hdots & 1 \\
1 & 1 & 1 & 1 & \hdots & 1 & K & 1 & 1 & \hdots & 1
\end{array}
\right]
\end{equation}
Due to  \cite[Section 2.2, Problem 7]{matrix_theory_basic_results_and_techniques},
\begin{equation}
\rank(\mathbf{H}_{P\cup S})  =\rank(\mathbf{H}_{J'' \cup S}) = \rank(\mathbf{H}_{J''}) +1 = m'' + 1
\end{equation}
which completes the proof.


\begin{thebibliography}{10}

\bibitem{xie_ulukus_isit_2013_mac}
J.~Xie and S.~Ulukus.
\newblock Secure degrees of freedom of the {G}aussian multiple access wiretap
  channel.
\newblock In {\em IEEE International Symposium on Information Theory},
  Istanbul, Turkey, July 2013.

\bibitem{xie_sdof_networks_in_prepare}
J.~Xie and S.~Ulukus.
\newblock Secure degrees of freedom of one-hop wireless networks.
\newblock {\it IEEE Trans. on Information Theory}, to appear. Also available at
  [arXiv:1209.5370].

\bibitem{xie_ulukus_isit_2013_kic}
J.~Xie and S.~Ulukus.
\newblock Unified secure {D}o{F} analysis of ${K}$-user {G}aussian interference
  channels.
\newblock In {\em IEEE International Symposium on Information Theory},
  Istanbul, Turkey, July 2013.

\bibitem{xie_unified_kic}
J.~Xie and S.~Ulukus.
\newblock Secure degrees of freedom of ${K}$-user {G}aussian interference
  channels: A unified view.
\newblock Submitted to {\it IEEE Trans. on Information Theory}, May 2013. Also
  available at [arXiv:1305.7214].

\bibitem{Shannon:1949}
C.~E. Shannon.
\newblock Communication theory of secrecy systems.
\newblock {\em Bell Syst. Tech. J.}, 28(4):656--715, October 1949.

\bibitem{wyner}
A.~D. Wyner.
\newblock The wiretap channel.
\newblock {\em Bell Syst. Tech. J.}, 54(8):1355--1387, January 1975.

\bibitem{csiszar}
I.~Csiszar and J.~Korner.
\newblock Broadcast channels with confidential messages.
\newblock {\em IEEE Trans. Inf. Theory}, 24(3):339--348, May 1978.

\bibitem{gaussian}
S.~K. Leung-Yan-Cheong and M.~E. Hellman.
\newblock {G}aussian wiretap channel.
\newblock {\em IEEE Trans. Inf. Theory}, 24(4):451--456, July 1978.

\bibitem{secrecy_ic}
R.~Liu, I.~Maric, P.~Spasojevic, and R.~D. Yates.
\newblock Discrete memoryless interference and broadcast channels with
  confidential messages: secrecy rate regions.
\newblock {\em IEEE Trans. Inf. Theory}, 54(6):2493--2507, June 2008.

\bibitem{xu_bounds_bc_cm_it_09}
J.~Xu, Y.~Cao, and B.~Chen.
\newblock Capacity bounds for broadcast channels with confidential messages.
\newblock {\em IEEE Trans. Inf. Theory}, 55(10):4529--4542, October 2009.

\bibitem{fading1}
A.~Khisti, A.~Tchamkerten, and G.~W. Wornell.
\newblock Secure broadcasting over fading channels.
\newblock {\em IEEE Trans. Inf. Theory}, 54(6):2453--2469, June 2008.

\bibitem{bagherikaram_bc_2008}
G.~Bagherikaram, A.~S. Motahari, and A.~K. Khandani.
\newblock Secure broadcasting: The secrecy rate region.
\newblock In {\em 46th Annual Allerton Conference on Communications, Control
  and Computing}, Monticello, IL, September 2008.

\bibitem{ersen_bc_asilomar_08}
E.~Ekrem and S.~Ulukus.
\newblock On secure broadcasting.
\newblock In {\em 42nd Asilomar Conference on Signals, Systems and Computers},
  Pacific Grove, October 2008.

\bibitem{ersen_eurasip_2009}
E.~Ekrem and S.~Ulukus.
\newblock Secrecy capacity of a class of broadcast channels with an
  eavesdropper.
\newblock {\em EURASIP Journal on Wireless Communications and Networking,
  Special Issue on Wireless Physical Layer Security}, 2009(824235), March 2009.

\bibitem{he_outerbound_gic_cm_ciss_09}
X.~He and A.~Yener.
\newblock A new outer bound for the {G}aussian interference channel with
  confidential messages.
\newblock In {\em 43rd Annual Conference on Information Sciences and Systems},
  Baltimore, MD, March 2009.

\bibitem{koyluoglu_ic_external}
O.~O. Koyluoglu and H.~{El Gamal}.
\newblock Cooperative encoding for secrecy in interference channels.
\newblock {\em {IEEE Trans. Inf. Theory}}, 57(9):5681--5694, September 2011.

\bibitem{tekin_gmac_w}
E.~Tekin and A.~Yener.
\newblock The {G}aussian multiple access wire-tap channel.
\newblock {\em IEEE Trans. Inf. Theory}, 54(12):5747--5755, December 2008.

\bibitem{cooperative_jamming}
E.~Tekin and A.~Yener.
\newblock The general {G}aussian multiple-access and two-way wiretap channels:
  Achievable rates and cooperative jamming.
\newblock {\em IEEE Trans. Inf. Theory}, 54(6):2735--2751, June 2008.

\bibitem{ersen_mac_allerton_08}
E.~Ekrem and S.~Ulukus.
\newblock On the secrecy of multiple access wiretap channel.
\newblock In {\em 46th Annual Allerton Conference on Communication, Control and
  Computing}, Monticello, IL, September 2008.

\bibitem{liang_mac_cm_08}
Y.~Liang and H.~V. Poor.
\newblock Multiple-access channels with confidential messages.
\newblock {\em IEEE Trans. Inf. Theory}, 54(3):976--1002, March 2008.

\bibitem{ersen_mac_book_chapter}
E.~Ekrem and S.~Ulukus.
\newblock Cooperative secrecy in wireless communications.
\newblock {\em Securing Wireless Communications at the Physical Layer}, W.
  Trappe and R. Liu, Eds., Springer-Verlag, 2009.

\bibitem{wiretap_channel_with_one_helper}
X.~Tang, R.~Liu, P.~Spasojevic, and H.V. Poor.
\newblock The {G}aussian wiretap channel with a helping interferer.
\newblock In {\em IEEE International Symposium on Information Theory}, Toronto,
  Canada, July 2008.

\bibitem{relay_1}
Y.~Oohama.
\newblock Relay channels with confidential messages.
\newblock {\it IEEE Trans. Inf. Theory, Special issue on Information Theoretic
  Security}, submitted Nov 2006. Also available at [arXiv:cs/0611125v7].

\bibitem{relay_2}
L.~Lai and H.~{El Gamal}.
\newblock The relay-eavesdropper channel: cooperation for secrecy.
\newblock {\em IEEE Trans. Inf. Theory}, 54(9):4005--4019, September 2008.

\bibitem{relay_3}
M.~Yuksel and E.~Erkip.
\newblock The relay channel with a wiretapper.
\newblock In {\em 41st Annual Conference on Information Sciences and Systems},
  Baltimore, MD, March 2007.

\bibitem{relay_4}
M.~Bloch and A.~Thangaraj.
\newblock Confidential messages to a cooperative relay.
\newblock In {\em IEEE Information Theory Workshop}, Porto, Portugal, May 2008.

\bibitem{he_untrusted_relay}
X.~He and A.~Yener.
\newblock Cooperation with an untrusted relay: A secrecy perspective.
\newblock {\em IEEE Trans. Inf. Theory}, 56(8):3807--3827, August 2010.

\bibitem{ersen_crbc_2011}
E.~Ekrem and S.~Ulukus.
\newblock Secrecy in cooperative relay broadcast channels.
\newblock {\em IEEE Trans. Inf. Theory}, 57(1):137--155, January 2011.

\bibitem{compound_wiretap_channel}
Y.~Liang, G.~Kramer, H.~V. Poor, and S.~Shamai (Shitz).
\newblock Compound wiretap channels.
\newblock {\em EURASIP Journal on Wireless Communications and Networking,
  Special Issue on Wireless Physical Layer Security}, 2009(142374), March 2009.

\bibitem{ersen_ulukus_degraded_compound}
E.~Ekrem and S.~Ulukus.
\newblock Degraded compound multi-receiver wiretap channels.
\newblock {\em IEEE Trans. Inf. Theory}, 58(9):5681--5698, September 2012.

\bibitem{he_k_gic_cm_09}
X.~He and A.~Yener.
\newblock ${K}$-user interference channels: Achievable secrecy rate and degrees
  of freedom.
\newblock In {\em IEEE Information Theory Workshop on Networking and
  Information Theory}, Volos, Greece, June 2009.

\bibitem{koyluoglu_k_user_gic_secrecy}
O.~O. Koyluoglu, H.~{El Gamal}, L.~Lai, and H.~V. Poor.
\newblock Interference alignment for secrecy.
\newblock {\em {IEEE Trans. Inf. Theory}}, 57(6):3323--3332, June 2011.

\bibitem{xie_k_user_ia_compound}
J.~Xie and S.~Ulukus.
\newblock Real interference alignment for the ${K}$-user {G}aussian
  interference compound wiretap channel.
\newblock In {\em 48th Annual Allerton Conference on Communication, Control and
  Computing}, Monticello, IL, September 2010.

\bibitem{secrecy_ia_new}
X.~He and A.~Yener.
\newblock Providing secrecy with structured codes: Two-user {G}aussian
  channels.
\newblock {\em IEEE Trans. Inf. Theory}, 60(4):2121--2138, April 2014.

\bibitem{xiang_he_thesis}
X.~He.
\newblock {\em Cooperation and information theoretic security in wireless
  networks}.
\newblock Ph.{D}. dissertation, Pennsylvania State University, 2010.

\bibitem{secrecy_ia5}
G.~Bagherikaram, A.~S. Motahari, and A.~K. Khandani.
\newblock On the secure degrees-of-freedom of the multiple-access-channel.
\newblock {\it IEEE Trans. Inf. Theory}, submitted March 2010. Also available
  at [arXiv:1003.0729].

\bibitem{raef_mac_it_12}
R.~Bassily and S.~Ulukus.
\newblock Ergodic secret alignment.
\newblock {\em IEEE Trans. Inf. Theory}, 58(3):1594--1611, March 2012.

\bibitem{secrecy_ia1}
T.~Gou and S.~A. Jafar.
\newblock On the secure {D}egrees of {F}reedom of wireless {X} networks.
\newblock In {\em 46th Annual Allerton Conference on Communication, Control and
  Computing}, Monticello, IL, September 2008.

\bibitem{interference_alignment_compound_channel}
A.~Khisti.
\newblock Interference alignment for the multiantenna compound wiretap channel.
\newblock {\em IEEE Trans. Inf. Theory}, 57(5):2976--2993, May 2011.

\bibitem{xie_gwch_allerton}
J.~Xie and S.~Ulukus.
\newblock Secure degrees of freedom of the {G}aussian wiretap channel with
  helpers.
\newblock In {\em 50th Annual Allerton Conference on Communication, Control and
  Computing}, Monticello, IL, October 2012.

\bibitem{xie_blind_cj_ciss_2013}
J.~Xie and S.~Ulukus.
\newblock Secure degrees of freedom of the {G}aussian wiretap channel with
  helpers and no eavesdropper {CSI}: Blind cooperative jamming.
\newblock In {\em Conference on Information Sciences and Systems}, Baltimore,
  MD, March 2013.

\bibitem{xie_layered_network_journal}
J.~Xie and S.~Ulukus.
\newblock Sum secure degrees of freedom of two unicast layered wireless
  networks.
\newblock {\em {IEEE Jour. on Selected Areas in Comm.}}, 31(9):1931--1943,
  September 2013.

\bibitem{khisti_arti_noise_alignment}
A.~Khisti and D.~Zhang.
\newblock Artificial-noise alignment for secure multicast using multiple
  antennas.
\newblock {\em IEEE Communications Letters}, 17(8):1568--1571, August 2013.

\bibitem{mohamed_yener_allerton_2013}
M.~Nafea and A.~Yener.
\newblock How many antennas does a cooperative jammer need for achieving the
  degrees of freedom of multiple antenna {G}aussian channels in the presence of
  an eavesdropper?
\newblock In {\em 51st Annual Allerton Conference on Communication, Control and
  Computing}, Monticello, IL, October 2013.

\bibitem{mohamed_yener_globalsip_2013}
M.~Nafea and A.~Yener.
\newblock Degrees of freedom of the single antenna gaussian wiretap channel
  with a helper irrespective of the number of antennas at the eavesdropper.
\newblock In {\em IEEE GlobalSIP Symposium on Cyber-Security and Privacy},
  Austin, TX, December 2013.

\bibitem{real_inter_align_exploit}
A.~S. Motahari, S.~Oveis-Gharan, M.~A. Maddah-Ali, and A.~K. Khandani.
\newblock Real interference alignment: Exploiting the potential of single
  antenna systems.
\newblock {\it IEEE Trans. Inf. Theory}, submitted November 2009. Also
  available at [arXiv:0908.2282].

\bibitem{real_inter_align}
A.~S. Motahari, S.~Oveis-Gharan, and A.~K. Khandani.
\newblock Real interference alignment with real numbers.
\newblock {\it IEEE Trans. Inf. Theory}, submitted August 2009. Also available
  at [arXiv:0908.1208].

\bibitem{tse_polymatroid}
D.~Tse and S.~V. Hanly.
\newblock Multiaccess fading channels{-}{P}art {I}: Polymatroid structure,
  optimal resource allocation and throughput capacities.
\newblock {\em IEEE Trans. Inf. Theory}, 44(7):2796--2815, November 1998.

\bibitem{convex_polytopes}
B.~Grunbaum.
\newblock {\em Convex Polytopes}.
\newblock Springer, second edition, 2003.

\bibitem{multiplexing_gain_of_networks}
A.~Host-Madsen and A.~Nosratinia.
\newblock The multiplexing gain of wireless networks.
\newblock In {\em IEEE International Symposium on Information Theory},
  Adelaide, Australia, September 2005.

\bibitem{interference_alignment}
V.~R. Cadambe and S.~A. Jafar.
\newblock Interference alignment and degrees of freedom of the ${K}$-user
  interference channel.
\newblock {\em IEEE Trans. Inf. Theory}, 54(8):3425--3441, August 2008.

\bibitem{linear_optimization_and_extension}
M.~Padberg.
\newblock {\em Linear Optimization and Extensions}.
\newblock Springer, second edition, 1999.

\bibitem{etkin_discontinuous_2009}
R.~H. Etkin and E.~Ordentlich.
\newblock The degrees-of-freedom of the {$K$}-user {G}aussian interference
  channel is discontinuous at rational channel coefficients.
\newblock {\em IEEE Trans. Inf. Theory}, 55(11):4932--4946, November 2009.

\bibitem{matrix_theory_basic_results_and_techniques}
F.~Zhang.
\newblock {\em Matrix Theory: Basic Results and Techniques}.
\newblock Springer, second edition, 2011.

\end{thebibliography}
\end{document}